\documentclass[]{elsarticle}
\usepackage[dvipsnames]{xcolor}

\makeatletter
\def\ps@pprintTitle{%
 \let\@oddhead\@empty
 \let\@evenhead\@empty
 \def\@oddfoot{}%
 \let\@evenfoot\@oddfoot}
\makeatother

\usepackage{moreverb}

\usepackage[colorlinks,bookmarksopen,bookmarksnumbered,citecolor=red,urlcolor=red]{hyperref}

\newcommand\BibTeX{{\rmfamily B\kern-.05em \textsc{i\kern-.025em b}\kern-.08em
\kern-.1667em\lower.7ex\hbox{E}\kern-.125emX}}

\usepackage{geometry}

\newcommand{\tdprod}%
     {\,{\scriptscriptstyle \stackrel{3}{\bullet}}\,}

\renewcommand{\max} {\ensuremath{\operatorname{max}}}
\renewcommand{\min} {\ensuremath{\operatorname{min}}}

\newcommand{\x} {\ensuremath{\vec{x}}}



\makeatletter
\DeclareFontFamily{OMX}{MnSymbolE}{}
\DeclareSymbolFont{myLargesymbols}  {OMX}{MnSymbolE}{m}{n}
\SetSymbolFont{myLargesymbols}{bold}{OMX}{MnSymbolE}{b}{n}
\DeclareFontShape{OMX}{MnSymbolE}{m}{n}{
    <-6>  MnSymbolE5
   <6-7>  MnSymbolE6
   <7-8>  MnSymbolE7
   <8-9>  MnSymbolE8
   <9-10> MnSymbolE9
  <10-12> MnSymbolE10
  <12->   MnSymbolE12}{}
\DeclareFontShape{OMX}{MnSymbolE}{b}{n}{
    <-6>  MnSymbolE-Bold5
   <6-7>  MnSymbolE-Bold6
   <7-8>  MnSymbolE-Bold7
   <8-9>  MnSymbolE-Bold8
   <9-10> MnSymbolE-Bold9
  <10-12> MnSymbolE-Bold10
  <12->   MnSymbolE-Bold12}{}
  
\DeclareMathSymbol{\downbrace}    {\mathord}{myLargesymbols}{'251}
\DeclareMathSymbol{\downbraceg}   {\mathord}{myLargesymbols}{'252}
\DeclareMathSymbol{\downbracegg}  {\mathord}{myLargesymbols}{'253}
\DeclareMathSymbol{\downbraceggg} {\mathord}{myLargesymbols}{'254}
\DeclareMathSymbol{\downbracegggg}{\mathord}{myLargesymbols}{'255}
\DeclareMathSymbol{\upbrace}      {\mathord}{myLargesymbols}{'256}
\DeclareMathSymbol{\upbraceg}     {\mathord}{myLargesymbols}{'257}
\DeclareMathSymbol{\upbracegg}    {\mathord}{myLargesymbols}{'260}
\DeclareMathSymbol{\upbraceggg}   {\mathord}{myLargesymbols}{'261}
\DeclareMathSymbol{\upbracegggg}  {\mathord}{myLargesymbols}{'262}
\DeclareMathSymbol{\braceld}      {\mathord}{myLargesymbols}{'263}
\DeclareMathSymbol{\bracelu}      {\mathord}{myLargesymbols}{'264}
\DeclareMathSymbol{\bracerd}      {\mathord}{myLargesymbols}{'265}
\DeclareMathSymbol{\braceru}      {\mathord}{myLargesymbols}{'266}
\DeclareMathSymbol{\bracemd}      {\mathord}{myLargesymbols}{'267}
\DeclareMathSymbol{\bracemu}      {\mathord}{myLargesymbols}{'270}
\DeclareMathSymbol{\bracemid}     {\mathord}{myLargesymbols}{'271}

\def\horiz@expandable#1#2#3#4#5#6#7#8{%
  \@mathmeasure\z@#7{#8}%
  \@tempdima=\wd\z@
  \@mathmeasure\z@#7{#1}%
  \ifdim\noexpand\wd\z@>\@tempdima
    $\m@th#7#1$%
  \else
    \@mathmeasure\z@#7{#2}%
    \ifdim\noexpand\wd\z@>\@tempdima
      $\m@th#7#2$%
    \else
      \@mathmeasure\z@#7{#3}%
      \ifdim\noexpand\wd\z@>\@tempdima
        $\m@th#7#3$%
      \else
        \@mathmeasure\z@#7{#4}%
        \ifdim\noexpand\wd\z@>\@tempdima
          $\m@th#7#4$%
        \else
          \@mathmeasure\z@#7{#5}%
          \ifdim\noexpand\wd\z@>\@tempdima
            $\m@th#7#5$%
          \else
           #6#7%
          \fi
        \fi
      \fi
    \fi
  \fi}

\def\overbrace@expandable#1#2#3{\vbox{\m@th\ialign{##\crcr
  #1#2{#3}\crcr\noalign{\kern2\p@\nointerlineskip}%
  $\m@th\hfil#2#3\hfil$\crcr}}}
\def\overbrace@#1#2#3{\vbox{\m@th\ialign{##\crcr
  #1#2\crcr\noalign{\kern2\p@\nointerlineskip}%
  $\m@th\hfil#2#3\hfil$\crcr}}}
  
\def\underbrace@expandable#1#2#3{\vtop{\m@th\ialign{##\crcr
  $\m@th\hfil#2#3\hfil$\crcr
  \noalign{\kern2\p@\nointerlineskip}%
  #1#2{#3}\crcr}}}
\def\underbrace@#1#2#3{\vtop{\m@th\ialign{##\crcr
  $\m@th\hfil#2#3\hfil$\crcr
  \noalign{\kern2\p@\nointerlineskip}%
  #1#2\crcr}}}

\def\bracefill@#1#2#3#4#5{$\m@th#5#1\leaders\hbox{$#4$}\hfill#2\leaders\hbox{$#4$}\hfill#3$}

\def\downbracefill@{\bracefill@\braceld\bracemd\bracerd\bracemid}
\DeclareRobustCommand{\downbracefill}{\downbracefill@\textstyle}

\def\upbracefill@{\bracefill@\bracelu\bracemu\braceru\bracemid}
\DeclareRobustCommand{\upbracefill}{\upbracefill@\textstyle}

\def\upbrace@expandable{%
  \horiz@expandable
    \upbrace
    \upbraceg
    \upbracegg
    \upbraceggg
    \upbracegggg
    \upbracefill@}
\def\downbrace@expandable{%
  \horiz@expandable
    \downbrace
    \downbraceg
    \downbracegg
    \downbraceggg
    \downbracegggg
    \downbracefill@}
    
\DeclareRobustCommand{\overbrace}[1]{\mathop{\mathpalette{\overbrace@expandable\downbrace@expandable}{#1}}\limits}
\DeclareRobustCommand{\underbrace}[1]{\mathop{\mathpalette{\underbrace@expandable\upbrace@expandable}{#1}}\limits}
\makeatother

\font\bigtenrm=cmr12 scaled 1200
\newcommand{\eexp}[1]{{\hbox{$\textfont1=\bigtenrm e$}}^{\raise3pt\hbox{$#1$}}}

\newcommand{\expnumber}[2]{{#1}\mathrm{e}{#2}}

\newcommand{\NarrowBand}{\ensuremath{\mathcal{N}}}
\newcommand{\Ball}{\ensuremath{\mathcal{B}}}
\newcommand{\Cell}{\ensuremath{\Omega}}
\newcommand{\CellStencil}{\ensuremath{\mathcal{S}}}

\newcommand{\Indicator}{\ensuremath{\chi}}
\newcommand{\IndicatorAppr}{\ensuremath{\tilde{\chi}}}
\newcommand{\InterfaceAppr}{\ensuremath{\tilde{\Sigma}}}
\newcommand{\MeshPoints}{\ensuremath{P_h}}
\newcommand{\MeshFaces}{\ensuremath{F_h}}
\newcommand{\normal}{\boldsymbol{n}} 
\newcommand{\PlicFraction}{\ensuremath{\VolFrac_c}}
\newcommand{\OmegaMinusAppr}{\ensuremath{\tilde{\Omega}^-}}
\newcommand{\sd}{\SignedDistance}

\newcommand{\SignedDistance}{\ensuremath{\phi}}

\newcommand{\Tetrahedron}{\ensuremath{T}}
\newcommand{\Triangle}{\ensuremath{\mathcal{T}}}

\newcommand{\VolFrac}{\ensuremath{\alpha}}
\newcommand{\VolFractions}{\ensuremath{\{\alpha_c}\}_{c \in C}}

\renewcommand{\x}{\ensuremath{\mathbf{x}}}

\newcommand{\Pvof}{SMCA}
\newcommand{\Lmax}{\ensuremath{l_\text{max}}}
\newcommand{\N}{\ensuremath{\mathbb{N}}}
\newcommand{\R}{\ensuremath{\mathbb{R}}}
\newcommand{\Ltet}{L_\text{tet}}
\newcommand{\Ltri}{L_\text{tri}}
\newcommand{\Ecdc}{E_\text{cdc}}
\newcommand{\conv}{\operatorname{conv}}

\graphicspath{{figures/}}
\usepackage{subcaption}
\usepackage{algorithm}
\usepackage{algpseudocode}
\usepackage[numbers]{natbib}
\usepackage{import} 
\usepackage{siunitx}
\usepackage[disable,textsize=scriptsize]{todonotes}
\usepackage{amssymb}
\usepackage{amsthm}
\usepackage{amsmath}
\usepackage{booktabs}
\usepackage{longtable}
\usepackage{cleveref}
\usepackage{lineno}

\colorlet{Reviewer1}{black}
\colorlet{Reviewer2}{black}

\begin{document}




\title{triSurfaceImmersion: Computing volume fractions and signed distances from triangulated surfaces immersed in unstructured meshes}

\author{Tobias Tolle} 
\ead{tolle@mma.tu-darmstadt.de}

\author{Dirk Gr\"{u}nding} 
\ead{gruending@mma.tu-darmstadt.de}

\author{Dieter Bothe} 
\ead{bothe@mma.tu-darmstadt.de}

\author{Tomislav Mari\'{c}\,\corref{corr}} 
\cortext[corr]{Corresponding author}
\ead{maric@mma.tu-darmstadt.de}

\address{Mathematical Modeling and Analysis Institute, Mathematics department, TU Darmstadt,\\
Alarich-Weiss-Stra\ss e 10, 64287 Darmstadt, Germany}


\pagenumbering{arabic}
\frenchspacing

\begin{abstract}

\textcolor{Reviewer1}{We propose a numerical method that enables the calculation of volume fractions from triangulated surfaces immersed in unstructured meshes. First, the signed distances are calculated geometrically near the triangulated surface. For this purpose, the computational complexity has been reduced by using an octree space subdivision. Second, an approximate solution of the Laplace equation is used to propagate the inside/outside information from the surface into the solution domain. Finally, volume fractions are computed from the signed distances in the vicinity of the surface. The volume fraction calculation utilizes either geometrical intersections or a polynomial approximation based on signed distances. An adaptive tetrahedral decomposition of polyhedral cells ensures a high absolute accuracy. The proposed method extends the admissible shape of the fluid interface (surface) to triangulated surfaces that can be open or closed, disjoint, and model objects of technical geometrical complexity.}

\textcolor{Reviewer1}{Current results demonstrate the effectiveness of the proposed algorithm for two-phase flow simulations of wetting phenomena, but the algorithm has broad applicability. For example, the calculation of volume fractions is crucial for achieving numerically stable simulations of surface tension-driven two-phase flows with the unstructured Volume-of-Fluid method. The method is applicable as a discrete phase-indicator model for the unstructured hybrid Level Set / Front Tracking method.} 

\noindent The implementation is available on GitLab \citep{argo}.

\noindent This a pre-print of the accepted article \href{https://doi.org/10.1016/j.cpc.2021.108249}{https://doi.org/10.1016/j.cpc.2021.108249}, when citing, please refer to the accepted article.

\end{abstract}


\maketitle

\noindent{\bf PROGRAM SUMMARY}

\begin{small}
\noindent
{\em Program Title:} argo/triSurfaceImmersion                                          \\
{\em CPC Library link to program files:} (to be added by Technical Editor) \\
{\em Developer's repository link:} https://gitlab.com/leia-methods/argo \\
{\em Code Ocean capsule:} (to be added by Technical Editor)\\
{\em Licensing provisions:} GPLv3 \\
{\em Programming language:} C++\\
{\em Nature of problem:}\\
  Computing volume fractions and signed distances from triangulated surfaces immersed in unstructured meshes. \\
{\em Solution method:}\\
  First, the algorithm computes minimal signed distances between mesh points (cell centers and cell corner-points) and the triangulated surface, in the close vicinity of the surface. 
The sign is computed with respect to the surface normal orientation. Afterwards, the sign is propagated throughout the unstructured volume mesh by an approximate solution of a diffusion equation. 
The bulk cells' volume fractions are set, and interface cells are identified based on the signed distances. 
Volume fractions in cells intersected by the triangulated surface mesh are either computed by geometric intersections between surface triangles and a cell or by an approximation of the volume fraction approximation from signed distances, coupled with tetrahedral cell decomposition and refinement.\\
{\em Additional comments including restrictions and unusual features:}\\
  The volume mesh can consist of cells of arbitrary shape. The surface mesh normal vectors need to be oriented consistently.
\end{small}

\section{Introduction}
\label{sec:intro}

We present a new numerical algorithm that calculates initial conditions for simulations of two-phase flow problems for fluid interfaces of complex shapes. The initial conditions are calculated in the form of signed distances and volume fractions from fluid interfaces approximated as arbitrarily shaped triangular surfaces immersed into unstructured meshes.
The signed distances are relevant as initial conditions for the Level Set method \citep{Sussman1998, Sussman1999ad} for multiphase flow simulation.
Volume fractions on unstructured meshes are required for the unstructured Volume-of-Fluid (VOF) method (cf. \citep{Maric2020vofrev} for a recent review). In fact, we have applied the proposed algorithms to model experimental fluid interfaces from wetting experiments \citep{Hartmann2021}, which was not possible using available contemporary approaches that model fluid interfaces using (compositions of) implicit functions or parameterized surfaces. The proposed algorithm approximates the surfaces using triangle meshes that are omnipresent in Computer-Aided Design (CAD) because of their versatility: they can approximate basic surfaces such as spheres and ellipsoids, but also surfaces of mechanical parts, disjoint surfaces in mechanical assemblies, or surfaces resulting from imaging scans.

\begin{figure}[ht]
 \begin{center}
 \def\svgwidth{.6\textwidth} 
 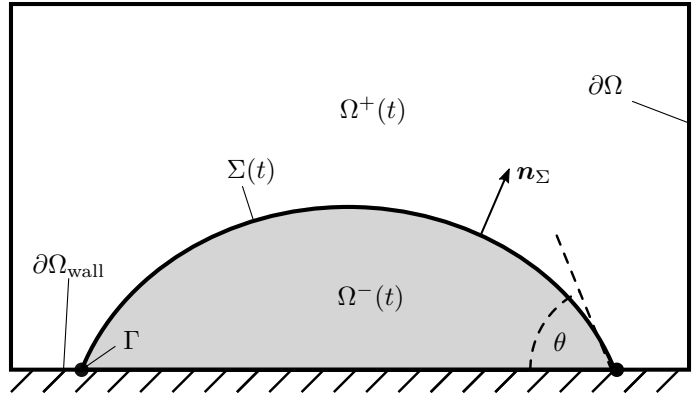 
 \caption{The different domains for a liquid ($-$) drop on a solid surface surrounded by a gas ($+$) phase.} 
  \label{fig:domains} 
 \end{center}  
\end{figure}

\begin{figure}[!h]
    \captionsetup{justification=centering}
    \centering
    \def\svgwidth{0.7\textwidth}
    {\footnotesize
     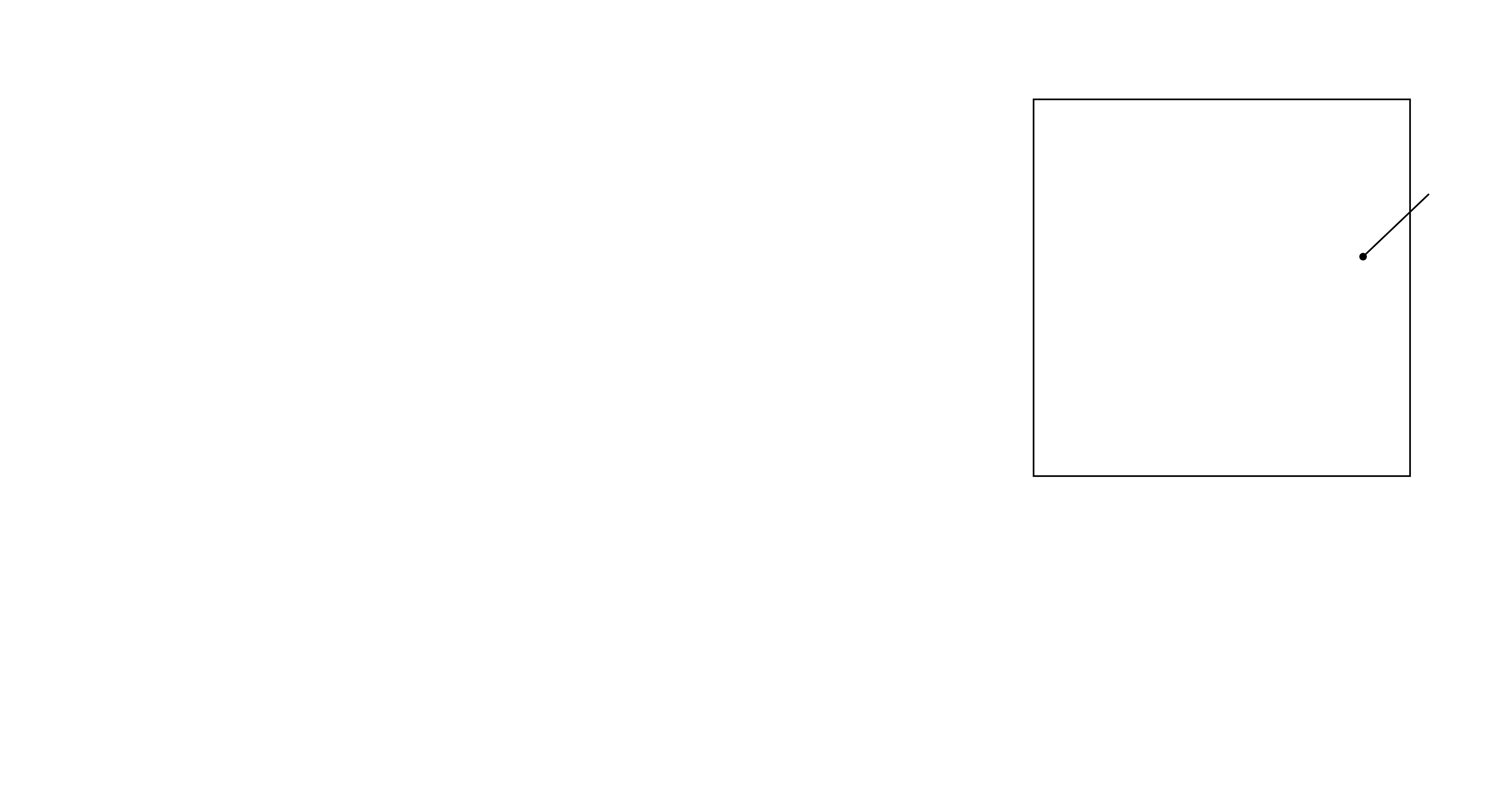
    }
    \caption{Calculating volume fractions of a circular interface by numerical integration.}
  \label{fig:volfrac-integrate}
\end{figure}

The overall simulation domain $\Omega \subset \mathbb{R}^3$ is separated into two subdomains $\Omega = \Omega^+(t) \cup \Omega^-(t)$, representing phase $1$ and phase $2$, respectively, as illustrated for a liquid drop on a surface in \cref{fig:domains}. At the contact line $\Gamma:= \partial \Omega \cap \overline{\Omega^+} \cap \overline{\Omega^-}$, the liquid-gas interface $\Sigma$ encloses a contact angle $\theta$ with the solid surface $\partial \Omega_\text{wall}$. Furthermore, the normal vector $\normal_\Sigma$ of the interface $\Sigma$ is oriented such that it points into the gas phase.
Typically, a continuum mechanical model is used for the description of such fluid mechanical problems. This description is often based on a sharp interface model, as depicted in \cref{fig:domains}. With this model, the liquid-gas interface can be described using an indicator function
\begin{equation}
  \chi(\x,t) := 
    \begin{cases}
      1, & \x \in \Omega^- \subset \mathbb{R}^3 \\ 
      0, & \text{otherwise}. 
    \end{cases}
  \label{eqn:indicator}
\end{equation}
An approximate solution of this model requires a decomposition of the solution domain into volumes that have no volume overlaps, the closed \emph{cells} $\Cell_c$, denoted by 
\begin{equation}
    \Omega \approx \tilde{\Omega} = \{ \Cell_c \}_{c \in C}
    \label{eqn:omegah}
\end{equation}
where $C = \{1,2,3, \dots , N_c\}$ is a set of indices to mesh cells.
As can be seen in \cref{fig:volfrac-integrate}, the mesh is a set of non-overlapping subsets (\emph{cells}) $\Cell_c \subset \tilde{\Omega}$. With non-overlapping, we mean that the volume of an intersection between any two cells is zero. \emph{Index sets} represent the unstructured mesh data \cite{Ghali2008}. We consider a set of cell corner-points $\MeshPoints$ where each point in $\MeshPoints$ is an element of $\mathbb{R}^3$. Geometrically, each cell $\Cell_c$ is a volume bounded by polygons, so-called \emph{faces}. A global set of faces $\MeshFaces$ is defined, and each face is a sequence of \emph{indices} of points in $\MeshPoints$. In this context, we define a cell set $C_c$ as a set of indices of faces in the set of mesh faces $\MeshFaces$. Therefore, when referring to a volume defined by the cell, we use $\Omega_c$ and its magnitude is then $|\Omega_c|$, and when we refer to the cell as an unordered index set, we use $C_c$ and its magnitude $|C_c|$ is the number of faces that bound the cell.  

Solutions of continuum mechanical problems in geometrically complex solution domains significantly benefit from unstructured meshes. For example, gradients of solution variables are resolved at geometrically complex boundaries by employing mesh boundary layers, strongly reducing the number of cells required to achieve specific accuracy. Hence, this approx reduces the overall required computational resources.

As the phase indicator $\Indicator(\x,t)$ given by \cref{eqn:indicator} contains a jump discontinuity, it poses difficulties for numerical simulations of two-phase flows. With Volume of fluid (VOF) methods, this non-continuous description is discretized by introducing the so-called \emph{volume fraction}   
\begin{equation}
  \PlicFraction = \dfrac{1}{|\Cell_c|}\int_{\Cell_c} \Indicator(\x,t) dx.
  \label{eqn:volfrac}
\end{equation}
The unstructured VOF methods \citep{Maric2020vofrev} rely on the volume fraction field $\PlicFraction$ to track interface with the advecting velocity obtained from the solution of two-phase Navier-Stokes equations in a single-field formulation. All multiphase flow simulation methods that utilize the single-field formulation of Navier-Stokes equations approximate the phase-indicator function similarly to \cref{eqn:volfrac}. The phase-indicator approximation utilizes signed distances in the Level Set \citep{Sussman1998,Sussman1999,Sussman1999ad} method, the volume fractions approximate the phase indicator for the Volume-of-Fluid \citep{DeBar1974,Noh1976,Hirt1981,Rider1998} method. 

Various methods exist that compute the volume fraction $\PlicFraction$ based on the exact phase indicator $\Indicator(\x,t)$. The majority of methods calculate the integral in \cref{eqn:volfrac} numerically, as schematically shown in \cref{fig:volfrac-integrate}, using numerical quadrature. 

Different approaches are \textcolor{Reviewer1}{outlined} below with increasing complexity in terms of admissible shapes of the fluid interface. The admissible shapes range from analytic descriptions of basic geometric shapes such as spheres and ellipsoids to implicit functions (or their combinations) and more general shapes approximated with volume meshes.

\citet{Strobl2016} propose an exact intersection between a sphere and a tetrahedron, a wedge, or a hexahedron. The proposed algorithm is exact and fast, though it is limited to the spherical interface shape.

\citet{Fries2016} represent the fluid interface as a level set and propose a higher-order quadrature for the integral on the right-hand side of \cref{eqn:volfrac}. The parametrization of the surface uses roots of the implicit function found by the closest-point algorithm. Results are presented for hexahedral and tetrahedral unstructured meshes that may also be strongly deformed. \citet[fig. 52, fig. 53]{Fries2016} also show results with higher-order ($>2$) convergence for the volume integration of an arbitrary non-linear function on hexahedral and tetrahedral meshes. However, the volume and area integration error is reported for a single function. While a relative global volume error between $\expnumber{1}{-08}$ and $\expnumber{1}{-06}$ is reported, no information about the required CPU times is provided. In the approach proposed by \citet{Fries2016}, fluid interfaces with complex shapes are modeled as a composition of implicit functions.

\citet{Kromer2019} propose an efficient third-order accurate quadrature for the eq. (3). Contrary to \citet{Jones2019}, who decompose cells into tetrahedrons, \citet{Kromer2019} locally approximate the hypersurface by a paraboloid based on the principal curvatures. Applying the Gaussian divergence theorem to eq. (3) then yields contributions from the cell boundary and the approximated hypersurface patch. Using the surface divergence theorem, \citet{Kromer2019} reformulate the contribution from the hypersurface patch into a set of line integrals, where the associated integrand emerges from the solution of a Laplace-Beltrami-type problem. The method of Kromer and Bothe [22] is directly applicable to unstructured meshes. However, locally, i.e., within a cell, the fluid interface must be \textcolor{Reviewer1}{$C^2$} and simply connected.

\citet{Aulisa2007} and \citet{Bna2015,Bna2016} calculate the volume fraction by representing the indicator function as a height function inside cubic cells, using the structure of the underlying Cartesian mesh. Numerical integration of the height function is illustrated by \cref{fig:volfrac-integrate}. However, extending this approach to unstructured meshes raises many questions. First, constructing a height function in a specific direction is complex and computationally expensive \citep{Owkes2015}. Second, the orientation of the interface in the chosen coordinate system may easily make the problem ill-conditioned. Finally, required mesh-search operations are complicated as the face normals of polyhedral cells are typically not aligned with the coordinate axes.  


\textcolor{Reviewer1}{The calculation of the volume fraction given by $\alpha_c = \frac{|\Omega^- \cap \Omega_c|}{|\Omega_c|}$ can be reformulated into the integration of a function $f = 1$ within $\Omega^- \cap \Omega_c$. Since $\partial \Omega_c$ consists of piecewise-planar surfaces (faces), the complexity lies in the non-planar part of the surface $\partial \Omega^- \cap \Omega_c = \Sigma(t) \cap \Omega_c$. Trimmed isogeometric analysis can be used to integrate $f=1$ within $\Omega^- \cap \Omega_c$ by representing $\partial \Omega^- \cap \Omega_c$ using a trimmed NURBS surface, effectively resulting in $\alpha_c = \frac{|\Omega^- \cap \Omega_c|}{|\Omega_c|}$ for complex non-linear CAD surfaces. Although not yet applied to volume fraction calculation ($f=1$ integration), trimmed isogeometric analysis has been applied to solving PDEs in solution domains bounded by NURBS surfaces \citep{Kim2009,Schmidt2012,Nitti2020}. Similarly, the immersed isogeometric analysis (e.g. \citep{Divi2020}) requires function integration in cut cells, where the integration of $f=1$ in the cut cell is equivalent to computing $|\Omega^- \cap \Omega_c|$ used in volume fraction calculation. Although it is a potentially interesting alternative approach for computing volume fractions from CAD surfaces, the isogeometric analysis requires NURBS trimming, octree refinement, and higher-order quadratures. These efforts are worthwhile for the goal of achieving higher-order solutions for PDEs in complex solution domains. However, as demonstrated in the results section, our proposed algorithms achieve sufficient accuracy for signed distances and volume fractions on unstructured meshes while relying on straightforward second-order accurate discretization.} 

The signed distances in the Level Set Method require re-distancing (correction). The re-distancing methods are usually based on approximate solutions of Partial Differential Equations (PDEs) that ensure the signed-distance property \citep{Russo2000}. Contrary to this approach, the unstructured Level Set / Front Tracking method \citep{Maric2015,Tolle2020} \textcolor{Reviewer1}{\emph{geometrically}} computes minimal signed distances from $\InterfaceAppr$. This calculation is relatively straightforward on structured meshes \citep{Shin2002,Shin2011}, but significantly more complex on unstructured meshes \citep{Maric2015,Tolle2020}. Here we significantly extend the calculation of signed distances from \citep{Maric2015, Tolle2020} by introducing an efficient approximate propagation of the inside/outside information from $\InterfaceAppr$. 

Volume fraction calculation methods outlined so far model the fluid interface using exact functions and handle more complex interface shapes via combinations of these functions. A combination of exact functions cannot accurately capture the shape of the fluid interface in many cases. For example, when the interface shape is prescribed experimentally \citet{Hartmann2021}. 

\begin{figure}[!h]
    \captionsetup{justification=centering}
    \centering
    \def\svgwidth{0.7\textwidth}
    {\footnotesize
\begingroup%
  \makeatletter%
  \providecommand\color[2][]{%
    \errmessage{(Inkscape) Color is used for the text in Inkscape, but the package 'color.sty' is not loaded}%
    \renewcommand\color[2][]{}%
  }%
  \providecommand\transparent[1]{%
    \errmessage{(Inkscape) Transparency is used (non-zero) for the text in Inkscape, but the package 'transparent.sty' is not loaded}%
    \renewcommand\transparent[1]{}%
  }%
  \providecommand\rotatebox[2]{#2}%
  \newcommand*\fsize{\dimexpr\f@size pt\relax}%
  \newcommand*\lineheight[1]{\fontsize{\fsize}{#1\fsize}\selectfont}%
  \ifx\svgwidth\undefined%
    \setlength{\unitlength}{4778.47595215bp}%
    \ifx\svgscale\undefined%
      \relax%
    \else%
      \setlength{\unitlength}{\unitlength * \real{\svgscale}}%
    \fi%
  \else%
    \setlength{\unitlength}{\svgwidth}%
  \fi%
  \global\let\svgwidth\undefined%
  \global\let\svgscale\undefined%
  \makeatother%
  \begin{picture}(1,0.46735345)%
    \lineheight{1}%
    \setlength\tabcolsep{0pt}%
    \put(0,0){\includegraphics[width=\unitlength,page=1]{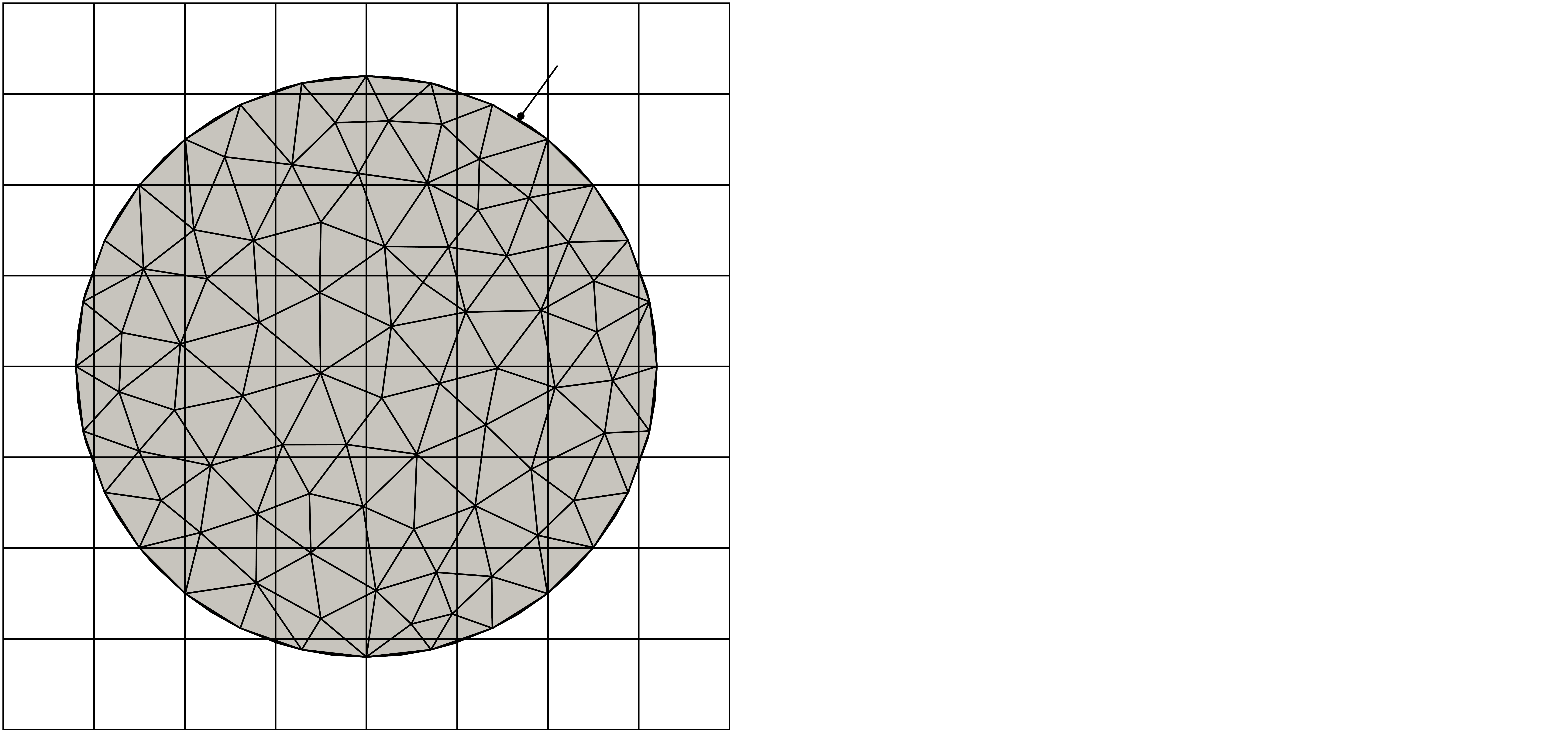}}%
    \put(0.35633776,0.42562831){\color[rgb]{0,0,0}\makebox(0,0)[lt]{\lineheight{0}\smash{\begin{tabular}[t]{l}$\InterfaceAppr$\end{tabular}}}}%
    \put(0.13540713,0.42906969){\color[rgb]{0,0,0}\makebox(0,0)[lt]{\lineheight{0}\smash{\begin{tabular}[t]{l}$\Cell_c$\end{tabular}}}}%
    \put(0,0){\includegraphics[width=\unitlength,page=2]{cylinder2Dmesh.pdf}}%
    \put(0.814188,0.35960982){\makebox(0,0)[lt]{\lineheight{1.25}\smash{\begin{tabular}[t]{l}$\Omega_c$\end{tabular}}}}%
    \put(0.7652519,0.28879644){\makebox(0,0)[lt]{\lineheight{1.25}\smash{\begin{tabular}[t]{l}$\tilde{\Omega}^-_l$\end{tabular}}}}%
    \put(0.07075099,0.37767849){\color[rgb]{0,0,0}\makebox(0,0)[lt]{\lineheight{0}\smash{\begin{tabular}[t]{l}$\tilde{\Omega}^-_l$\end{tabular}}}}%
  \end{picture}%
\endgroup%

    }
    \caption{Calculating volume fractions from a circular interface by volume mesh intersection.}
  \label{fig:volfrac-meshintersect}
\end{figure}

One approach exists that can handle arbitrarily complex interface shapes. In this approach, the fluid interface encloses a volumetric mesh as its boundary surface mesh. This mesh given by the fluid interface is intersected with a "background" mesh that stores volume fractions. This approach is called \emph{volume mesh intersection}. An example for such an intersection between $\tilde{\Omega}$ and cells from $\OmegaMinusAppr$ is shown in \cref{fig:volfrac-meshintersect}. In principle, this approach is relatively straightforward, provided an accurate geometrical intersection of tetrahedrons is available. However, geometrical operations based on floating-point numbers are not stable and can lead to severe errors \cite[chap. 45]{Toth2017}. 

\citet{Ahn2007} have initialized volume fractions by volume mesh intersection as shown in \cref{fig:volfrac-meshintersect}. In this approach, the approximated phase  $\tilde{\Omega}^-(t)$ is decomposed into volumes (an unstructured mesh), equivalently to the decomposition $\tilde{\Omega}$ given by \cref{eqn:omegah}. The boundary $\partial \Omega^-$ is the fluid interface $\Sigma(t)$, and it is approximated as a polygonal surface mesh, leading to 
\begin{equation}
    \Omega^- \approx \tilde{\Omega}^- := \{\tilde{\Omega}^-_l\}_{l  \in L},
\end{equation}
i.e.\ an approximation of $\Omega^-$. Generally, as shown in the detail in \cref{fig:volfrac-meshintersect}, a cell $\Omega_c$ of the background mesh $\tilde{\Omega}$ may overlap with multiple cells $\Omega_l$ from the $\OmegaMinusAppr$ mesh, and vice versa. We define a set of indices $l$ of cells $\OmegaMinusAppr_l$ in $\OmegaMinusAppr$ that overlap with the cell $\Omega_c$: the so-called \emph{cell stencil} of $\Omega_c$ in $\OmegaMinusAppr_l$, namely
\begin{equation}
    \mathcal{S}(\Cell_c, \OmegaMinusAppr) = \{ l \in L : \Cell_c \cap \OmegaMinusAppr_l \ne \emptyset, \text{where}~ \Omega_c \in \tilde{\Omega}, \OmegaMinusAppr_l \in \OmegaMinusAppr \},
    \label{eqn:inter-stencil}
\end{equation}
where $L$ is an index set, containing indices of cells from $\tilde{\Omega}^-$. Volume fractions $\VolFractions$ can then be calculated by performing the intersection 
\begin{equation}
     \VolFrac_c = \frac{|\cup_{l \in \CellStencil(\Omega_c, \OmegaMinusAppr)} \Omega_c \cap \OmegaMinusAppr_l|}{|\Omega_c|}.
     \label{eq:volfrac-volintersect}
\end{equation}
Since each $\OmegaMinusAppr_l$ overlaps with at least a one cell from $\tilde{\Omega}$, and we can approximate the number of cells from $\tilde{\Omega}$ that intersect each cell from $\tilde{\Omega}^-$ as
\begin{equation}
    N(\OmegaMinusAppr,\tilde{\Omega}) \approx |\OmegaMinusAppr| \underset{l \in L}{\text{mean}}(|\CellStencil(\OmegaMinusAppr_l,\tilde{\Omega})|),
\end{equation}
where $|\OmegaMinusAppr|$ denotes the number of cells in the mesh $\OmegaMinusAppr$. The average number of cells $\Omega_c$ overlapping $\OmegaMinusAppr_l$, $\underset{l \in L}{\text{mean}}|C(\OmegaMinusAppr_l,\tilde{\Omega})|$, depends on the mesh densities of both meshes, $\tilde{\Omega}$ and $\OmegaMinusAppr$. However, we do know that $\underset{l \in L}{\text{mean}}|C(\OmegaMinusAppr_l,\tilde{\Omega})|>1$. Next, we know that $|\OmegaMinusAppr|$ grows quadratically in $2D$ and cubically in $3D$ with a uniform increase in mesh resolution, taken as the worst case scenario. It grows linearly in $2D$ and quadratically in $3D$ if $\OmegaMinusAppr$ is refined only near the interface $\tilde{\Sigma} := \partial \OmegaMinusAppr$. Consequently, the computational complexity of the volume mesh intersection algorithm in terms of cell/cell intersections is quadratic in $2D$ and cubic in $3D$ in the worst case, and linear in $2D$ and quadratic in $3D$ if local refinement is used to increase the resolution of $\tilde{\Sigma}$. The quadratic complexity in $3D$ is a serious drawback of this algorithm, especially for large simulations where $|\tilde{\Omega}^-|$ easily reaches hundred thousand cells per CPU core. 
\citet{Menon2011} have extended the volume mesh intersection algorithm from \citet{Ahn2007} to perform a volume conservative remapping of variables in the collocated Finite Volume Method (FVM) with second-order accuracy on unstructured meshes. Their results confirm the polynomial computational complexity in terms of absolute CPU times for this volume mesh intersection algorithm \citep[table 3]{Menon2011}.

\citet{Lopez2019} propose a volume truncation algorithm for non-convex cells and apply it to the initialization of volume fractions from exact functions on unstructured meshes. Cell-subdivision is introduced to handle cases for which the interface crosses an edge of a cell twice. Non-planar truncated volumes are triangulated \citep[fig 18]{Lopez2019}, and second-order accuracy is demonstrated in terms of the relative global volume error for a uniform resolution and a higher-order accuracy when locally refined sub-grid meshes are used. 

\citet{Ivey2015} initialize volume fractions on unstructured meshes using tetrahedral decomposition of non-convex cells and perform geometrical intersections with a similar approach as the \textcolor{Reviewer1}{approach} from \citet{Ahn2007}. Unlike \citet{Ahn2007}, \citet{Ivey2015} compute volume fractions of intersected tetrahedrons by intersecting them with exact signed distance functions that are used to model the fluid interface. Therefore, this algorithm cannot directly utilize arbitrarily shaped interfaces. However, their approach utilizes a linear interpolation of intersection points between the tetrahedron and the signed-distance function and yields second-order accuracy. Accuracy is further increased using adaptive mesh refinement. 

The approaches reviewed so far require an exact representation of the interface using explicit analytic expressions, which hinders the direct application of such algorithms to initial conditions resulting from experiments as these are typically not available as function compositions. The volume mesh intersection algorithm \citep{Ahn2007} is flexible but computationally expensive, and it requires highly accurate and robust geometrical intersections. 

The following sections outline the proposed algorithm that uses an unstructured surface mesh $\tilde{\Sigma}$ to compute signed distances and volume fractions on unstructured meshes. Relying on unstructured surface meshes retains the ability to handle arbitrary-shaped surfaces while avoiding computationally expensive cell/cell intersections. Of course, using surface meshes to approximate the fluid interface renders the proposed algorithm second-order accurate; however, \textcolor{Reviewer1}{sufficient absolute accuracy is achievable with second-order accurate methods using local mesh refinement on the background mesh \citep{Cummins2005,Francois2006}. Applying local mesh refinement on the background mesh in the close vicinity of the triangulated surface increases the accuracy and limits it to the resolution of the surface mesh, not the background mesh that stores volume fractions and signed distances.} The proposed algorithm geometrically computes signed distances near the fluid interface. These signed distances (so-called \emph{narrow-band} signed-distances) are then propagated throughout $\tilde{\Omega}$ by an approximate solution of a diffusion equation. The propagated signed distances determine the value of the phase indicator $\Indicator(\x,t)$ in those cells that are either completely empty $(\VolFrac_c = 0)$, or completely full $(\VolFrac_c = 1)$. Finally, second-order accurate volume fraction values are calculated in intersected cells $(0 < \VolFrac_c < 1)$. This work enables the calculation of complex initial conditions for different multiphase simulation methods. These include in particular geometric \citep{Jofre2014,Ivey2015,Owkes2017,Maric2018,Scheufler2019} and algebraic VOF methods \citep{Ubbink1997,Desphande2012}. The calculation of volume fractions from a surface mesh (marker points in 2D) was done in the mixed markers / VOF method by \citet{Aulisa2003a}: the proposed algorithm significantly extends this idea towards an accurate and fast volume fraction model for Front Tracking methods \citep{Tryggvason2001}, as well as the hybrid Level Set / Front Tracking methods on structured \citep{Shin2002,Shin2011} or unstructured \citep{Maric2015,Tolle2020} meshes. 
Signed distances and the respective inside-outside information from triangulated surfaces are available for unstructured Level Set and Immersed Boundary methods.

\section{Surface mesh / cell intersection algorithm}
\label{sec:alphainit}
\begin{figure}[!h]
    \captionsetup{justification=centering}
  \centering
    \begin{subfigure}[t]{0.29\textwidth}
        \centering
        \def\svgwidth{\textwidth}
        {\footnotesize
\begingroup%
  \makeatletter%
  \providecommand\color[2][]{%
    \errmessage{(Inkscape) Color is used for the text in Inkscape, but the package 'color.sty' is not loaded}%
    \renewcommand\color[2][]{}%
  }%
  \providecommand\transparent[1]{%
    \errmessage{(Inkscape) Transparency is used (non-zero) for the text in Inkscape, but the package 'transparent.sty' is not loaded}%
    \renewcommand\transparent[1]{}%
  }%
  \providecommand\rotatebox[2]{#2}%
  \newcommand*\fsize{\dimexpr\f@size pt\relax}%
  \newcommand*\lineheight[1]{\fontsize{\fsize}{#1\fsize}\selectfont}%
  \ifx\svgwidth\undefined%
    \setlength{\unitlength}{2284.5bp}%
    \ifx\svgscale\undefined%
      \relax%
    \else%
      \setlength{\unitlength}{\unitlength * \real{\svgscale}}%
    \fi%
  \else%
    \setlength{\unitlength}{\svgwidth}%
  \fi%
  \global\let\svgwidth\undefined%
  \global\let\svgscale\undefined%
  \makeatother%
  \begin{picture}(1,1.03578464)%
    \lineheight{1}%
    \setlength\tabcolsep{0pt}%
    \put(0.46123009,0.60082719){\color[rgb]{0,0,0}\makebox(0,0)[lt]{\lineheight{0}\smash{\begin{tabular}[t]{l}$r_p$\end{tabular}}}}%
    \put(0,0){\includegraphics[width=\unitlength,page=1]{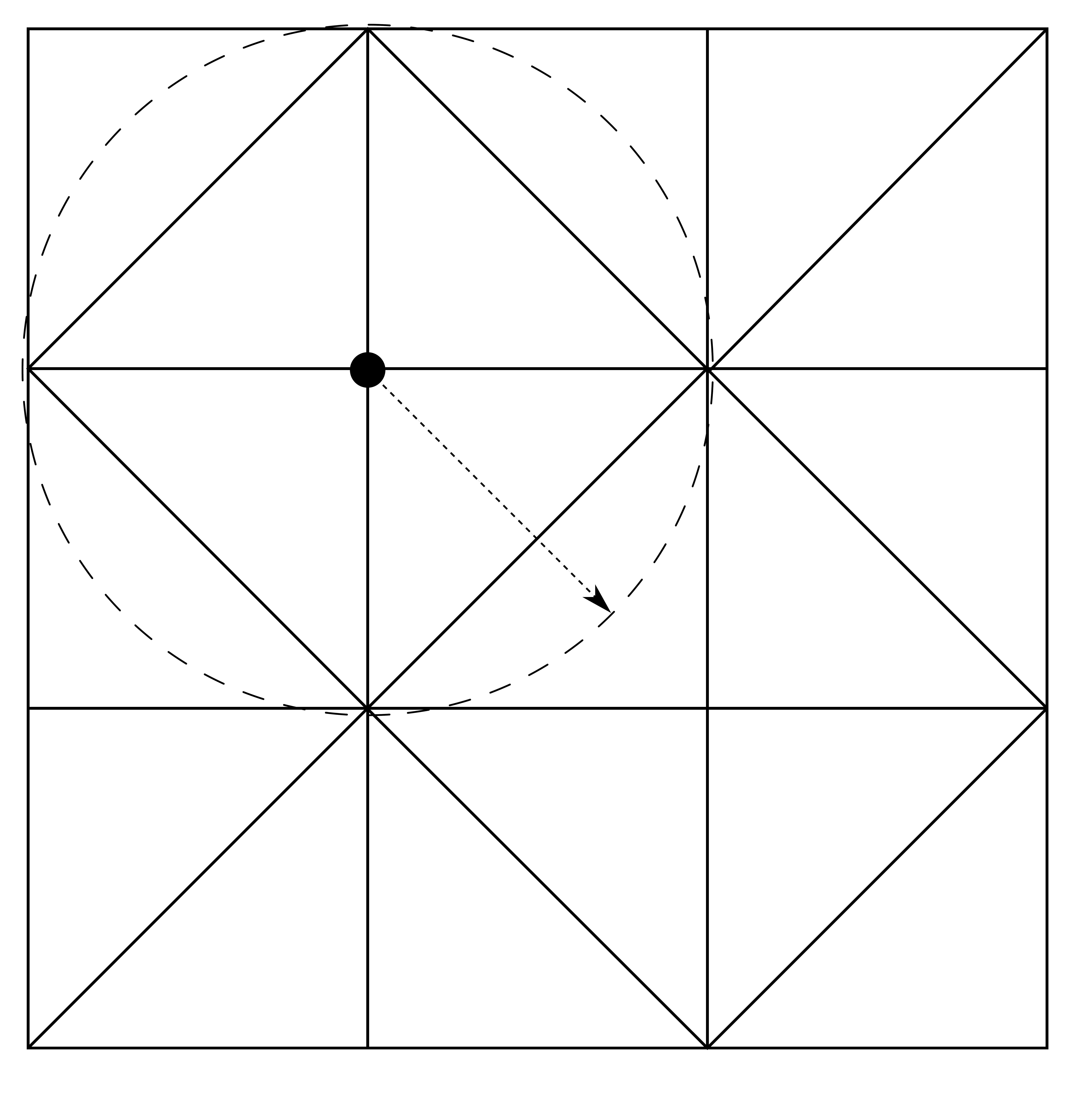}}%
    \put(0.35504435,0.71886674){\color[rgb]{0,0,0}\makebox(0,0)[lt]{\lineheight{0}\smash{\begin{tabular}[t]{l}$\x_p \in \MeshPoints$\end{tabular}}}}%
    \put(0,0){\includegraphics[width=\unitlength,page=2]{search-distance.pdf}}%
    \put(0.66399581,0.33096251){\color[rgb]{0,0,0}\makebox(0,0)[lt]{\lineheight{0}\smash{\begin{tabular}[t]{l}$\x_c$\end{tabular}}}}%
    \put(0,0){\includegraphics[width=\unitlength,page=3]{search-distance.pdf}}%
    \put(0.57043397,0.23182146){\color[rgb]{0,0,0}\makebox(0,0)[lt]{\lineheight{0}\smash{\begin{tabular}[t]{l}$r_c$\end{tabular}}}}%
  \end{picture}%
\endgroup%

        }
        \caption{Calculation of search radii.} 
        \label{fig:searchradii}
    \end{subfigure}
    \hspace{1.5em}
    \begin{subfigure}[t]{0.29\textwidth}
        \captionsetup{singlelinecheck = false, justification=justified}
        \centering
        \def\svgwidth{\textwidth}
        {\footnotesize
\begingroup%
  \makeatletter%
  \providecommand\color[2][]{%
    \errmessage{(Inkscape) Color is used for the text in Inkscape, but the package 'color.sty' is not loaded}%
    \renewcommand\color[2][]{}%
  }%
  \providecommand\transparent[1]{%
    \errmessage{(Inkscape) Transparency is used (non-zero) for the text in Inkscape, but the package 'transparent.sty' is not loaded}%
    \renewcommand\transparent[1]{}%
  }%
  \providecommand\rotatebox[2]{#2}%
  \newcommand*\fsize{\dimexpr\f@size pt\relax}%
  \newcommand*\lineheight[1]{\fontsize{\fsize}{#1\fsize}\selectfont}%
  \ifx\svgwidth\undefined%
    \setlength{\unitlength}{2284.5bp}%
    \ifx\svgscale\undefined%
      \relax%
    \else%
      \setlength{\unitlength}{\unitlength * \real{\svgscale}}%
    \fi%
  \else%
    \setlength{\unitlength}{\svgwidth}%
  \fi%
  \global\let\svgwidth\undefined%
  \global\let\svgscale\undefined%
  \makeatother%
  \begin{picture}(1,1.03578464)%
    \lineheight{1}%
    \setlength\tabcolsep{0pt}%
    \put(0,0){\includegraphics[width=\unitlength,page=1]{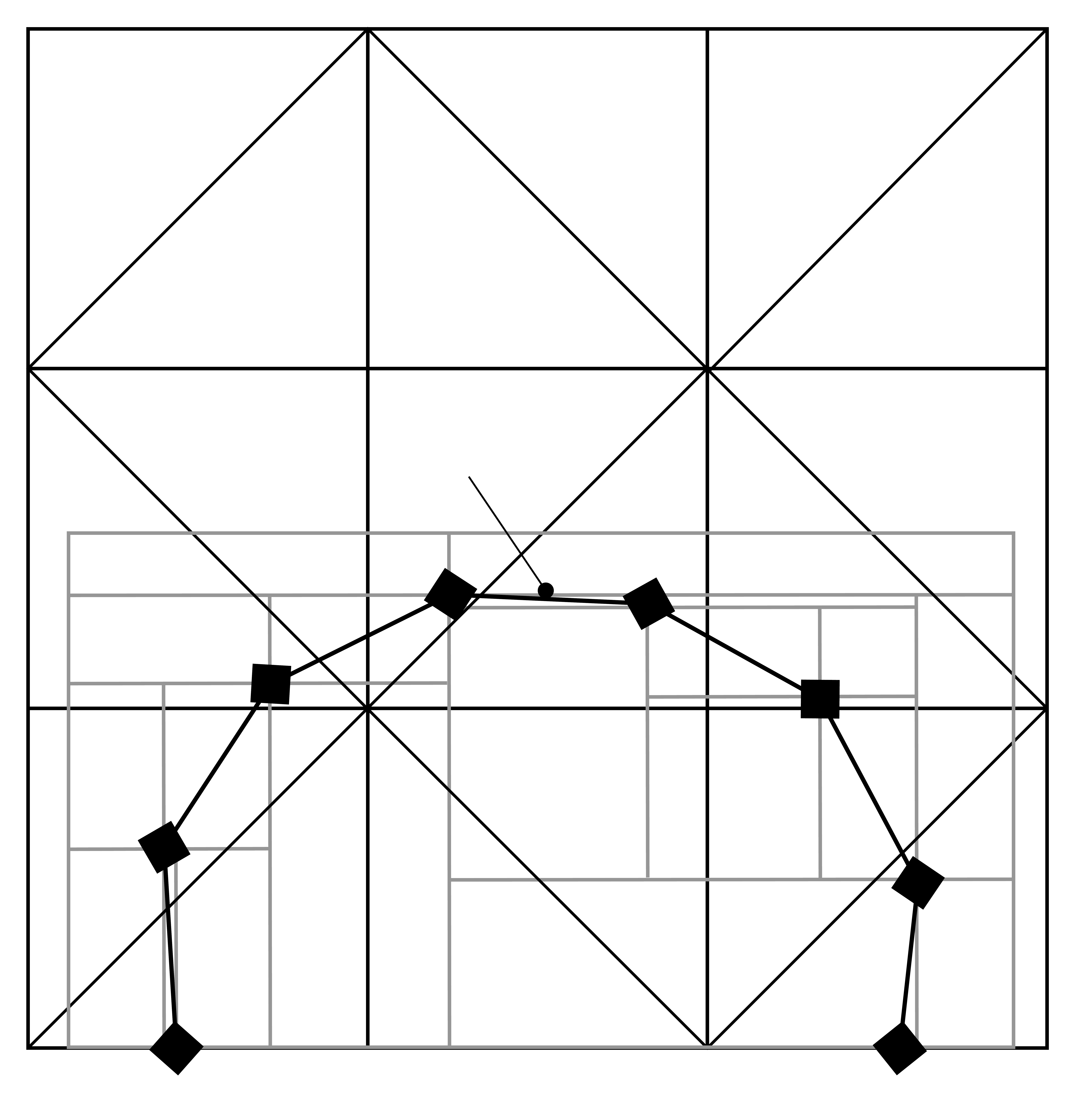}}%
    \put(0.39380396,0.60550386){\makebox(0,0)[lt]{\lineheight{1.25}\smash{\begin{tabular}[t]{l}$\InterfaceAppr$\end{tabular}}}}%
    \put(0,0){\includegraphics[width=\unitlength,page=2]{octree.pdf}}%
    \put(0.76842885,0.64858927){\makebox(0,0)[lt]{\lineheight{1.25}\smash{\begin{tabular}[t]{l}octree\end{tabular}}}}%
    \put(0,0){\includegraphics[width=\unitlength,page=3]{octree.pdf}}%
    \put(0.22625784,0.57837675){\makebox(0,0)[lt]{\lineheight{1.25}\smash{\begin{tabular}[t]{l}$\mathbf{n}_{\InterfaceAppr}$\end{tabular}}}}%
  \end{picture}%
\endgroup%

        }
        \subcaption{Octree sub-division of the surface mesh $\InterfaceAppr$ bounding-box.} 
        \label{fig:octree}
    \end{subfigure}
    \hspace{1.5em}
    \begin{subfigure}[t]{0.29\textwidth}
        \captionsetup{singlelinecheck = false, justification=justified}
        \centering
        \def\svgwidth{\textwidth}
        {\footnotesize
\begingroup%
  \makeatletter%
  \providecommand\color[2][]{%
    \errmessage{(Inkscape) Color is used for the text in Inkscape, but the package 'color.sty' is not loaded}%
    \renewcommand\color[2][]{}%
  }%
  \providecommand\transparent[1]{%
    \errmessage{(Inkscape) Transparency is used (non-zero) for the text in Inkscape, but the package 'transparent.sty' is not loaded}%
    \renewcommand\transparent[1]{}%
  }%
  \providecommand\rotatebox[2]{#2}%
  \newcommand*\fsize{\dimexpr\f@size pt\relax}%
  \newcommand*\lineheight[1]{\fontsize{\fsize}{#1\fsize}\selectfont}%
  \ifx\svgwidth\undefined%
    \setlength{\unitlength}{2284.5bp}%
    \ifx\svgscale\undefined%
      \relax%
    \else%
      \setlength{\unitlength}{\unitlength * \real{\svgscale}}%
    \fi%
  \else%
    \setlength{\unitlength}{\svgwidth}%
  \fi%
  \global\let\svgwidth\undefined%
  \global\let\svgscale\undefined%
  \makeatother%
  \begin{picture}(1,1.03578464)%
    \lineheight{1}%
    \setlength\tabcolsep{0pt}%
    \put(0,0){\includegraphics[width=\unitlength,page=1]{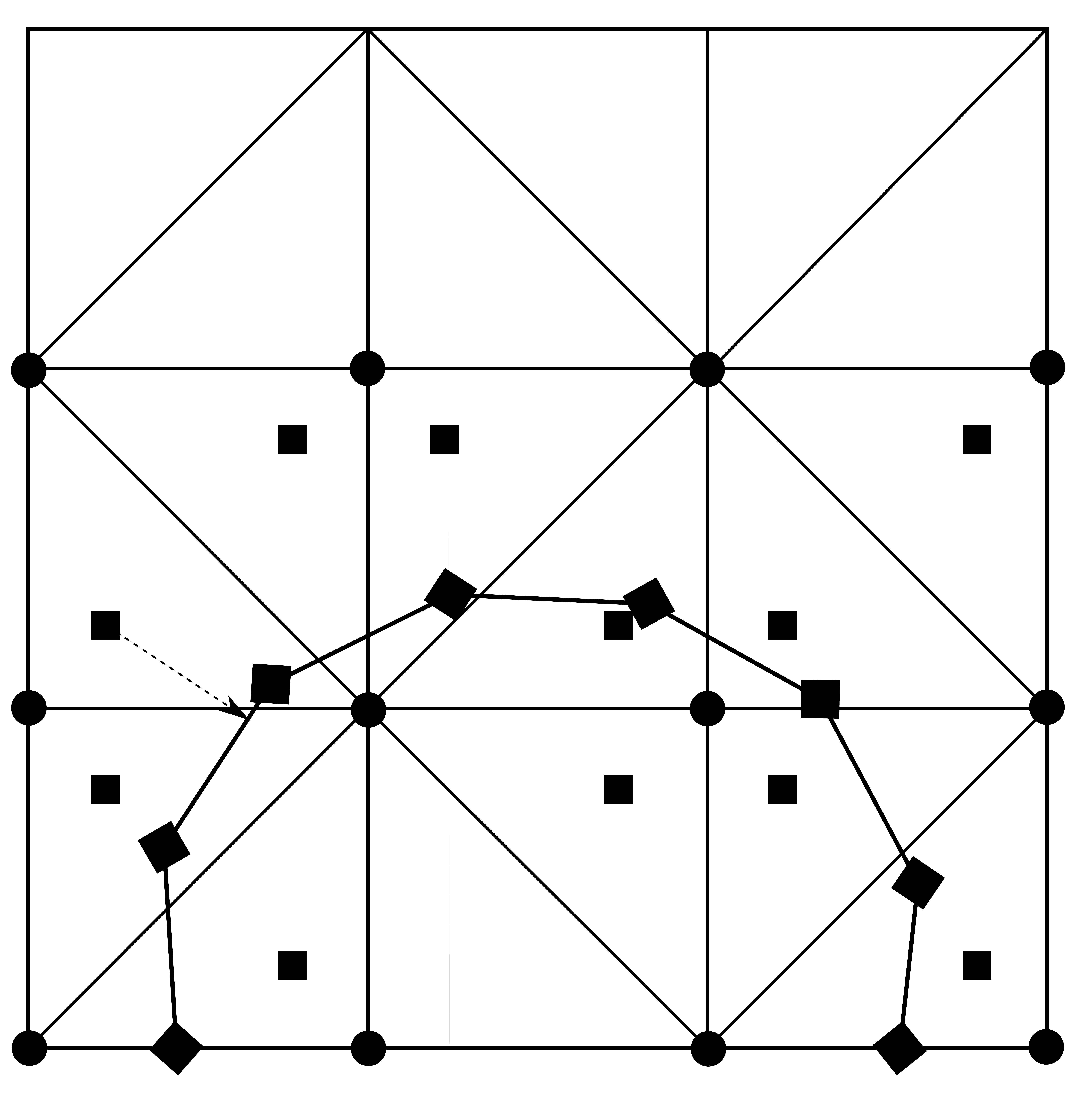}}%
    \put(0.08015303,0.48800188){\makebox(0,0)[lt]{\lineheight{1.25}\smash{\begin{tabular}[t]{l}$\phi_c^+$\end{tabular}}}}%
    \put(0,0){\includegraphics[width=\unitlength,page=2]{narrow-band-distance.pdf}}%
    \put(0.2343753,0.58003789){\makebox(0,0)[lt]{\lineheight{1.25}\smash{\begin{tabular}[t]{l}$\mathbf{n}_{\InterfaceAppr}$\end{tabular}}}}%
    \put(0.25568304,0.16701719){\makebox(0,0)[lt]{\lineheight{1.25}\smash{\begin{tabular}[t]{l}$\phi_c^-$\end{tabular}}}}%
    \put(0,0){\includegraphics[width=\unitlength,page=3]{narrow-band-distance.pdf}}%
    \put(0.71258918,0.71334194){\makebox(0,0)[lt]{\lineheight{1.25}\smash{\begin{tabular}[t]{l}$\phi_p^+$\end{tabular}}}}%
    \put(0,0){\includegraphics[width=\unitlength,page=4]{narrow-band-distance.pdf}}%
    \put(0.52959055,0.01124818){\makebox(0,0)[lt]{\lineheight{1.25}\smash{\begin{tabular}[t]{l}$\phi_p^-$\end{tabular}}}}%
  \end{picture}%
\endgroup%

        }
        \subcaption{Narrow-band signed distances from the search radii and the octree.}
        \label{fig:nbanddist}
    \end{subfigure}
    \begin{subfigure}[t]{0.29\textwidth}
        \captionsetup{singlelinecheck = false, justification=justified}
        \centering
        \def\svgwidth{\textwidth}
        {\footnotesize
          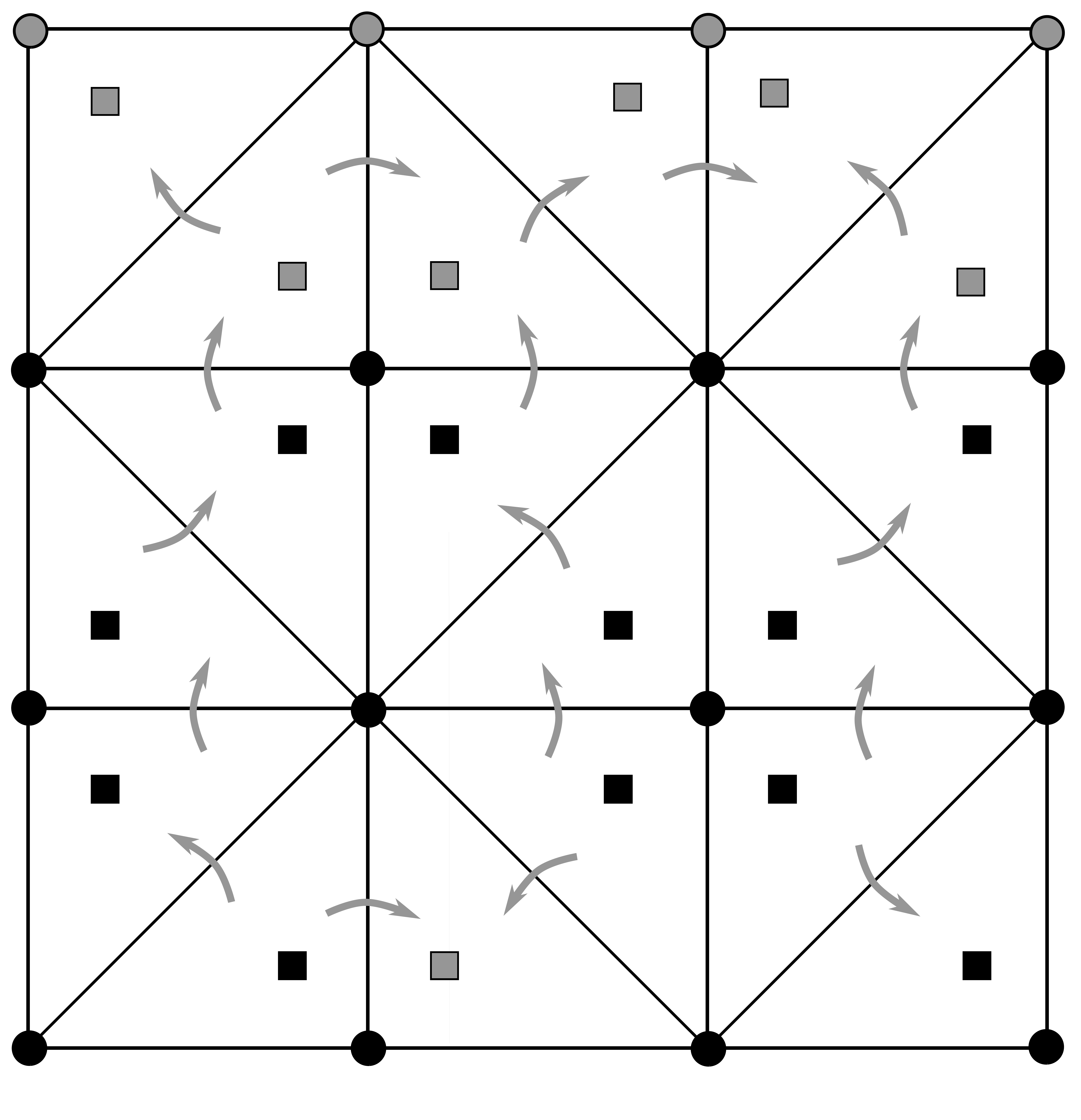
        }
        \subcaption{Positive and negative sign diffusion throughout $\tilde{\Omega}$.}
        \label{fig:distdiff}
    \end{subfigure}
    \hspace{1.5em}
    \begin{subfigure}[t]{0.29\textwidth}
        \captionsetup{justification=justified}
        \centering
        \def\svgwidth{\textwidth}
        {\footnotesize
          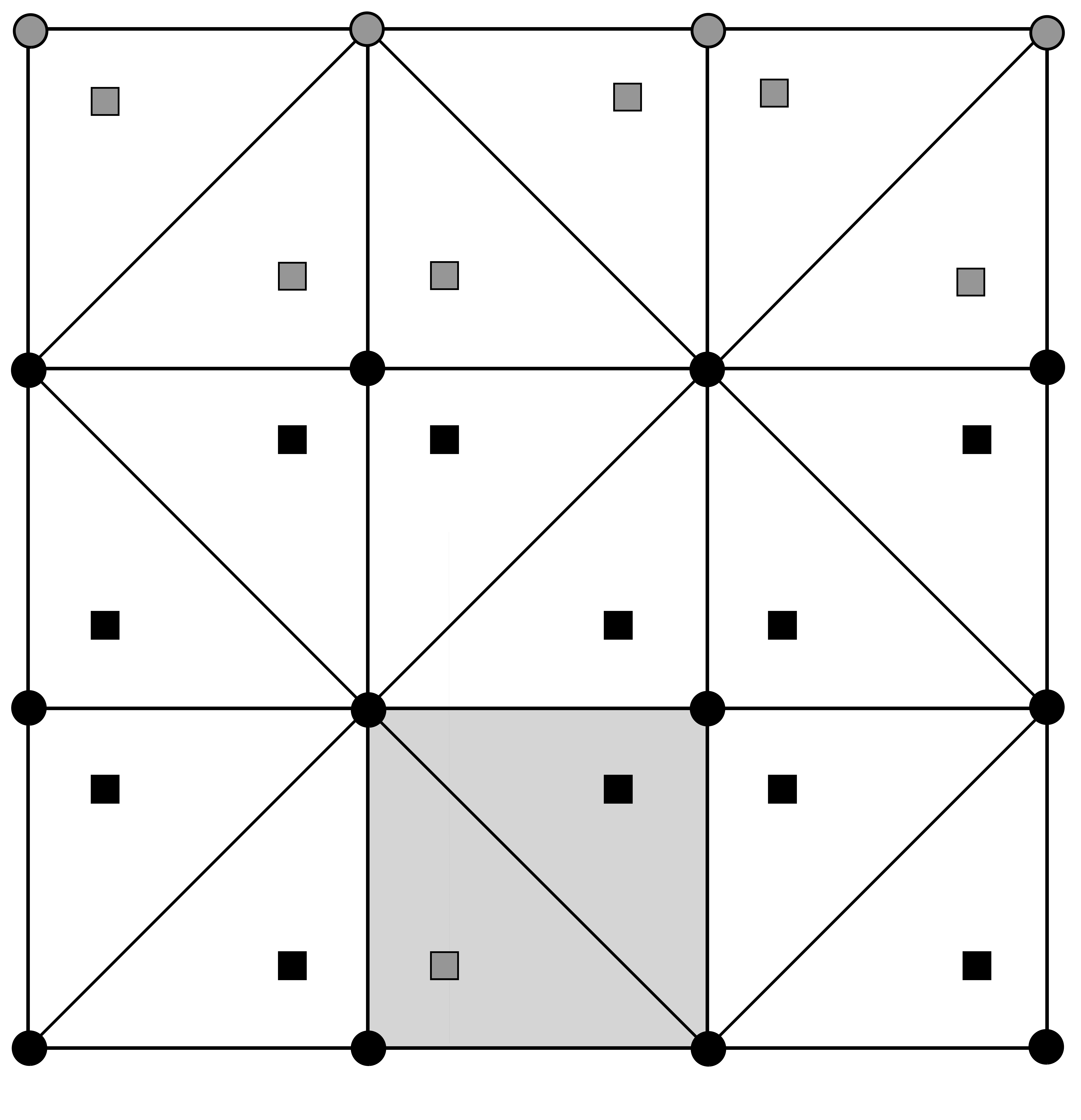
        }
        \subcaption{Phase indicator $\IndicatorAppr$ is $1$ or $0$ in cells strictly inside/outside of $\InterfaceAppr$, respectively.}
        \label{fig:indicatorinside}
    \end{subfigure}
    \hspace{1.5em}
    \begin{subfigure}[t]{0.29\textwidth}
        \captionsetup{singlelinecheck = false, justification=justified}
        \centering
        \def\svgwidth{\textwidth}
        {\footnotesize
          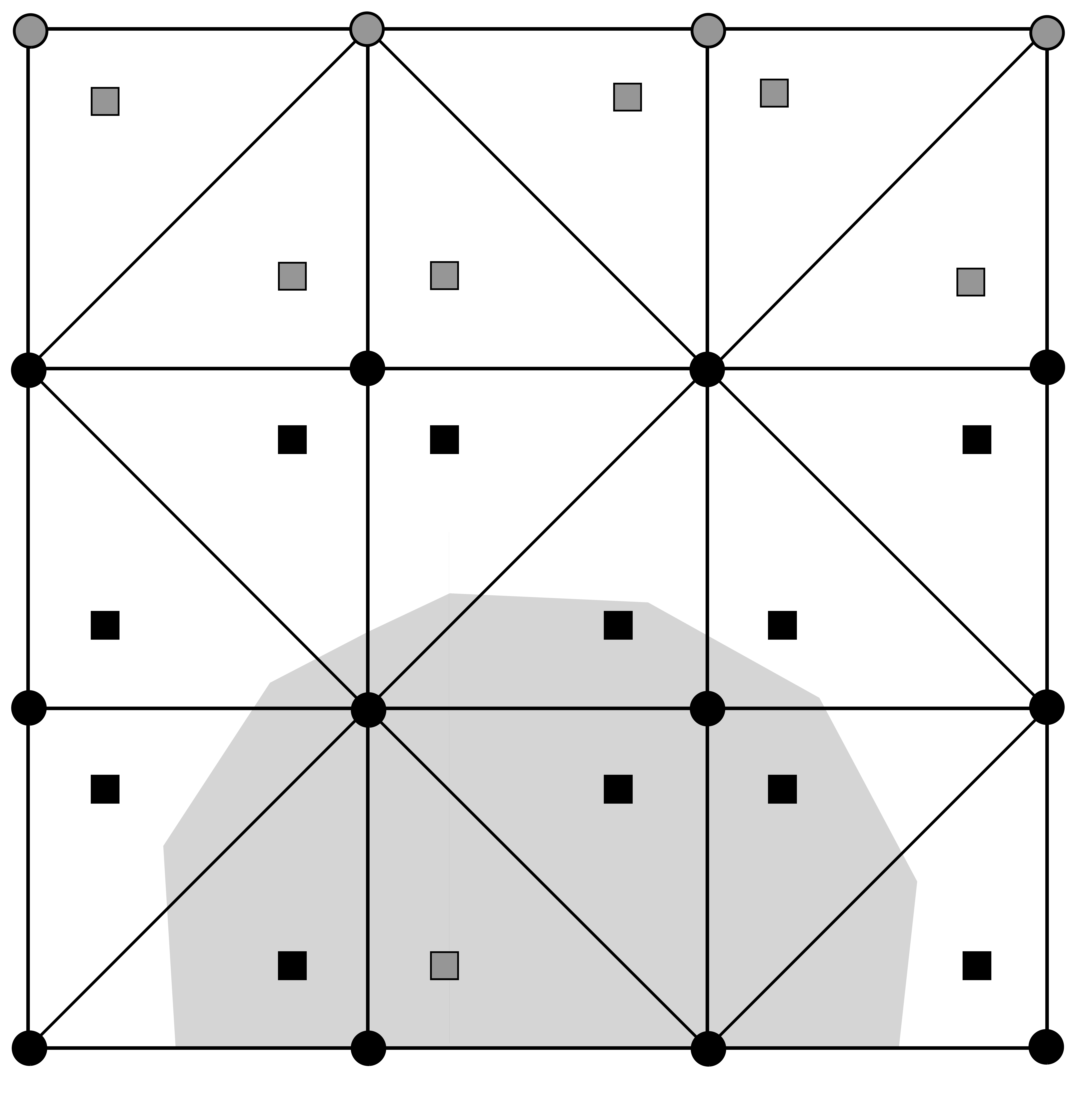
        }
        \subcaption{Computing the approximated phase indicator $\IndicatorAppr$ in intersected cells.}
        \label{fig:indicatorcut}
    \end{subfigure}
  \caption{Steps of the Surface Mesh Intersection / Approximation (SMCI/A) algorithms.}
  \label{fig:smcia}
\end{figure}

The calculation of volume fractions by the proposed Surface Mesh Cell Intersection/Approximation (SMCI/A) algorithm, outlined in \cref{fig:smcia}, requires signed distances to the interface at cell centres and cell corner points. As a naive computation is computationally \textcolor{Reviewer1}{expensive} (\cref{subsec:sigdistcalc}), we employ an octree based approach to the calculation of signed distances. Starting point of the octree based search is the calculation of search radii at the relevant points.

\subsection{Calculation of search radii}
\label{sub:searchradii}
In the first step, a search radius $r_c$
and $r_p$
is calculated at each cell center and cell-corner point, respectively. This is illustrated in \cref{fig:searchradii}. Here, the cell search radius $r_c$ is defined by
\begin{equation}
    r_c = 
    \lambda_s \min_{f \in F_c} \| \x_{f,O} - \x_{f,N} \|_2,
    \label{eq:rc}
\end{equation}
where $\x_c$ is the cell center, $\lambda_s > 0$  is the \emph{search radius factor} detailed below and $\x_{f,O}$, $\x_{f,N}$ are the cell centers of two cells that share the face with index $f$ of the cell $\Cell_c$ ($O$ for owner cell with a smaller cell index than the neighbor cell $N$). Here, the index set $F_c$ contains the indices of those faces that form the boundary of $\Omega_c$. Based on \eqref{eq:rc}, the corner-point search radius $r_p$ is defined by  
\begin{equation}
    r_p = 
    \lambda_s \min_{c \in C_{p}(\x_p)} r_{c}, 
    \label{eq:rp}
\end{equation}
where $\x_p$ is the cell-corner point, while the \emph{point-cell stencil} is the index set $\CellStencil(\x_p, \tilde{\Omega})$, that contains indices of all cells from $\tilde{\Omega}$ whose corner-point is $\x_p$.

The search radii introduced above are used to define search balls in $3D$ (circles in $2D$), which are used to reduce the number of calculations to determine signed distances
between the cell corner points $\x_p$ and the cell centers $\x_c$ with respect to the provided surface mesh $\InterfaceAppr$.

\subsection{Octree decomposition of the surface mesh and signed distance calculation}
\label{subsec:sigdistcalc}
In contrast to various other approaches for volume fraction initialization, \textcolor{Reviewer1}{the fluid interface is not represented by the proposed algorithm using a function, but as a surface mesh, consisting of triangles}.
To define the interface $\InterfaceAppr$, we first denote the convex hull of a set of $n$ points $P^n=\{\x_1,\ldots,\x_n\},\x_i \in \R^3$ by
\begin{equation}
  \conv(P^n):= \left\{\x \in \R^3 : \x = \sum_{\x_i \in P^n} \gamma_i\x_i,
    \sum_{i=1}^n \gamma_i = 1\right\}.
    \label{eq:convexcombination}
\end{equation}
Using this, a triangle is defined as the convex hull of a point triple: $\Triangle := \conv(P^3)$. Consequently, the surface mesh is defined as
\begin{equation}
  \InterfaceAppr := \{\Triangle_1, \Triangle_2,\ldots,
    \Triangle_n\}.
  \label{eq:interfacedef}
\end{equation}
With the structure of $\InterfaceAppr$ in mind, we want to emphasize why an octree based approach is the key to obtaining reasonable computation times. Consider the case where a minimal distance between a point $\x$ and $\InterfaceAppr$ would be calculated
for each cell center $\x_c$ and cell-corner point $\x_p$. 
The need for the spatial subdivision and search operations becomes obvious, as this would require a distance computation between each point of the interface mesh and each cell centers and cell corner points of the background mesh. 
Consequently, this would require $|C||\InterfaceAppr|$ operations to compute the geometric signed distances at cell centers and additional computations for evaluating signed distances at cell-corner points. 
For our computations below, the number $|C|$ often reaches the order of $1e05$ per CPU core, while $|\InterfaceAppr|$ is typically on the order of $1e04$ per CPU core. 
Aiming at redistancing computations for a dynamic setting in multiphase flows where $\InterfaceAppr=\InterfaceAppr(t)$, such a large number of distance computations makes such a brute force redistancing approach prohibitively expensive.

The first step of the signed distance calculation is the computation of an Axis-Aligned Bounding Box (AABB) from the surface mesh $\InterfaceAppr$.
The AABB is used to build an octree data structure, illustrated as a $2D$ quadtree subdivision in \cref{fig:octree}, which is used to access $\InterfaceAppr$. The octree data structure enables fast search queries involving cell centers and cell corner-points that are close to the surface mesh $\InterfaceAppr$, with a logarithmic computational complexity with respect to the number of vertices in $\InterfaceAppr$ \citep{Meagher1982,Mehta2004}. The structure of the octree depends on the ordering of vertices in $\InterfaceAppr$: since $\InterfaceAppr$ is an unstructured surface mesh, its vertices are generally sufficiently unordered, which makes the octree well-balanced.
Once the octree has been constructed, it can be used to find the closest points $\x \in \InterfaceAppr$ to cell centres $\x_c$ and cell corner points $\x_p$.
Note that this is only true for those $\x_c,\x_p$ which are sufficiently close to $\InterfaceAppr$ in terms of their search radius $r_c,r_p$.
\todo{Check N used for sets above. Tobi: this is fine. Above $N$ is used to denote some set, while here $\mathcal{N}$ is used.} Thus, the search radii define a so-called \emph{narrow band} around $\InterfaceAppr$, where the nearest distances are calculated geometrically. We denote the narrow band of $\InterfaceAppr$ with $\NarrowBand(\InterfaceAppr)$, and the closed ball $\Ball(\x^*,r) := \{\x \in \R^3 |\, \| \x-\x^*\|_2 \leq r\}$ with a radius $r$ around a point $\x$. Then
\todo{check whether we can define this as subset of the cell corner points and cell centers}
\begin{equation}
    \NarrowBand(\InterfaceAppr) : = \left\{\x \in \mathbb{R}^3 \vert ~ \exists ~\Triangle \in \InterfaceAppr \text { such that } \Triangle \cap \Ball(\x,r) \ne \emptyset \right\},
    \label{eq:narrowband}
\end{equation}
where $r$ is either $r_p$ or $r_c$.

For a point $\x \in \NarrowBand(\InterfaceAppr)$, the octree provides the
closest point $\x_\text{min} \in \Triangle_\text{min}$ for some $\Triangle \in \InterfaceAppr$ and the corresponding triangle $\Triangle_\text{min}$ itself. While the absolute distance can be directly computed as $\|\x - \x_\text{min}\|_2$, care must be taken 
when computing the sign with respect to the orientation of $\InterfaceAppr$. 
Directly using the triangle normals $\mathbf{n}_\Triangle$ may lead to false signs and consequently, to erroneous volume fractions. Thus, we follow the work 
of \citep{Thuerrner1998,Baerentzen2005} and compute \emph{angle weighted normals}
\begin{equation}
  \mathbf{n}_{\x_v} = \frac{\sum_{\Triangle \in \text{ngh}(\x_v)}\beta_\Triangle\mathbf{n}_\Triangle}
    {\sum_{\Triangle \in \text{ngh}(\x_v)} \beta_\Triangle
  }
  \label{eq:vertexnormal}
\end{equation}
at the vertices $\x_v$ of $\InterfaceAppr$. Here, $\text{ngh}(\x_v)$ denotes 
the set of all triangles containing $\x_v$, $\mathbf{n}_\Triangle$ a triangle 
normal and $\beta_\Triangle$ the inner angle of  $\Triangle$ at $\x_v$.
\citet{Baerentzen2005} propose a classification of the point $\x_\text{min}$ 
whether it is located within a triangle, on an edge, or a vertex and base the choice of the normal on this classification. While such a classification is simple in theory, a robust implementation is difficult due to the limited precision of floating point arithmetic. Thus, we opt for a linear 
interpolation of $\mathbf{n}_{\x_v}$ within $\Triangle_\text{min}$ to $\x_\text{min}$, 
denoted $\mathbf{n}_I(\x_\text{min}, \Triangle_\text{min})$. With this normal computation, the signed distance between $\x$ and $\x_\text{min}$ is calculated by
\begin{equation}
  \SignedDistance^g(\x, \InterfaceAppr) =
    \text{sign}((\x - \x_\text{min}) \cdot \mathbf{n}_I(\x_\text{min}, \Triangle_\text{min}))\|\x - \x_\text{min}\|_2.
    \label{eq:tridist}
\end{equation}
where the supindex $g$ indicates a geometrical construction. This procedure is illustrated in \cref{fig:nbanddist}. The robustness of this approach with regard to inside/outside classification is 
demonstrated in \cref{subsec:cad}.

Using the spatial subdivision provided by the octree, the computational complexity for finding the minimal distances between mesh points and $\InterfaceAppr$ is reduced severely, as the vast majority of cell centers $\x_c$ are not even considered for calculation as no triangle $\Triangle \in \InterfaceAppr$ exists within the corresponding search ball. The closest triangles of those points $\x_c$, whose ball $\mathcal{B}(\x_c, r_c)$ intersects $\InterfaceAppr$ are found with logarithmic search complexity with respect to $|\InterfaceAppr|$. This significant reduction of complexity can potentially enable a future application of the proposed algorithm on moving interfaces $\InterfaceAppr(t)$ as a geometrically exact marker field model for unstructured Front Tracking methods. Therefore, it is crucial to understand that the $\min_{\Triangle \in \InterfaceAppr}$ operation in \cref{eq:tridist} throughout this text relies on the octree spatial subdivision and search queries. 

\subsection{Signed distance propagation} 
\label{subsec:sigdistpropagation}
After the calculation of geometric signed distances in the narrow band around $\InterfaceAppr$, the signed distances are propagated to the bulk of different phases, as shown in \cref{fig:distdiff}. In \citep{Maric2015,Tolle2020}, the geometric signed distances are set to large positive numbers throughout the domain, and a graph-traversal algorithm is used to iteratively correct the signs of signed distances using face-cell and point-point graph connectivity provided by the unstructured mesh. Graph-traversal is computationally expensive and complicated to implement in parallel. Here we propose a straightforward alternative that instantaneously propagates signs of signed distances through the solution domain and is parallelized easily. We rely on the diffusion equation for the signed distances, namely 
\begin{equation}
  \begin{aligned}
    - \Delta \SignedDistance &= 0, \\
    \nabla \phi  &= 0, \quad \text{for}\quad  \x \in \partial \Omega
  \end{aligned}
  \label{eq:distdiff}
\end{equation}
and its discretization using the unstructured finite volume method in OpenFOAM \cite{Jasak1996,Juretic2005,Moukalled2016}, giving a linear system of equations.
The key idea to sign propagation is to apply a few iterations ($<5$) of an 
iterative linear solver to this system. In our case a Conjugate Gradient approach with an incomplete lower upper preconditioner has been used. With the initial field set to
\begin{equation}
    \begin{aligned}
    \phi (\x) & = \begin{cases}
        \phi_g(\x,\InterfaceAppr), & \quad \text{if } \x \in \NarrowBand(\InterfaceAppr) \\
    0, & \quad \text{otherwise,}
    \end{cases}
    \end{aligned}
    \label{eq:distinitial}
\end{equation}
this small number of iterations suffices to properly propagate
$\text{sign}(\SignedDistance)$ with respect to the orientation of 
$\InterfaceAppr$ throughout $\tilde{\Omega}$. Prerequisite for this approach to work 
is that the narrow band has a certain minimum width in interface normal 
direction. At least four cells on each side of the interface are required to 
ensure a robust propagation. This is achieved by setting a global search radius factor $\lambda_s :=4$ in \cref{eq:rc} used to calculate $r_c$ at cell centers. Note that increasing $\lambda_s$ beyond this value only increases computational costs, and does not impact the accuracy of the proposed algorithm, as with a larger value of $\lambda_s$ the narrow band $\NarrowBand(\Sigma)$ becomes wider and consequently the geometrical signed distances are calculated at more points $\x_c,\x_p$, using \cref{eq:phic,eq:phip}, respectively. 

Two aspects have to be considered when solving the linear system of equations
resulting from the discretization of \cref{eq:distdiff}. First, cells for which 
$\x_c \in \NarrowBand(\InterfaceAppr)$ have to be excluded from the vector of 
unknowns as $\SignedDistance^g(\x_c)$ is already known for those. Second, for
cells away
from $\NarrowBand(\InterfaceAppr)$ the only relevant information is 
$\text{sign}(\SignedDistance_c)$ indicating $\Cell_c \in \Omega^-$ or 
$\Cell_c \in \Omega^+$, respectively. A few iterations of a linear solver suffice to 
reliably propagate $\text{sign}(\SignedDistance_c)$ to the entire domain.
The resulting field is
\begin{equation}
    \SignedDistance_{c} = \begin{cases} 
        \phi^g_{c},& \text{if } \x_c \in \NarrowBand(\InterfaceAppr), \ \\
        \phi^a_c, & \text {otherwise,}
    \end{cases}
    \label{eq:phic}
\end{equation}
with $\SignedDistance^g_c$ denoting geometric signed distances and 
$\SignedDistance^a_c$ approximate values from the solution of \cref{eq:distdiff}
carrying inside/outside information but without geometric meaning.

Once the cell-centered signed distances $\SignedDistance_c$ are computed, they are used to calculate the signed distances at cell corner-points via
\begin{equation}
    \SignedDistance^I_{p} = \sum_{c \in C_p} w_{p,c} \SignedDistance_{c},  
    \label{eq:phipidw}
\end{equation}
where $C_p$ is the index set of cells that contain the cell corner point $\x_p$
and the supindex $I$ indicating interpolation. Furthermore, $w_{p,c}$ is the \emph{inverse-distance weighted}
(IDW) interpolation weight
\begin{equation}
    w_{p,c} = \frac{\|\x_c - \x_p\|_2^{-1}}{\sum_{\tilde{c} \in C_p} \|\x_{\tilde{c}} - \x_p\|_2^{-1}}.
\end{equation}
As with $\SignedDistance_{c}$, the accuracy of $\SignedDistance_{p}$ is irrelevant outside of the narrow band of $\InterfaceAppr$, only the sign of the signed distance is important in the bulk. To correct for the error introduced by the IDW-interpolation in \cref{eq:phipidw}, signed distances at cell-corner points of intersected cells are calculated geometrically 
\begin{equation}
    \SignedDistance_{p} = \begin{cases} 
        \phi^g_p, & \quad \text{if } \x_p \in \NarrowBand(\InterfaceAppr), \\
        \phi^I_p, & \quad \text{otherwise.}
    \end{cases}
    \label{eq:phip}
\end{equation}

\Cref{eq:phic,eq:phip} define the final signed distances at cell centers and cell-corner points, respectively. These quantities will have the value of a geometrical distance to $\InterfaceAppr$ in the narrow band, while outside of the narrow band only the correct sign resulting from the approximative solution of \cref{eq:distdiff} \textcolor{Reviewer1}{is relevant}.  


\subsection{Volume fraction calculation}
\label{subsec:vof-geom}

\begin{figure}[!htb]
    \centering
    \begin{subfigure}[h]{0.35\textwidth}
      \def\svgwidth{\textwidth}
      {\footnotesize
\begingroup%
  \makeatletter%
  \providecommand\color[2][]{%
    \errmessage{(Inkscape) Color is used for the text in Inkscape, but the package 'color.sty' is not loaded}%
    \renewcommand\color[2][]{}%
  }%
  \providecommand\transparent[1]{%
    \errmessage{(Inkscape) Transparency is used (non-zero) for the text in Inkscape, but the package 'transparent.sty' is not loaded}%
    \renewcommand\transparent[1]{}%
  }%
  \providecommand\rotatebox[2]{#2}%
  \newcommand*\fsize{\dimexpr\f@size pt\relax}%
  \newcommand*\lineheight[1]{\fontsize{\fsize}{#1\fsize}\selectfont}%
  \ifx\svgwidth\undefined%
    \setlength{\unitlength}{537.54321481bp}%
    \ifx\svgscale\undefined%
      \relax%
    \else%
      \setlength{\unitlength}{\unitlength * \real{\svgscale}}%
    \fi%
  \else%
    \setlength{\unitlength}{\svgwidth}%
  \fi%
  \global\let\svgwidth\undefined%
  \global\let\svgscale\undefined%
  \makeatother%
  \begin{picture}(1,0.9753131)%
    \lineheight{1}%
    \setlength\tabcolsep{0pt}%
    \put(0,0){\includegraphics[width=\unitlength,page=1]{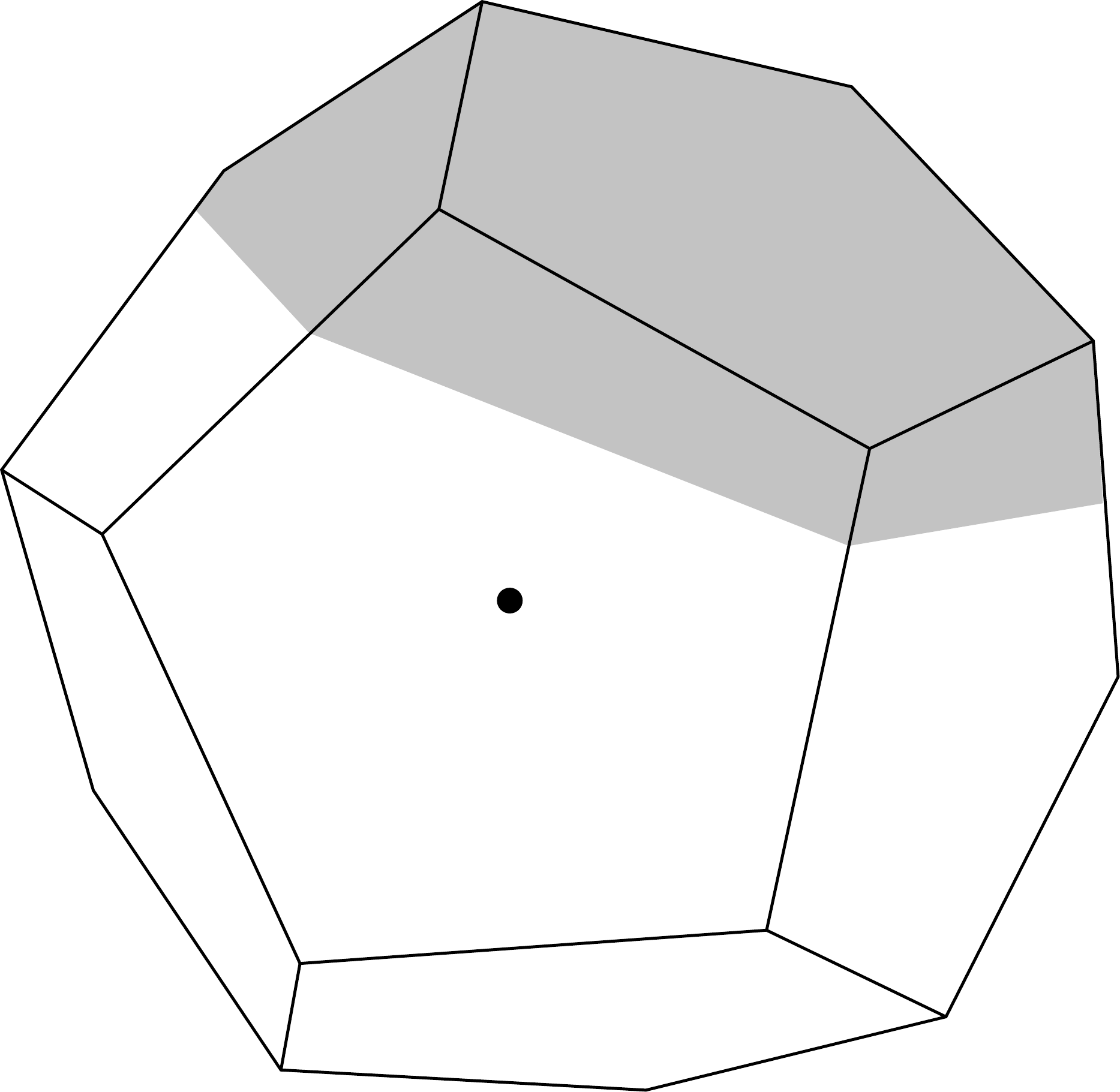}}%
    \put(0.46933083,0.40167952){\color[rgb]{0,0,0}\makebox(0,0)[lt]{\lineheight{1.25}\smash{\begin{tabular}[t]{l}$\mathbf{x}_c$\end{tabular}}}}%
  \end{picture}%
\endgroup%

      }
      \caption{A cell $\Cell_c$ intersected by $\InterfaceAppr$.}
      \label{fig:interfacecell}
    \end{subfigure}
    \hspace{5em}
    \begin{subfigure}[h]{0.35\textwidth}
      \def\svgwidth{\textwidth}
      {\footnotesize
        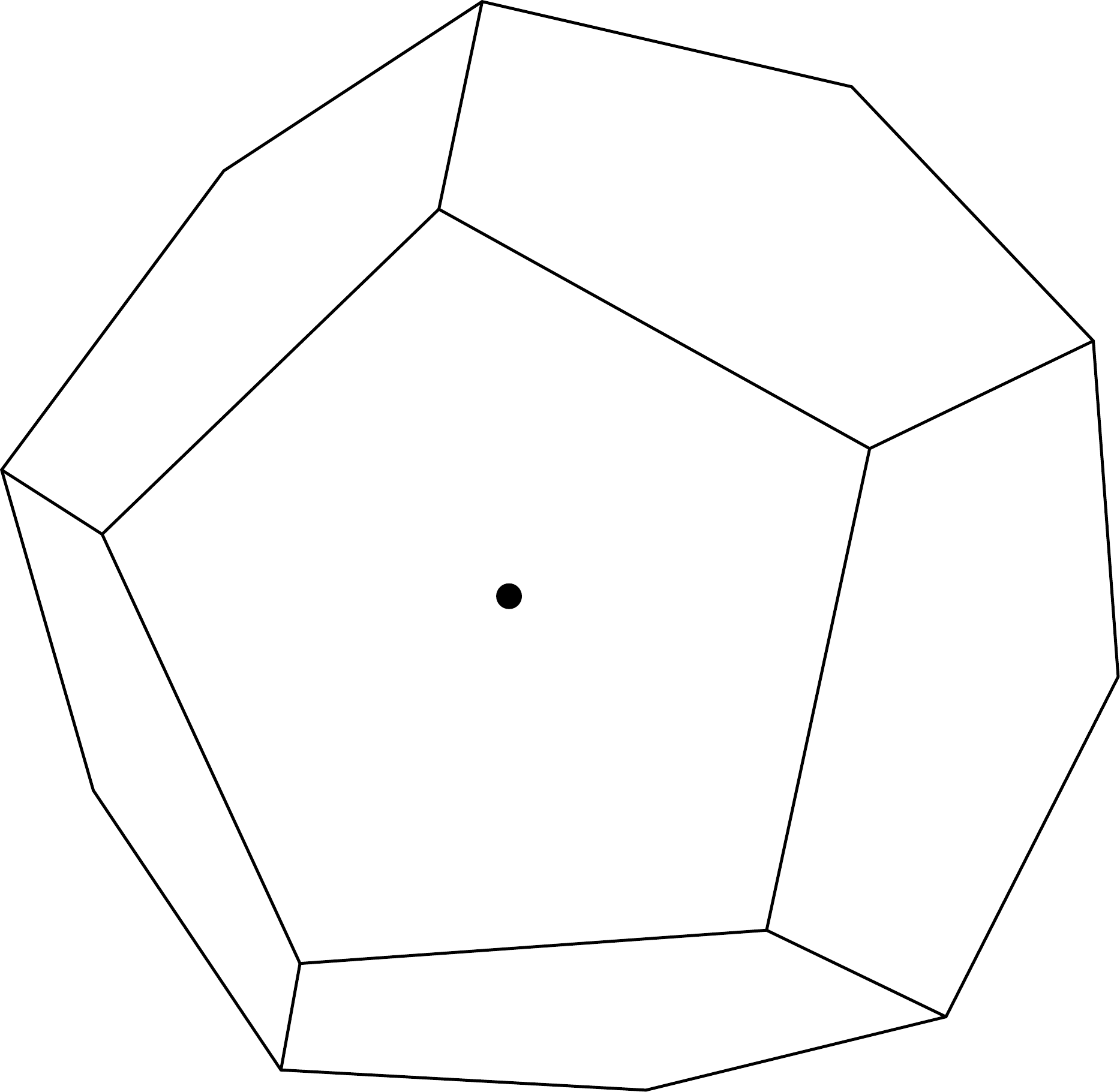
      }
      \caption{Tetrahedral cell decomposition.}
      \label{fig:smcivolcalc}
    \end{subfigure}
    \caption{Centroid decomposition of an interface cell into tetrahedra and 
      calculation of $\VolFrac_c$ using the SMCI/A algorithms.
    }
    \label{fig:centroid-decomposition}
\end{figure}

Once the signed distances at cell centers $\{\phi_c\}_{c = 1,2,\dots, |\tilde{\Omega}|}$ and cell corner points $\{\phi_p\}_{p = 1,2, \dots |P_h|}$ are calculated as outlined in the previous section, the SMCI algorithm calculates the volume fractions in a straightforward way. 
The volume fraction calculation is shown schematically for the SMCI algorithm in \cref{fig:smcivolcalc}. 
Each cell is decomposed into tetrahedrons, using the cell centroid $\x_c$ as the base point of the tetrahedron, the centroid of the face $\x_{c,f}$, and two successive points from the cell-face, $\x_{c,f,i},\x_{c,f,i+1}$. 
The resulting tetrahedron has the distance $\phi_c$ associated to the cell centroid, the distance $\phi_{c,f}$ associated to the face centroid, and and $(\phi_{c,f,i},\phi_{c,f,i+1})$ pair of distances associated with a pair of points that belong to the cell-face $(c,f)$, as shown in \cref{fig:smcivolcalc}.
If all the distances of the tetrahedron are negative, the tetrahedron lies in the negative halfspace with respect to $\InterfaceAppr$, and its total volume contributes to the sum of the volume of phase $1$ inside the volume $\Omega_c$.  
If a pair of distances in a tetrahedron has different signs, the tetrahedron is intersected by the interface approximated by the surface mesh $\InterfaceAppr$. 
The volume of this intersection is calculated by geometrically intersecting the tetrahedron with those triangles from $\InterfaceAppr$, that have a non-zero intersection with a ball $\mathcal{B}$ enclosing the tetrahedron. 
The center of the ball $\mathcal{B}_{c,f,i} := \mathcal{B}(\x_{c,f,i}, R{c,f,i})$ is the centroid of the tetrahedron $\x_{c,f,i} = 0.25(\x_c + \x_{c,f} + \x_{c,f,i} + \x_{c,f,\text{mod}(i+1, |F_{c,f}|)})$, where $i = 0,\dots, |F_{c,f}|-1$, and $F_f$ is the oriented set of indices of the points $\x$ (cf. \cref{fig:smcivolcalc}) that belong to the face $f$ of the cell $\Omega_c$. 
The radius of the tetrahedron-ball $\mathcal{B}_{c,f,i}$ is then
\begin{equation}
    R_{c, f, i} = \max(\|\x_c - \x_{c,f,i}\|,\|\x_{c,f} - \x_{c,f,i}\|,\|\x_{c,f,j} - \x_{c,f,i} \|, \|\x_{c,f,\text{mod}(j+1,|F_{c,f}|)} - \x_{c,f,i} \|), 
\end{equation}
$j = 0,\dots, |F_{c,f}|-1$.
This sub-set of $\InterfaceAppr$ is found using the octree data structure with logarithmic complexity with respect to $\InterfaceAppr$, as outlined in the previous section. 
For the example tetrahedron in the cell shown in \cref{fig:smcivolcalc}, the resulting intersection between the approximated interface $\InterfaceAppr$ and a tetrahedron from the cell $\Omega_c$ is shown as the shaded volume. 
The magnitude of this volume is computed by applying the Gauss divergence theorem using \cref{eq:Ve-expanded}.  
The phase-specific volumes from cell-tetrahedrons are summed up for the cell $\Omega_c$, into the total phase-specific volume of the phase $1$ within the cell $\Omega_c$, and the volume fraction is therefore computed as 
\begin{equation}
    \alpha_c = \dfrac{
        \sum_{f = 0, \dots |C_c|-1} \sum_{i = 0, \dots, |F_{c,f}|-1}
            | 
                T(\x_c, \x_{c,f}, \x_{c,f,{i}}, \x_{c,f,\text{mod}(i+1, |F_{c,f}|)}) \cap 
                (\mathcal{B}_{c,f,i} \cap 
                \InterfaceAppr)
            |
        }
        {|\Omega_c|}
        \label{eq:volfracsmci}
\end{equation}
with $\Tetrahedron := \{\x_1,\x_2,\x_3,\x_4\}$  denoting a tetrahedron.
The SMCI algorithm is summarized by \cref{alg:gvof:smci}. 

\begin{algorithm}[!htb]
    \centering
    \caption{The Surface-Mesh / Cell Intersection Algorithm (SMCI)} 
    \label{alg:gvof:smci}
    {\small
        \begin{algorithmic}[1]
        \State $\PlicFraction = 0$, $\phi_{c,p} = 0$
        \State Compute search radius for cell centers ${r_c}_{c \in C}$ using \cref{eq:rc}.
        \For{cell centroids $\{\x_c\}_{c \in C}$} 
            \State Place the vertices of $\InterfaceAppr$ into an octree (\cref{subsec:sigdistcalc}).
            \State Find the triangle $\Triangle_n \in \InterfaceAppr$ nearest to $\x_c$ within a ball $\mathcal{B}(\x_c, r_c)$.
            \State Set $\phi^g_c:=\phi^g(\x_c,\Triangle_n)$ using \cref{eq:tridist}.  
        \EndFor
        \State Approximately solve \cref{eq:distdiff} to propagate $sign(\phi_c)$.
        \State Compute search radius for cell corner points ${r_p}_{p \in P}$ using \cref{eq:rp}.
            \State Find all intersected cells $I = \{c, \quad \phi_c \phi_p < 0 \text{ for at least one } $p$ \}$.
        \State Use \cref{eq:phic} to correct $\phi_c$ within the narrow band.
        \State Compute $\phi_p$ in the bulk using \cref{eq:phipidw}.   
        \State Use \cref{eq:phip} to correct $\phi_p$ within the narrow band.
        \For{cells $\{\Omega_c\}_{c \in C}$} 
            \If {$\phi_c \le 0$ and all corner-point distances $\phi_p \le 0$} \Comment Cell is inside the negative $\InterfaceAppr$-halfspace.
                \State $\alpha_c = 1$ 
            \EndIf
            \If {cell $\Omega_c$ is intersected, $c \in I$} \Comment Cell is intersected by $\InterfaceAppr$.
                \State $\alpha_c$ given by \cref{eq:volfracsmci}.
            \EndIf
        \EndFor
      \end{algorithmic}
    }
\end{algorithm}

\section{Surface-Mesh / Cell Approximation algorithm}
\label{sec:algorithm-pvof}
This section presents an alternative approach to the computation of volume fractions presented in \cref{subsec:vof-geom}. While  \cref{subsec:vof-geom} details a method based on geometric intersections, this section introduces an algorithm based on volumetric reconstruction by adaptive mesh refinement.
\citet{Detrixhe2016} introduce a second order accurate approximation for the
volume fraction of a triangle (2D) or a tetrahedron (3D). Their model is an
algebraic expression taking the signed distances \SignedDistance{} of the
vertices as arguments.
In contrast, we propose a volume fraction initialization algorithm that employs
this model in combination with an adaptive tetrahedral
cell decomposition and the octree-based signed distance calculation described
in \cref{sec:alphainit}. We term this algorithm \emph{Surface-Mesh/Cell Approximation}
(\Pvof{}) and it is outlined below.

\begin{figure}[htb]
    \captionsetup{justification=centering}
  \centering
    \begin{subfigure}[t]{0.29\textwidth}
        \captionsetup{singlelinecheck = false, justification=justified}
        \centering
        \def\svgwidth{\textwidth}
        {\footnotesize
          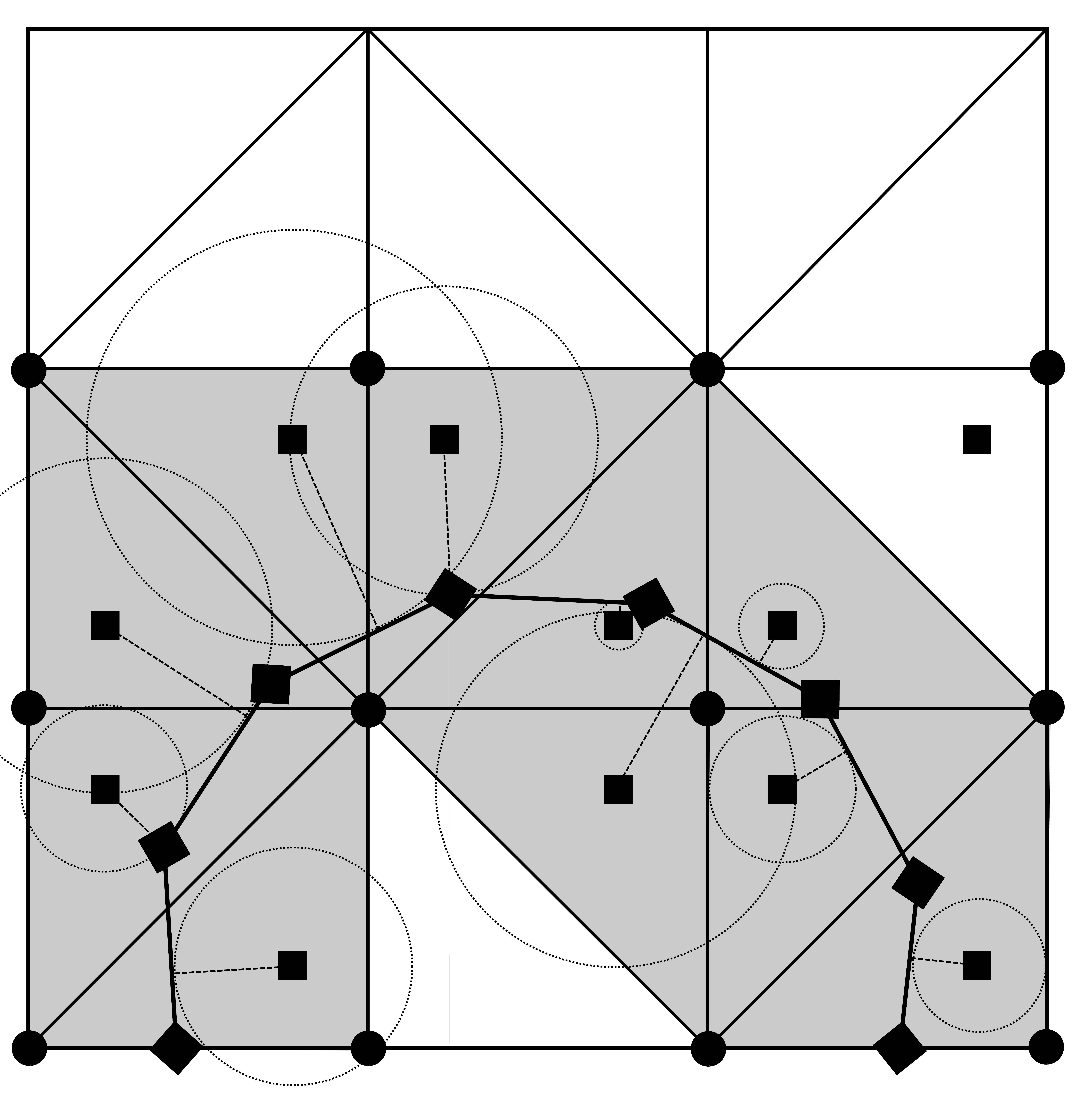
        }
        \subcaption{Identify potential interface cells (marked grey) using
         bounding ball criterion. Shown are circles with radii
         $|\SignedDistance_c|$.}
        \label{fig:alg-identifyinterfacecell}
    \end{subfigure}
    \hspace{1.5em}
    \begin{subfigure}[t]{0.29\textwidth}
        \captionsetup{singlelinecheck = false, justification=justified}
        \centering
        \def\svgwidth{\textwidth}
        {\footnotesize
          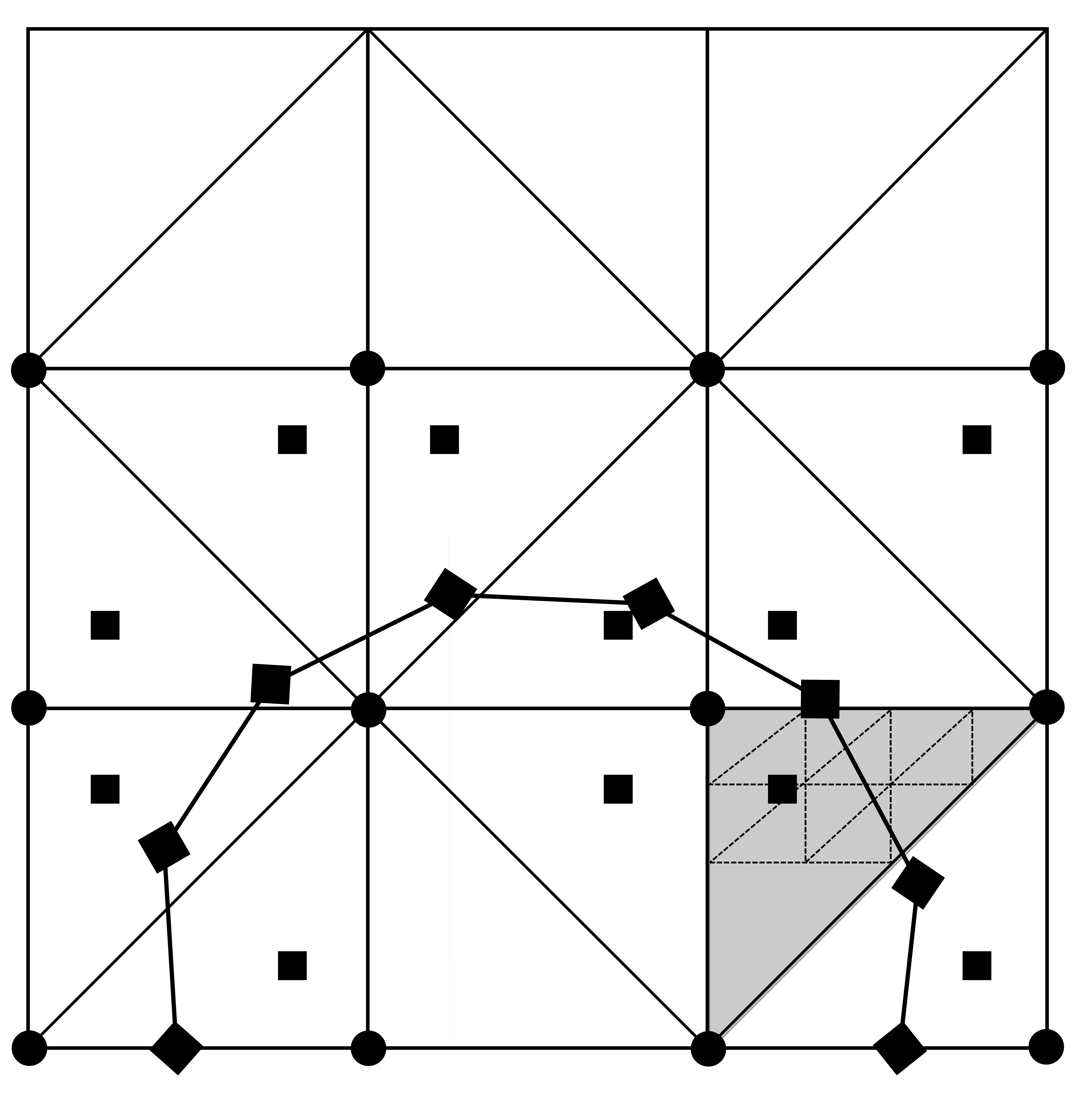
        }
        \subcaption{Adaptive, tetrahedral decomposition of interface cells.
             Compute $\SignedDistance$ at new vertices.}
        \label{fig:alg-decomposition}
    \end{subfigure}
    \hspace{1.5em}
    \begin{subfigure}[t]{0.29\textwidth}
        \captionsetup{singlelinecheck = false, justification=justified}
        \centering
        \def\svgwidth{\textwidth}
        {\footnotesize
          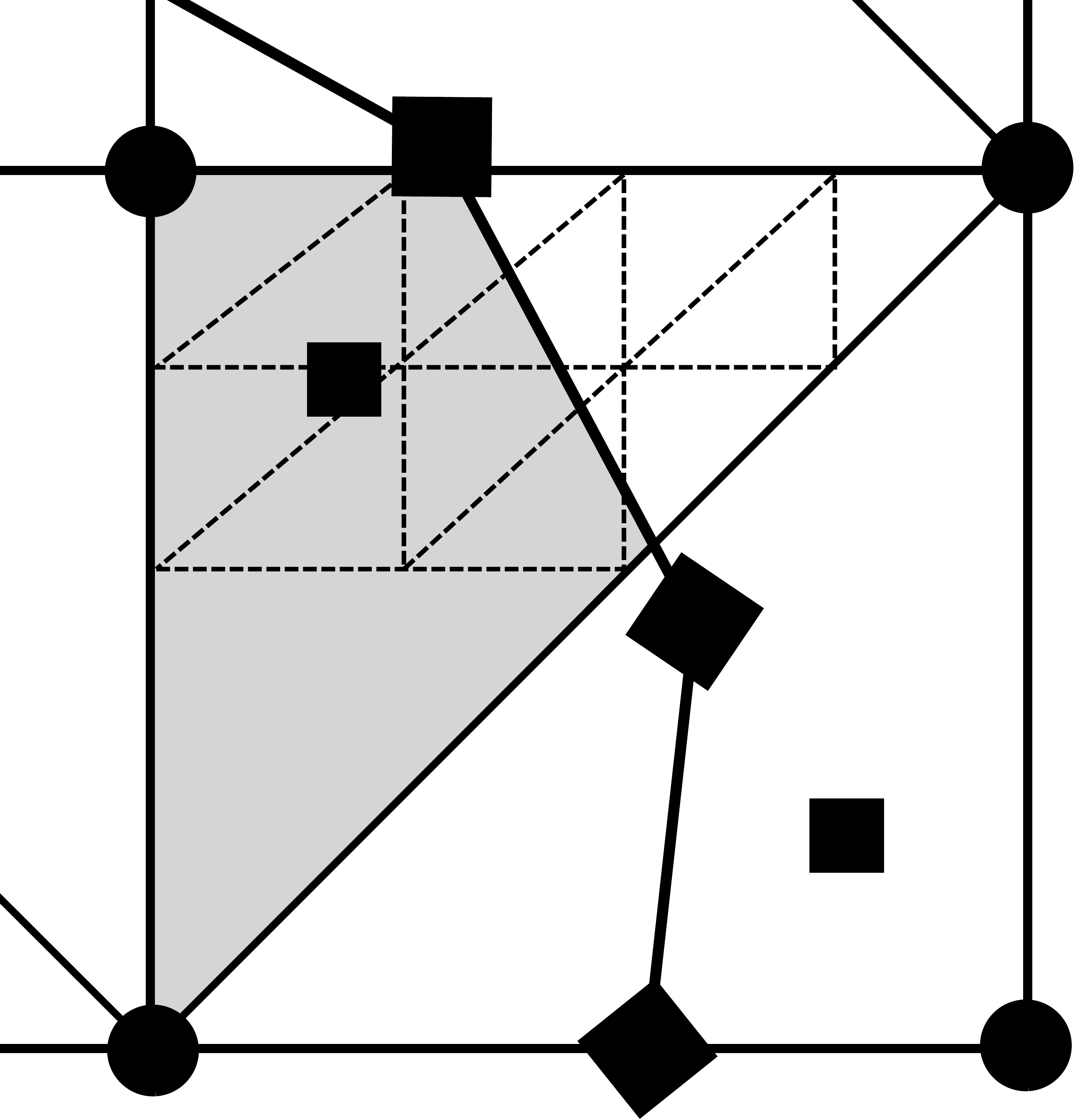
        }
        \subcaption{Compute the volume fraction $\VolFrac_c$ using
            the model of \citet{Detrixhe2016} (detail view).}
        \label{fig:alg-vof-calculation}
    \end{subfigure}
    \caption{Steps of the \Pvof{} algorithm following signed distance computation and
        inside/outside propagation.
    }
    \label{fig:pvof}
\end{figure}
\begin{figure}[htb]
    \centering
    \begin{subfigure}[t]{0.45\textwidth}
        \captionsetup{singlelinecheck = false, justification=justified}
        \centering
        \def\svgwidth{\textwidth}
        {\footnotesize
\begingroup%
  \makeatletter%
  \providecommand\color[2][]{%
    \errmessage{(Inkscape) Color is used for the text in Inkscape, but the package 'color.sty' is not loaded}%
    \renewcommand\color[2][]{}%
  }%
  \providecommand\transparent[1]{%
    \errmessage{(Inkscape) Transparency is used (non-zero) for the text in Inkscape, but the package 'transparent.sty' is not loaded}%
    \renewcommand\transparent[1]{}%
  }%
  \providecommand\rotatebox[2]{#2}%
  \newcommand*\fsize{\dimexpr\f@size pt\relax}%
  \newcommand*\lineheight[1]{\fontsize{\fsize}{#1\fsize}\selectfont}%
  \ifx\svgwidth\undefined%
    \setlength{\unitlength}{226.77165354bp}%
    \ifx\svgscale\undefined%
      \relax%
    \else%
      \setlength{\unitlength}{\unitlength * \real{\svgscale}}%
    \fi%
  \else%
    \setlength{\unitlength}{\svgwidth}%
  \fi%
  \global\let\svgwidth\undefined%
  \global\let\svgscale\undefined%
  \makeatother%
  \begin{picture}(1,0.75)%
    \lineheight{1}%
    \setlength\tabcolsep{0pt}%
    \put(0,0){\includegraphics[width=\unitlength,page=1]{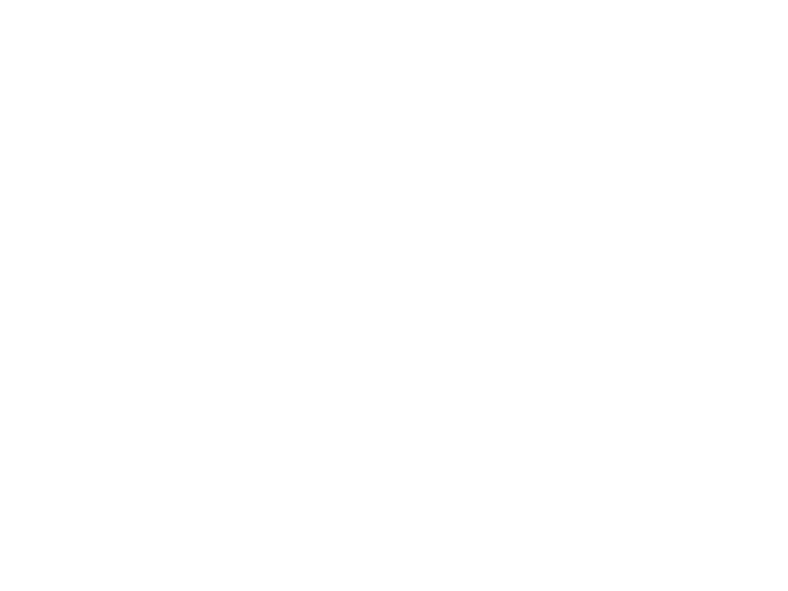}}%
    \put(0.05265631,0.18348832){\color[rgb]{0,0,0}\makebox(0,0)[lt]{\lineheight{1.25}\smash{\begin{tabular}[t]{l}$\Sigma$\end{tabular}}}}%
    \put(0,0){\includegraphics[width=\unitlength,page=2]{bounding-ball-concept.pdf}}%
    \put(0.71968278,0.35505689){\color[rgb]{0,0,0}\makebox(0,0)[lt]{\lineheight{1.25}\smash{\begin{tabular}[t]{l}$r_\text{bb}$\end{tabular}}}}%
    \put(0.60275607,0.24891612){\color[rgb]{0,0,0}\makebox(0,0)[lt]{\lineheight{1.25}\smash{\begin{tabular}[t]{l}$|\SignedDistance_c|$\end{tabular}}}}%
    \put(0,0){\includegraphics[width=\unitlength,page=3]{bounding-ball-concept.pdf}}%
    \put(0.64913057,0.43572864){\color[rgb]{0,0,0}\makebox(0,0)[lt]{\lineheight{1.25}\smash{\begin{tabular}[t]{l}$\x_c$\end{tabular}}}}%
    \put(0,0){\includegraphics[width=\unitlength,page=4]{bounding-ball-concept.pdf}}%
  \end{picture}%
\endgroup%

        }
        \subcaption{Bulk cell: the ball $\Ball(\x_c, |\SignedDistance_c|)$
            contains the cell bounding ball $\Ball(\x_c, r_{bb})$.}
        \label{fig:bs-concept}
    \end{subfigure}
    \hspace{1.5em}
    \begin{subfigure}[t]{0.45\textwidth}
        \captionsetup{singlelinecheck = false, justification=justified}
        \centering
        \def\svgwidth{\textwidth}
        {\footnotesize
\begingroup%
  \makeatletter%
  \providecommand\color[2][]{%
    \errmessage{(Inkscape) Color is used for the text in Inkscape, but the package 'color.sty' is not loaded}%
    \renewcommand\color[2][]{}%
  }%
  \providecommand\transparent[1]{%
    \errmessage{(Inkscape) Transparency is used (non-zero) for the text in Inkscape, but the package 'transparent.sty' is not loaded}%
    \renewcommand\transparent[1]{}%
  }%
  \providecommand\rotatebox[2]{#2}%
  \newcommand*\fsize{\dimexpr\f@size pt\relax}%
  \newcommand*\lineheight[1]{\fontsize{\fsize}{#1\fsize}\selectfont}%
  \ifx\svgwidth\undefined%
    \setlength{\unitlength}{226.77165354bp}%
    \ifx\svgscale\undefined%
      \relax%
    \else%
      \setlength{\unitlength}{\unitlength * \real{\svgscale}}%
    \fi%
  \else%
    \setlength{\unitlength}{\svgwidth}%
  \fi%
  \global\let\svgwidth\undefined%
  \global\let\svgscale\undefined%
  \makeatother%
  \begin{picture}(1,0.75)%
    \lineheight{1}%
    \setlength\tabcolsep{0pt}%
    \put(0,0){\includegraphics[width=\unitlength,page=1]{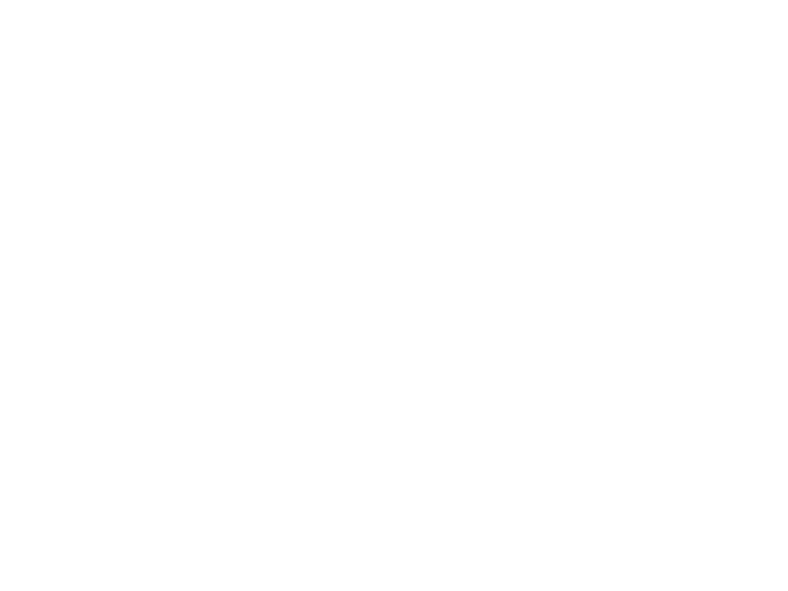}}%
    \put(0.05265631,0.18348832){\color[rgb]{0,0,0}\makebox(0,0)[lt]{\lineheight{1.25}\smash{\begin{tabular}[t]{l}$\Sigma$\end{tabular}}}}%
    \put(0,0){\includegraphics[width=\unitlength,page=2]{bounding-ball-false-positive.pdf}}%
    \put(0.52840222,0.23862466){\color[rgb]{0,0,0}\makebox(0,0)[lt]{\lineheight{1.25}\smash{\begin{tabular}[t]{l}$|\SignedDistance_c|$\end{tabular}}}}%
    \put(0,0){\includegraphics[width=\unitlength,page=3]{bounding-ball-false-positive.pdf}}%
    \put(0.57749135,0.31832239){\color[rgb]{0,0,0}\makebox(0,0)[lt]{\lineheight{1.25}\smash{\begin{tabular}[t]{l}$\x_c$\end{tabular}}}}%
    \put(0.47649284,0.35196569){\color[rgb]{0,0,0}\makebox(0,0)[lt]{\lineheight{1.25}\smash{\begin{tabular}[t]{l}$r_\text{bb}$\end{tabular}}}}%
    \put(0,0){\includegraphics[width=\unitlength,page=4]{bounding-ball-false-positive.pdf}}%
  \end{picture}%
\endgroup%

        }
        \subcaption{False positive: a bulk cell which is not detected by the bounding
            ball criterion as $\Ball(\x_c, r_{bb}) \nsubseteq \Ball(\x_c, |\SignedDistance_c|).$
        }
        \label{fig:bs-false-positive}
    \end{subfigure}
    \caption{Illustration of the idea of the bounding ball criterion in 2D for 
        clarity. The solid grey line represents $\Ball(\x_c, |\SignedDistance_c|)$,
        the grey dashed one $\Ball(\x_c, r_{bb})$.
    }
    \label{fig:boundingball}
\end{figure}

The SMCA-algorithm is based on the signed distance results of the SMCI-algorithm introduced in \cref{sec:alphainit}. The steps depicted in \cref{fig:searchradii} - \ref{fig:distdiff} of the SMCI/A
are used to compute $\SignedDistance_c,\SignedDistance_p$ in the narrow band
and propagate inside/outside information in the rest of the mesh points. Subsequent steps for the computation of volume fractions are displayed in
\cref{fig:pvof}. First, all cells intersected by $\InterfaceAppr$ are identified to reduce computational costs, as only these cells have intermediate values $0 < \VolFrac_c < 1$. This step is depicted in \cref{fig:alg-identifyinterfacecell}. Each cell for which
$\x_c \in \NarrowBand(\InterfaceAppr)$ is checked with the 
\emph{bounding ball criterion}. We define a bounding ball (bb) for a point $\x_\text{bb} \in \Cell_c$ using $r_{bb} = \max_{\x \in \Cell_c}\|\x - \x_{bb}\|_2$. This ball is the smallest ball that contains all points of $\Cell_c$. We compare this bounding ball to $\Ball(\x_\text{bb}, |\SignedDistance(\x_\text{bb}))$. These balls are shown in \cref{fig:boundingball}, where the bounding ball is illustrated by a dashed and the other ball by a continuous line.
As a general observation, if the bounding ball is contained in the ball with the radius $|\SignedDistance(\x_{bb})|$, i.e. $\Ball(\x_{bb}, r_{bb}) \subseteq \Ball(\x_{bb}, |\SignedDistance(\x_{bb})|)$, then such a cell is guaranteed to be a bulk cell. This cell can then be removed from the set of cells in the narrow band to reduce the number of cells which are considered for decomposition in the next step. 
If the criterion is not satisfied, the cell is considered an interface cell. Two remarks on this criterion: first, the existence of such a $\x_\text{bb}$ 
is not a necessary but a sufficient condition. Second, in a practical implementation evaluation of this criterion is only feasible for a small number of points when aiming to keep
computational costs reasonable. Thus, the actual check is performed by evaluating
\begin{equation}
    f_\text{bb}(\x, \SignedDistance_\x, \Cell_c) =
    \begin{cases}
        1, \quad \max_{\x_i \in \Cell_c} \|\x_i -\x\|_2 \leq
            |\SignedDistance_\x|, \\
        0, \quad \text{otherwise}
    \end{cases}
    \label{eq:bounding-ball}
\end{equation}
with $\x \in \Cell_c$. The evaluation of the $\max$-operator is based on a comparison to the corner points  $\x_i$ of the cell $\Cell_c$. 
For example, in our implementation this function is only evaluated at cell centres $\x_c$ (original mesh cells, see below) or cell corner points (tetrahedra resulting from 
decomposition). As a consequence, a few bulk cells are considered as interface cells (\cref{fig:bs-false-positive}).
We deem this acceptable as this only has a minor impact on the computational time,
but not on the computed volume fractions.

After identification of interface cells, the cell volume fractions are initialized 
according to the sign of $\SignedDistance_c$,
\begin{equation}
     \VolFrac_c = \begin{cases}
         1,\quad \SignedDistance_c \leq 0, \\
         0,\quad \text{otherwise}.
     \end{cases}
     \label{eq:vol-frac-bulk}
\end{equation}
This gives correct volume fractions for bulk cells, while the values of 
interface cells are updated as described below.
Each cell flagged as an interface cell by the method described above is decomposed into tetrahedra using its
centroid and cell face centroids as shown in \cref{fig:centroid-decomposition}.
Each resulting tetrahedron is further refined in an adaptive manner such that
resolution is only subsequently increased where a new tetrahedron is again intersected by the interface. To achieve this, a tetrahedron $T$ is checked with the bounding ball
criterion \cref{eq:bounding-ball}. The criterion is only evaluated at the vertex
$\x_\text{max} \in \Tetrahedron$ for which
$|\SignedDistance(\x_\text{max})| = \max_{\x \in \Tetrahedron}|\SignedDistance(\x)|$.
Only if $f_\text{bb}(\x_\text{max}, \SignedDistance, \Tetrahedron)~=~0$ (\cref{eq:bounding-ball}), $\Tetrahedron$
is considered for further decomposition.
\begin{figure}[htb]
    \centering
    \begin{subfigure}[t]{0.3\textwidth}
        \def\svgwidth{1.0\textwidth}
        {\footnotesize
            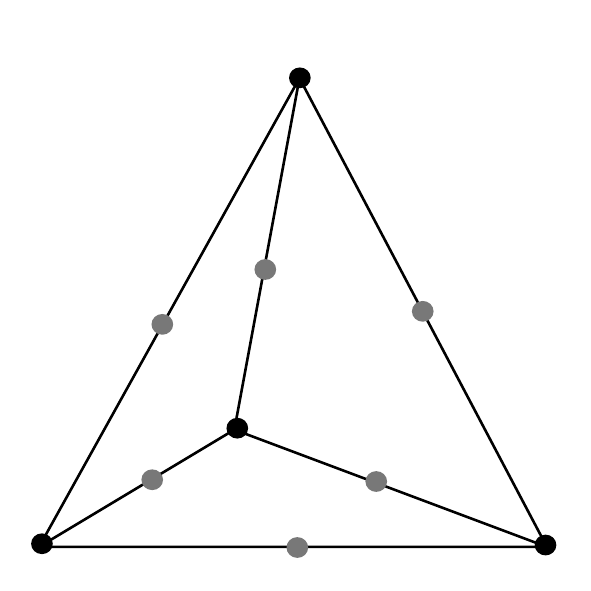
        }
        \caption{Original tetrahedron with vertices ($\x_i$, black) and
            edge midpoints ($\x_{ij}$, grey).
        }
        \label{subfig:tetdecomposition-a} 
    \end{subfigure}
    ~
    \begin{subfigure}[t]{0.3\textwidth}
        \def\svgwidth{1.0\textwidth}
        {\footnotesize
            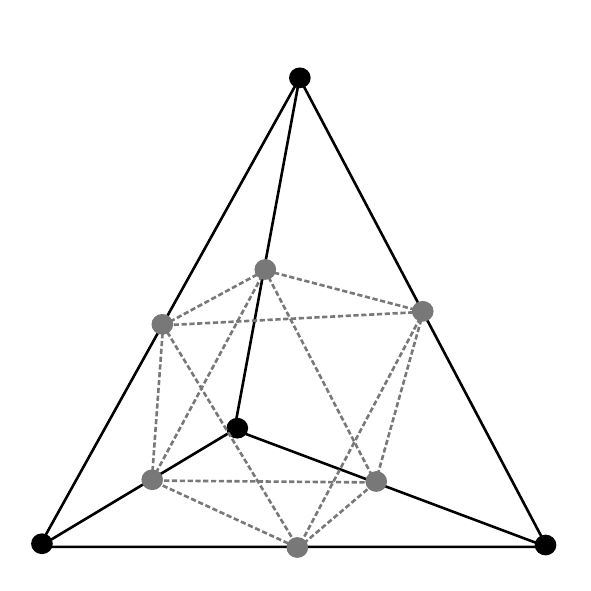
        }
        \caption{Four tetrahedra are created by combining each vertex with its 
            connected edge midpoints (indicated by dashed lines).
        }
        \label{subfig:tetdecomposition-b} 
    \end{subfigure}
    ~
    \begin{subfigure}[t]{0.3\textwidth}
        \def\svgwidth{1.0\textwidth}
        {\footnotesize
\begingroup%
  \makeatletter%
  \providecommand\color[2][]{%
    \errmessage{(Inkscape) Color is used for the text in Inkscape, but the package 'color.sty' is not loaded}%
    \renewcommand\color[2][]{}%
  }%
  \providecommand\transparent[1]{%
    \errmessage{(Inkscape) Transparency is used (non-zero) for the text in Inkscape, but the package 'transparent.sty' is not loaded}%
    \renewcommand\transparent[1]{}%
  }%
  \providecommand\rotatebox[2]{#2}%
  \newcommand*\fsize{\dimexpr\f@size pt\relax}%
  \newcommand*\lineheight[1]{\fontsize{\fsize}{#1\fsize}\selectfont}%
  \ifx\svgwidth\undefined%
    \setlength{\unitlength}{84.14639571bp}%
    \ifx\svgscale\undefined%
      \relax%
    \else%
      \setlength{\unitlength}{\unitlength * \real{\svgscale}}%
    \fi%
  \else%
    \setlength{\unitlength}{\svgwidth}%
  \fi%
  \global\let\svgwidth\undefined%
  \global\let\svgscale\undefined%
  \makeatother%
  \begin{picture}(1,1.0226096)%
    \lineheight{1}%
    \setlength\tabcolsep{0pt}%
    \put(0,0){\includegraphics[width=\unitlength,page=1]{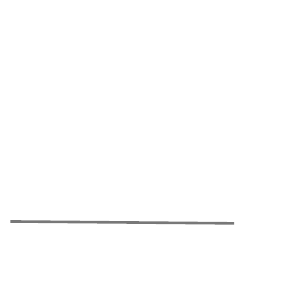}}%
    \put(0.58772377,-0.00123285){\color[rgb]{0,0,0}\makebox(0,0)[lt]{\lineheight{1.25}\smash{\begin{tabular}[t]{l}$\x_{12}$\end{tabular}}}}%
    \put(0.97021603,0.77143403){\color[rgb]{0,0,0}\makebox(0,0)[lt]{\lineheight{1.25}\smash{\begin{tabular}[t]{l}$\x_{23}$\end{tabular}}}}%
    \put(0.46030392,1.01850316){\color[rgb]{0,0,0}\makebox(0,0)[lt]{\lineheight{1.25}\smash{\begin{tabular}[t]{l}$\x_{34}$\end{tabular}}}}%
    \put(0.82953655,0.18337974){\color[rgb]{0,0,0}\makebox(0,0)[lt]{\lineheight{1.25}\smash{\begin{tabular}[t]{l}$\x_{24}$\end{tabular}}}}%
    \put(0.00905234,0.18558978){\color[rgb]{0,0,0}\makebox(0,0)[lt]{\lineheight{1.25}\smash{\begin{tabular}[t]{l}$\x_{14}$\end{tabular}}}}%
    \put(0.0213619,0.8558081){\color[rgb]{0,0,0}\makebox(0,0)[lt]{\lineheight{1.25}\smash{\begin{tabular}[t]{l}$\x_{13}$\end{tabular}}}}%
    \put(0,0){\includegraphics[width=\unitlength,page=2]{tet-decomposition-c.pdf}}%
  \end{picture}%
\endgroup%

        }
        \caption{Decompose octahedron into four tetrahedra by combining each
            grey edge with the black line formed by two opposite points (here $\x_{12}$, $\x_{34}$).
        }
        \label{subfig:tetdecomposition-c} 
    \end{subfigure}
    \caption{Decomposition of a tetrahedron into eight tetrahedra using edge midpoints.}
    \label{fig:tetdecomposition}
\end{figure}
An obvious choice would be decomposition at the centroid of $T$.  However, repeated application of this approach results in 
increasingly flattened tetrahedra. To avoid this problem, we apply the decomposition shown in \cref{fig:tetdecomposition}. 
First,
from the vertices edge centres of the tetrahedron
\begin{equation}
    \x_{ij} =  \frac{1}{2}(\x_i + \x_j),\quad i,j \in \{1,2,3,4\},i\neq j
    \label{eq:edgemidpoint}
\end{equation}
are computed (\cref{subfig:tetdecomposition-a}).
By combining each vertex $\x_i$ with the three edge centres of the adjacent edges, four
new tetrahedra are created (\cref{subfig:tetdecomposition-b}). 
The remainder of
the original tetrahedron is an octahedron (\cref{subfig:tetdecomposition-b} grey dashed 
lines) constituted by the edge centres $\x_{ij}$.
This octahedron is decomposed into four additional tetrahedra by choosing two opposite edge
centres as shown by the black line in \cref{subfig:tetdecomposition-c}. The indices of vertices of such a line are the numbers one to four. \todo{Tobi improve description if possible based on whether the other edge centres are located on the one of the triangles that use the current edge in question } 
From the remaining four edge centres, point
pairs are created such that $\{\x_{mn},\x_{mo}\}$ or
$\{\x_{mn},\x_{on}\}$, yielding four pairs.
Combining each pair with $\{\x_{ij},\x_{kl}\}$ (e.g. black edge in 
\cref{subfig:tetdecomposition-c})
gives the aforementioned four tetrahedra.
Subsequently, $\SignedDistance$ is computed for the added vertices $\x_{ij}$. The decomposition is based on the pair of edge centres that have the smallest distance between each other.
Refinement is completed when a maximum refinement level $\Lmax$ is reached.
This can either be an arbitrary prescribed value or can be computed such that
the edge length of the refined tetrahedra is comparable to the edge length of 
surface triangles. In the latter case, 
\begin{equation}
    \Lmax = \min_{l \in \N}\left(\frac{\Ltet}{\Ltri} < 2^l\right)
    \label{eq:refinement-inequality}
\end{equation}
with $\Ltet$ and $\Ltri$ being cell specific reference lengths for
tetrahedra and surface triangles, respectively. Different choices for
$\Ltet$ and $\Ltri$ are possible. We choose
\begin{align*}
    \Ltet = \frac{1}{n_t}\sum_{\mathbf{e} \in \Ecdc}|\mathbf{e}|, \\
    \Ltri = \min_{\mathbf{e} \in E_{\InterfaceAppr,c}}|\mathbf{e}|
\end{align*}
with $\Ecdc$ denoting the set of edges resulting from tetrahedral decomposition
of a cell $\Cell_c$ at its centroid, $n_t$ the number of edges in $\Ecdc$ and
$E_{\InterfaceAppr,c}$ a subset of edges of $\InterfaceAppr$.
The set $E_{\InterfaceAppr,c}$ consists of all edges of $\Triangle \in 
\InterfaceAppr$ for which $\Triangle \cap \Ball(\x_\text{cp}, r_{\text{cp}}) \neq \emptyset$.
Here,
\begin{align*}
    \x_\text{cp} &= \frac{1}{|P_\text{cp}|}\sum_{\x_i \in P_\text{cp}}, \\
    P_\text{cp}  &:=\{\x \in \InterfaceAppr : \min_{\x_i \in \Cell_c}\|\x - \x_i\|_2\}
\end{align*}  
and the radius $r_\text{cp} = \max_{\x \in P_\text{cp}}\|\x - \x_\text{cp}\|_2$.

Finally, after computing a tetrahedral decomposition of each interface cell, the
volume fraction of a cell $\Cell_c$ is calculated as
\begin{equation}
    \VolFrac_c = \frac{1}{|\Cell_c|}\sum_{\Tetrahedron \in T_c}
        \VolFrac(\Tetrahedron) |\conv (\Tetrahedron)|
    \label{eq:volfrac-smca}
\end{equation}
where $T_c$ denotes the set of tetrahedra resulting from the decomposition of $\Cell_c$ and $|\conv(\Tetrahedron)|$ the volume of
$\Tetrahedron$. The volume fraction $\VolFrac(\Tetrahedron)$ is computed
with the approach of \citet{Detrixhe2016} (eq. 7), repeated here
\begin{equation}
    \VolFrac(\Tetrahedron) = \left\{
    \begin{aligned}
        &1, 
            & &\sd_4 \leq 0, \\
        &1 - \frac{\sd_4^3}{(\sd_4 - \sd_1)(\sd_4 - \sd_2)(\sd_4 -  \sd_3)},
            & &\sd_3 \leq 0 < \sd_4, \\
        &1 - \frac{\sd_1\sd_2(\sd_3^2 + \sd_3\sd_4 + \sd_4^2) + \sd_3\sd_4(\sd_3\sd_4 - (\sd_1 + \sd_2)(\sd_3 + \sd_4))}{(\sd_1 - \sd_3)(\sd_2 - \sd_3)(\sd_1 - \sd_4)(\sd_2 - \sd_4)},
            & &\sd_2 \leq 0 < \sd_3, \\
        &- \frac{\sd_1^3}{(\sd_2 - \sd_1)(\sd_3 - \sd_1)(\sd_4 - \sd_1)},
            & &\sd_1 \leq 0 < \sd_2, \\
        &0
            & &\sd_1 > 0,
    \end{aligned}
    \right.
    \label{eq:alphatet}
\end{equation}
where
$\SignedDistance_4 \geq \SignedDistance_3 \geq \SignedDistance_2 \geq \SignedDistance_1$ are the signed distances at the vertices $\x_i$ of
$\Tetrahedron$.
The overall approach is summarized in \cref{alg:pvof-smca}.

\begin{algorithm}[htb]
    \centering
    \caption{The Surface-Mesh / Cell Approximation Algorithm (SMCA)} 
    \label{alg:pvof-smca}
    {\small
      \begin{algorithmic}[1]
        \State Follow \cref{alg:gvof:smci} up to step 13.
        \State Identify interface cells (\cref{eq:bounding-ball})
        \State Set bulk $\VolFrac_c$ (\cref{eq:vol-frac-bulk})
        \State Centroid decomposition of cells into tetrahedra (\cref{fig:centroid-decomposition})
        \For{$l \in \{1,\ldots,\Lmax\}$}
            \State Flag tetrahedra for further refinement (\cref{eq:bounding-ball})
            \State Decompose flagged tetrahedra (\cref{fig:tetdecomposition})
            \State Compute $\SignedDistance$ for new points (\cref{eq:tridist})
        \EndFor
        \State Compute $\VolFrac_c$ for interface cells (\cref{eq:volfrac-smca})
      \end{algorithmic}
    }
\end{algorithm}
\section{Results}
\label{sec:results}

The software implementation is available on GitLab \citep{argo}: we refer to the specific version (git tag) used to generate results described below. 
Detailed information on how to build and use the software is provided in the \texttt{README.md} file in the root folder of the software repository. \\

We use the difference between the total volume given by the volume fraction calculated from the surface on the unstructured mesh, and the exact volume bounded by the surface, namely 
\begin{equation}
    E_v = \frac{1}{V_e} \left| V_e - \sum_{c \in C}\alpha_c |\Omega_c| \right|, 
    \label{eq:Ev}
\end{equation}
as the measure of accuracy of the proposed algorithms. Here, $V_e$ is the volume given by the exact surface function, or the volume that is bounded by a given surface mesh if an exact surface function is not available, e.g. in \cref{subsec:experiment,subsec:cad}. In these cases, we calculate $V_e$ using 
\begin{equation}
    V_e = \frac{1}{3} \left|\int_{V_e} \nabla \cdot \x \, dV\right| 
        = \frac{1}{3} \left|\int_{\partial V_e} \x \cdot \mathbf{n} \, dS\right|
    \label{eq:Ve}
\end{equation}
where $\partial V_e$ is the surface that bounds $V_e$. As this surface is triangluated, \cref{eq:Ve} can be expanded further  
\begin{align}
    V_e & = \frac{1}{3} \left|\sum_{t \in {1..N_{\InterfaceAppr}}} \int_{T_t} \x \cdot \mathbf{n} \, dS \right|= \frac{1}{3} \left|\sum_{t \in {1..N_{\InterfaceAppr}}} \int_{T_t} (\x - \x_t + \x_t) \cdot \mathbf{n} \, dS \right|
     = \frac{1}{3} \left|\sum_{t \in {1..N_{\InterfaceAppr}}} \x_t \cdot \mathbf{S}_t \right|
    \label{eq:Ve-expanded}
\end{align}
where $N_{\InterfaceAppr}$ is the number of triangles in $\InterfaceAppr$, $T_t \in \InterfaceAppr$ are triangles that form the interface mesh, and $\x_t, \mathbf{S_t}$ are their respective centroids and area normal vectors.

\begin{table}[h]
  \begin{center}
      {\scriptsize
        \begin{tabular}{lllll}
\toprule
Computing architecture & \\
\midrule
CPU \\         
  & vendor\_id	: AuthenticAMD \\
  & cpu family	: 23\\
  & model	: 49\\
  & model name	: AMD Ryzen Threadripper 3990X 64-Core Processor\\
  & frequency   : 2.90 GHz \\
Compiler \\
  & version : g++ (Ubuntu 10.2.0-5ubuntu1~20.04) 10.2.0 \\
  & optimization flags : -std=c++2a -O3 \\
\bottomrule
\end{tabular}

      }
  \end{center}
  \caption{Used computing architecture.}
  \label{tab:testingarch}
\end{table}

\Cref{tab:testingarch} contains the details on the computing architectures used to report the absolute CPU times in the result section. We have fixed the CPU frequency to 2.9GHz to stabilize the CPU time measurements. 

\subsection{Sphere and ellipsoid}
\label{subsec:sphere-ellips}

Exact initialization algorithms for spheres are available on unstructured meshes \citep{Strobl2016,Kromer2019}. We use the sphere and ellipsoid test cases to confirm the second-order convergence of SMCI/A algorithms and their applicability as a volume fraction model for the unstructured Level Set / Front Tracking method \citep{Maric2015,Tolle2020}. The sphere case consists of a sphere with a radius $R=0.15$, and the ellipsoid half-axes are $(0.4, 0.3, 0.2)$. 
Both the sphere and ellipsoid center are at $(0.5, 0.5, 0.5)$, in a unit box domain. Error convergence, CPU time and additional data are publicly available \citep{smcia-results}.

\subsubsection{SMCI Algorithm}

\begin{figure}[!htb]
    \centering
    \begin{subfigure}{0.49\textwidth}
        \includegraphics[width=\textwidth]{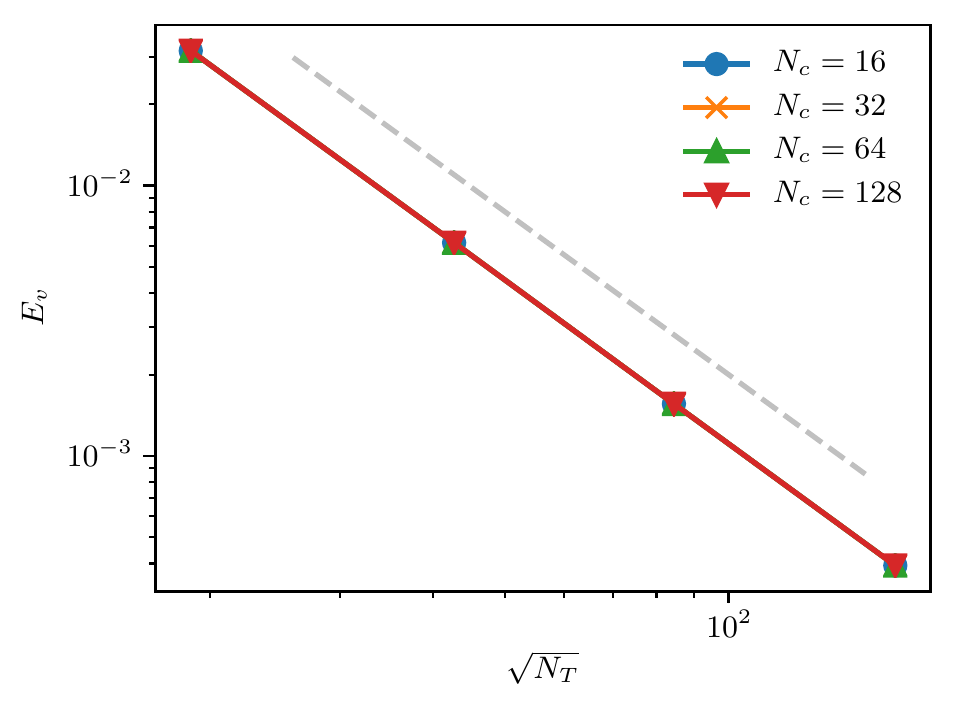}
        \caption{Equidistant mesh.}
        \label{fig:smci:err:sphere:cubic}
    \end{subfigure}
    \begin{subfigure}{0.49\textwidth}
        \includegraphics[width=\textwidth]{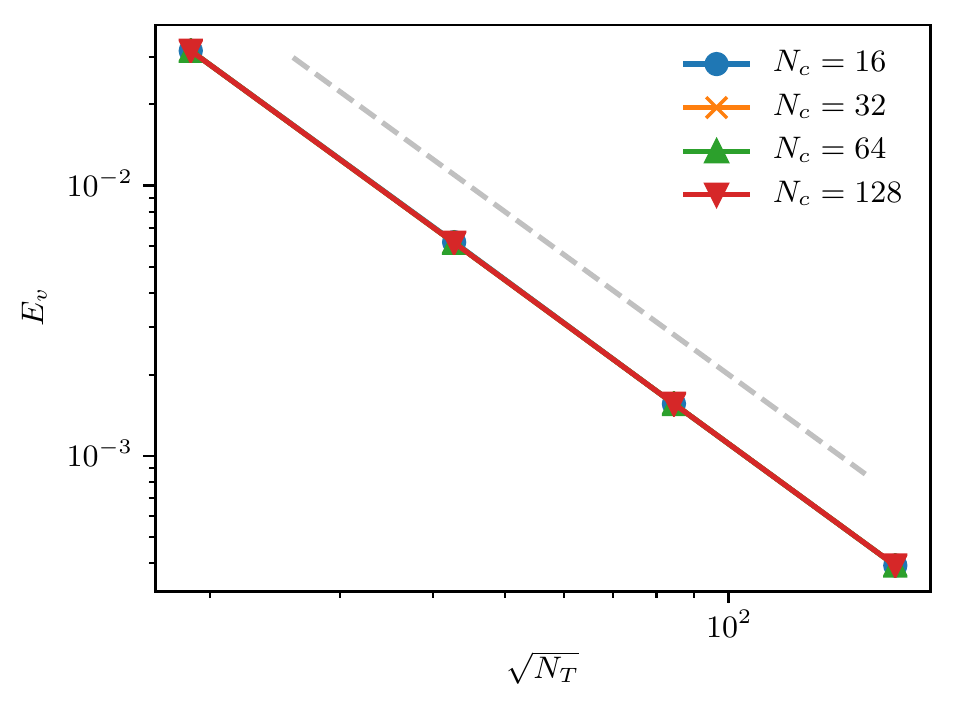}
        \caption{Irregular hexahedral mesh.}
        \label{fig:smci:err:sphere:hex}
    \end{subfigure}
    \caption{$E_v$ errors of the SMCI algorithm for the sphere. The grey dashed line indicates second order convergence.} 
    \label{fig:smci:err:sphere}
\end{figure}

 \Cref{fig:smci:err:sphere} shows the expected second-order convergence of the global error $E_v$ given by \cref{eq:Ev} on cubic \cref{fig:smci:err:sphere:cubic} and irregular hexahedral \cref{fig:smci:err:sphere:hex} unstructured meshes. In \cref{fig:smci:err:sphere}, $N_c$ is the number of cells used along each spatial dimension of $\tilde{\Omega}$ and $N_T$ is the number of triangles used to resolve the sphere. 

\begin{figure}[!htb]
    \centering
    \includegraphics{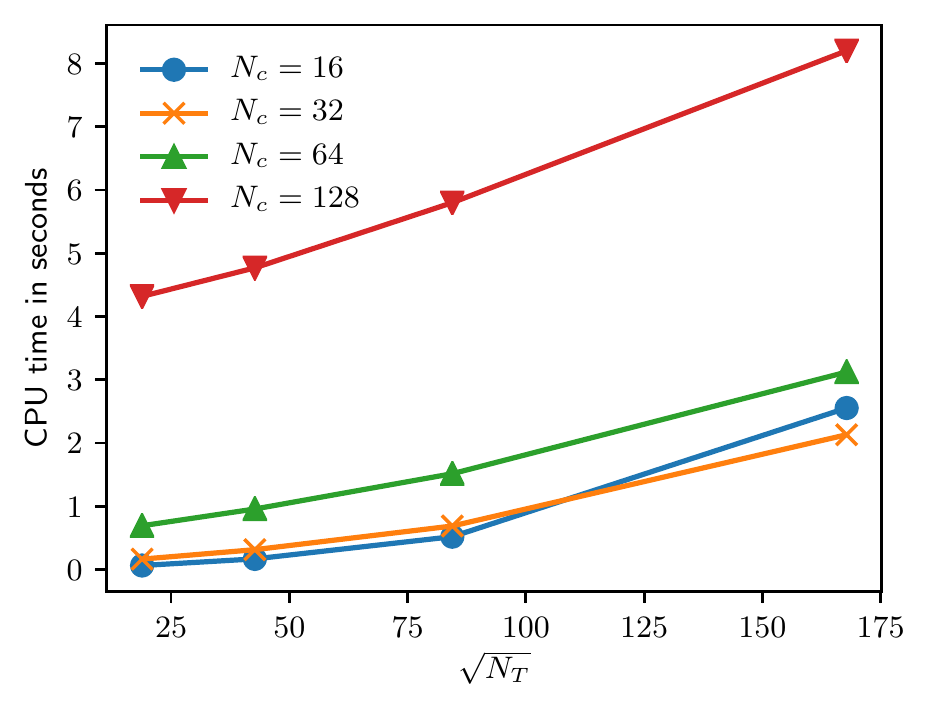}
    \caption{CPU times of the SMCI algorithm for the sphere initialized on a cubic unstructured mesh.} 
    \label{fig:smci:cpu:sphere}
\end{figure}

The CPU times reported in \cref{fig:smci:cpu:sphere} for the architecture A1 in \cref{tab:testingarch} show that the SMCI algorithm is a promising candidate for a volume fraction model for the unstructured Level Set / Front Tracking method. The complexity of the algorithm expressed in terms of the measured CPU time remains, linear for a constant ratio $\sqrt{N_T} / N_c$. The computational complexity increases to quadratic with an increasing number of triangles per cell $\sqrt{N_T} / N_c$: this happens when a very fine surface mesh is used to compute volume fractions on a very coarse volume mesh. An intersection between a highly resolved surface mesh and single cell of a relatively coarse mesh is shown in \cref{fig:smci:sphere-ellipsoid-detail}.

\todo{why are our plans relevant at this point? this would fit introduction or conclusion }This configuration is relevant for accurate initialization of volume fractions on coarse meshes, but irrelevant for calculating the phase indicator for Front Tracking, where only a small number of triangles per multimaterial cell ($\le 10$) is present. Therefore, linear complexity of the SMCI algorithm for small ratios $\sqrt{N_T} / N_c$ makes SMCI a potential candidate for a highly accurate geometrical volume fraction model for the unstructured Level Set / Front Tracking method. We will investigate this possibility in our future work. When considering the absolute CPU times, it is important to note that the SMCI algorithm has not yet been optimized for performance.

\todo{second order reference lines would be good}
The volume error $E_v$ for a sphere is shown in \cref{fig:smci:err:sphere:hex} for a perturbed hexahedral mesh. An example perturbed mesh from this parameter study is shown in \cref{fig:smci:sphere-ellipsoid}. The mesh is distorted by randomly perturbing cell corner points, using a length scale factor $\alpha_e \in [0,1]$ for the edges $e$ that surround the mesh point. We have used $\alpha_e=0.25$, resulting in perturbations that are of the size of $0.25 \, \times$ the edge length. This results in a severe perturbation of the mesh shown in \cref{fig:smci:sphere-ellipsoid}, as well as non-planarity of the faces of hexahedral cells. Still, as shown in \cref{fig:smci:err:sphere:hex}, SMCI retains second-order convergence, which is also the case for the initialization of the ellipsoid on the equidistant \cref{fig:smci:err:ellipsoid} and perturbed hexahedral mesh \cref{fig:smci:err:ellipsoid:hex}. 

\begin{figure}[!htb]
    \centering
    \begin{subfigure}[b]{0.34\textwidth}
        \includegraphics[width=\textwidth]{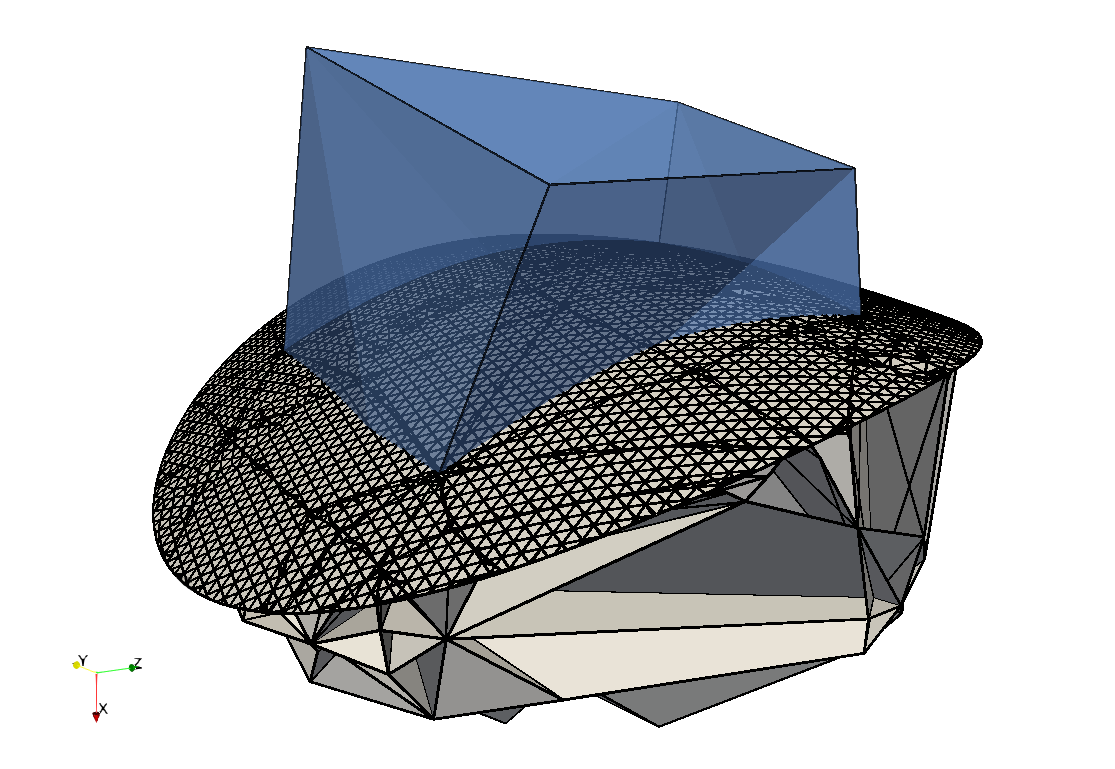}
        \caption{SMCI: intersected cell.} 
        \label{fig:smci:sphere-ellipsoid-detail}
    \end{subfigure}
    \begin{subfigure}[b]{0.65\textwidth}
        \def\svgwidth{\textwidth}
        {\footnotesize
\begingroup%
  \makeatletter%
  \providecommand\color[2][]{%
    \errmessage{(Inkscape) Color is used for the text in Inkscape, but the package 'color.sty' is not loaded}%
    \renewcommand\color[2][]{}%
  }%
  \providecommand\transparent[1]{%
    \errmessage{(Inkscape) Transparency is used (non-zero) for the text in Inkscape, but the package 'transparent.sty' is not loaded}%
    \renewcommand\transparent[1]{}%
  }%
  \providecommand\rotatebox[2]{#2}%
  \newcommand*\fsize{\dimexpr\f@size pt\relax}%
  \newcommand*\lineheight[1]{\fontsize{\fsize}{#1\fsize}\selectfont}%
  \ifx\svgwidth\undefined%
    \setlength{\unitlength}{344.06402828bp}%
    \ifx\svgscale\undefined%
      \relax%
    \else%
      \setlength{\unitlength}{\unitlength * \real{\svgscale}}%
    \fi%
  \else%
    \setlength{\unitlength}{\svgwidth}%
  \fi%
  \global\let\svgwidth\undefined%
  \global\let\svgscale\undefined%
  \makeatother%
  \begin{picture}(1,0.5386176)%
    \lineheight{1}%
    \setlength\tabcolsep{0pt}%
    \put(0,0){\includegraphics[width=\unitlength,page=1]{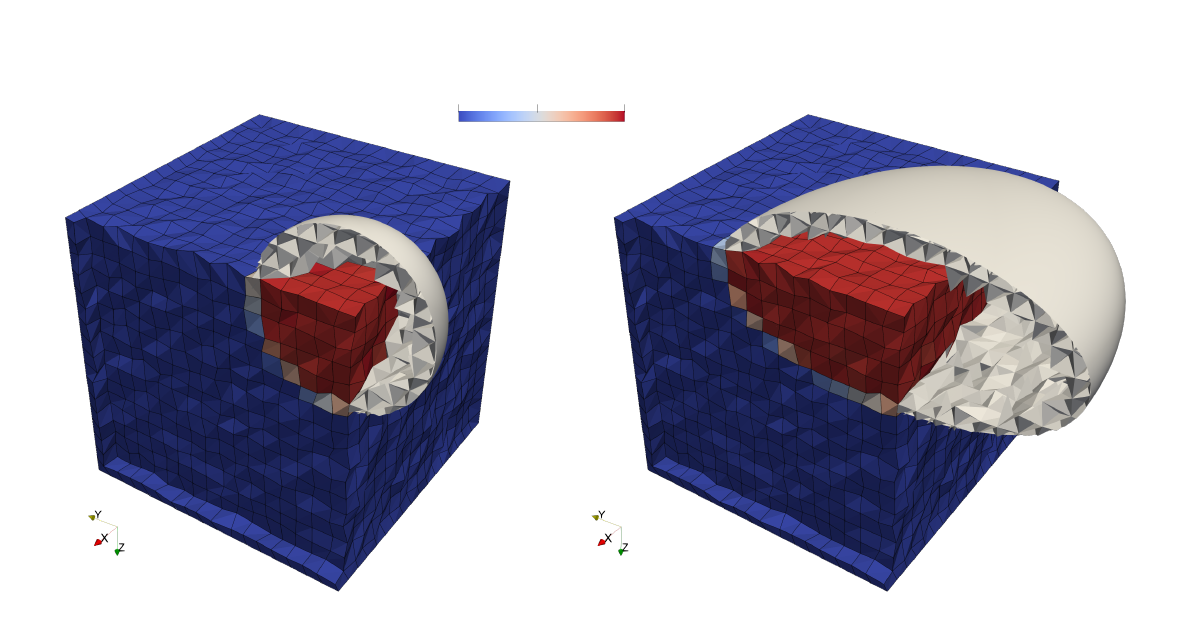}}%
    \put(0.44519674,0.49166846){\makebox(0,0)[lt]{\lineheight{1.25}\smash{\begin{tabular}[t]{l}$\alpha_c$\end{tabular}}}}%
    \put(0.38180693,0.45320472){\makebox(0,0)[lt]{\lineheight{1.25}\smash{\begin{tabular}[t]{l}0\end{tabular}}}}%
    \put(0.44502401,0.45283231){\makebox(0,0)[lt]{\lineheight{1.25}\smash{\begin{tabular}[t]{l}0.5\end{tabular}}}}%
    \put(0.52060307,0.4530645){\makebox(0,0)[lt]{\lineheight{1.25}\smash{\begin{tabular}[t]{l}1\end{tabular}}}}%
  \end{picture}%
\endgroup%

        }
        \caption{SMCI: sphere and ellipsoid volume fractions.} 
        \label{fig:smci:sphere-ellipsoid}
    \end{subfigure}
    \caption{SMCI algorithm used with a sphere and an ellipsoid on an unstructured hexahedral mesh.} 
\end{figure}
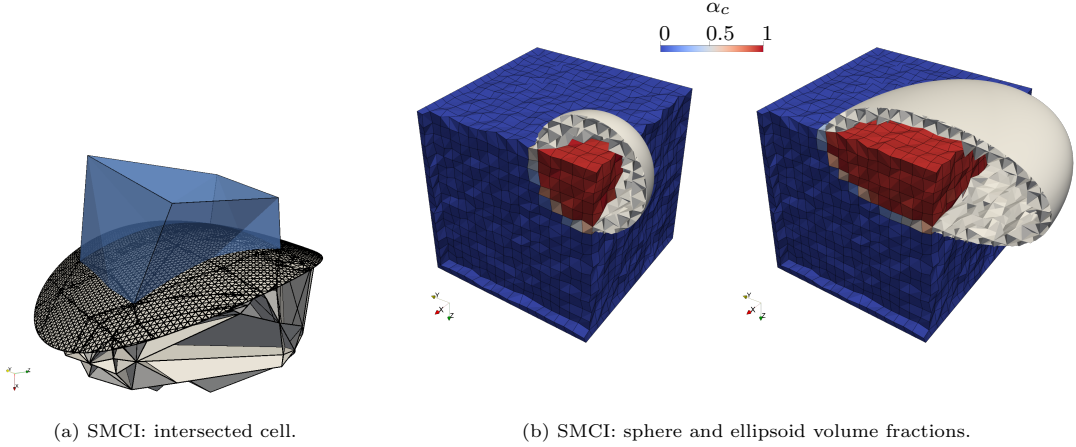
\begin{figure}[!htb]
    \centering
    \begin{subfigure}{0.49\textwidth}
        \includegraphics[width=\textwidth]{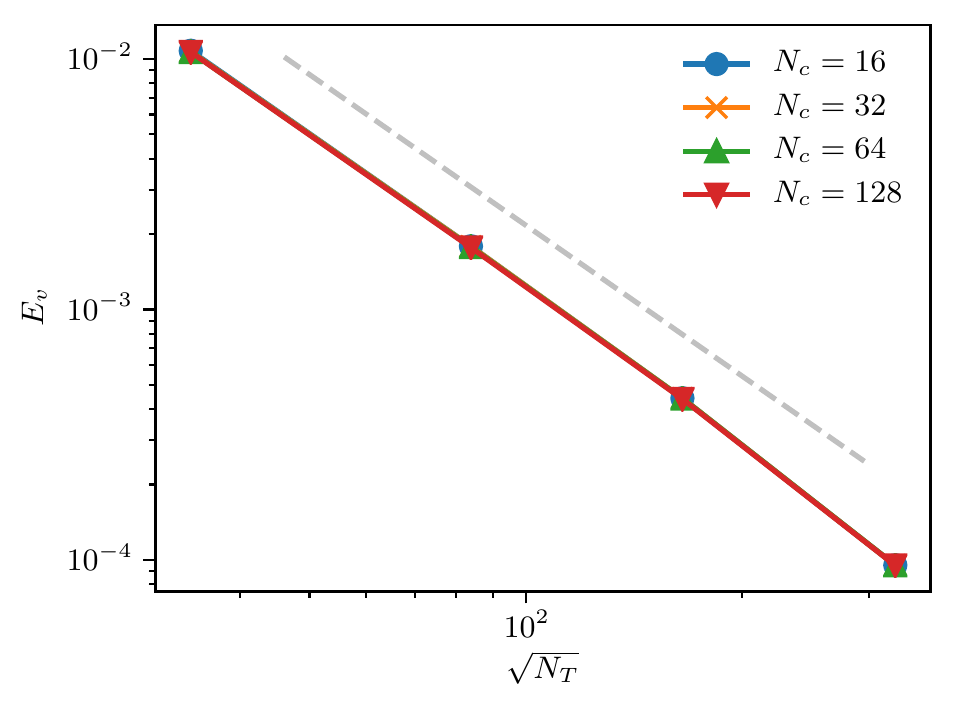}
        \caption{Equidistant mesh.}
        \label{fig:smci:err:ellipsoid:cubic}
    \end{subfigure}
    \begin{subfigure}{0.49\textwidth}
        \includegraphics[width=\textwidth]{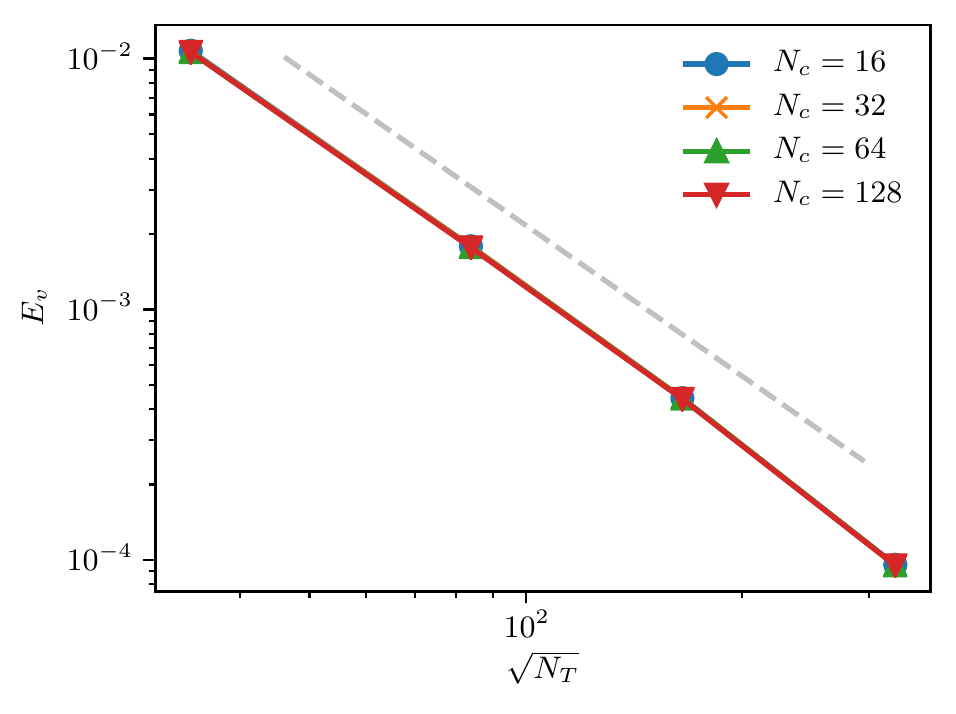}
        \caption{Irregular hexahedral mesh.}
        \label{fig:smci:err:ellipsoid:hex}
    \end{subfigure}
    \caption{$E_v$ errors of the SMCI algorithm for the ellipsoid. The grey dashed line indicates second order convergence.} 
    \label{fig:smci:err:ellipsoid}
\end{figure}

\subsubsection{\Pvof{} algorithm}
\begin{figure}[htb]
    \centering
    \includegraphics{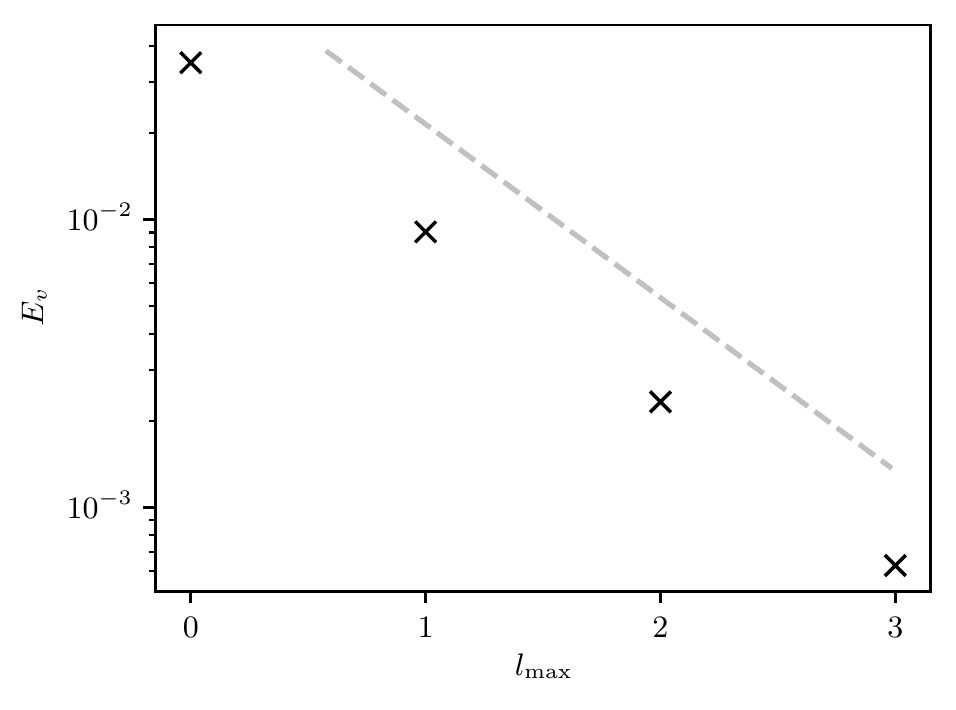}
    \caption{$E_v$ errors of the \Pvof{} algorithm using different refinement levels
        $\Lmax$ for a sphere. Resolution of volume and surface mesh are fixed to
        $N_c=16$ and $\sqrt{N_T}\approx410$. The grey dashed line indicates second order
        convergence.
    }
    \label{fig:smca:err:sphere:refinement}
\end{figure}

\begin{figure}[!hbt]
    \centering
    \includegraphics[width=0.6\textwidth]{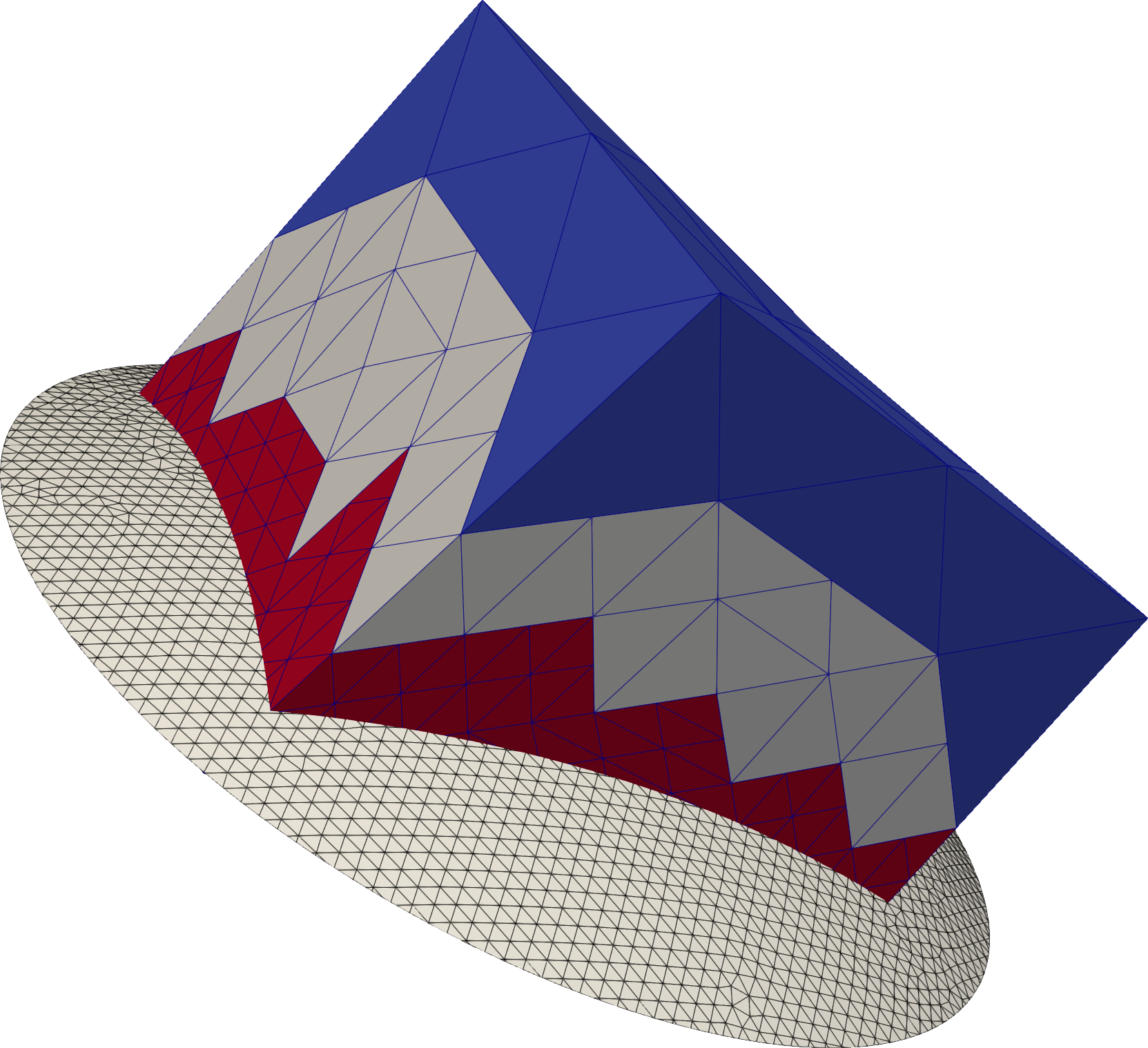}
    \caption{Tetrahedral decomposition of a perturbed hex cell used to approximate
        $\VolFrac_c$. Tetrahedra from different refinement levels are shown
        in different colors (level 1: blue, level 2: grey, level 3: red). Due to adaptivity, the
        highest refinement level is localized in the vicinity of the surface mesh..
    }
    \label{fig:smca:hex-cell-tet-decomposition}
\end{figure}

First, the effectiveness of the local adaptivity employed in the
\Pvof{} algorithm is examined with a spherical interface as described in \cref{subsec:sphere-ellips}. Resolution of the volume mesh
is fixed to $N_c=16$ cells in each 
direction while the sphere is resolved with $\sqrt{N_T}\approx410$ triangles.
Maximum refinement levels $\Lmax$ from $0$ to $3$ are manually prescribed.
In \cref{fig:smca:err:sphere:refinement}, the resulting global volume errors $E_v$ are
displayed. This test case confirms the expected second-order convergence of $E_v$ with
adaptive refinement.
An exemplary tetrahedral decomposition of a perturbed hex cell with a part of the 
the surface mesh is displayed in \cref{fig:smca:hex-cell-tet-decomposition}.
It demonstrates that the adaptive refinement based on the bounding ball criterion \cref{eq:bounding-ball} works as intended. Refinement is localized to the vicinity
around the interface. Yet, the approach ensures all tetrahedra intersected by the interface
are actually refined.
The effectiveness of the local adaptive refinement compared to a uniform
one becomes apparent when comparing the resulting number of tetrahedra.
Our adaptive approach yields around $2247$ tetrahedra per interface cell on average
for the spherical interface with $\sqrt{N_T}\approx410$, $N_c=16$ and
$\Lmax=3$. A uniform decomposition, on the contrary, would result in 
$M_i\times M_r^{\Lmax}=24\times8^3\approx47.9\times10^3$ tetrahedra, where $M_i$ denotes the
number of tetrahedra from initial cell decomposition and $M_r$ the number of 
tetrahedra from refining a tetrahedron. Thus, the local adaptive refinement reduces the required overall number of tetrahedra by a factor of $5.5$ in comparison to a uniform refinement, 
without affecting the accuracy.

\begin{figure}[!htb]
    \centering
    \begin{subfigure}{0.49\textwidth}
        \includegraphics[width=\textwidth]{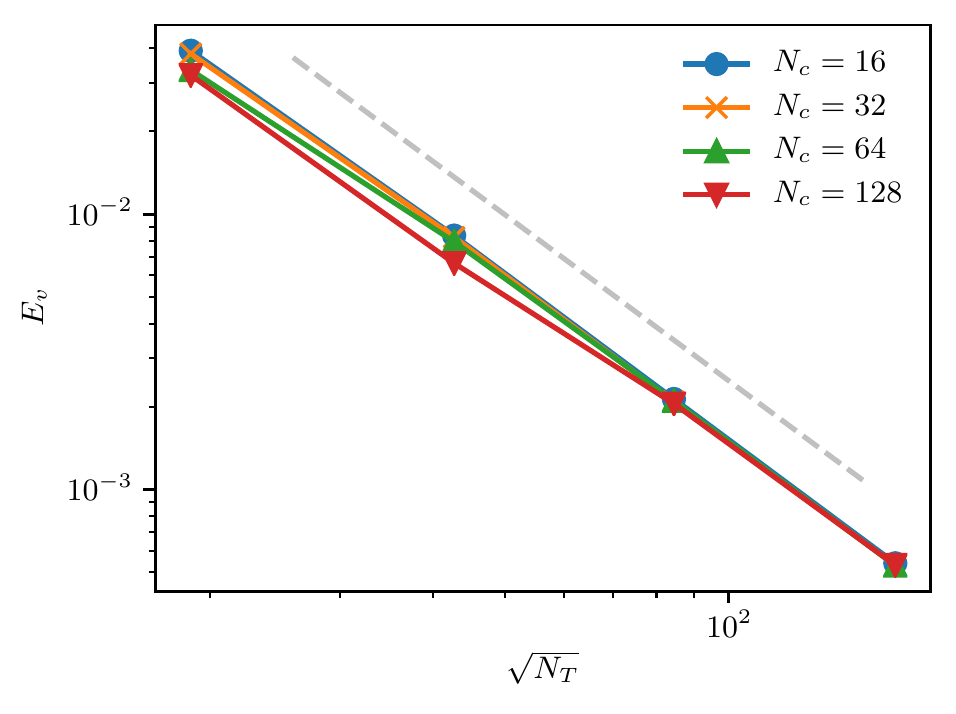}
        \caption{Equidistant mesh.}
        \label{fig:smca:err:sphere:cartesian}
    \end{subfigure}
    \begin{subfigure}{0.49\textwidth}
        \includegraphics[width=\textwidth]{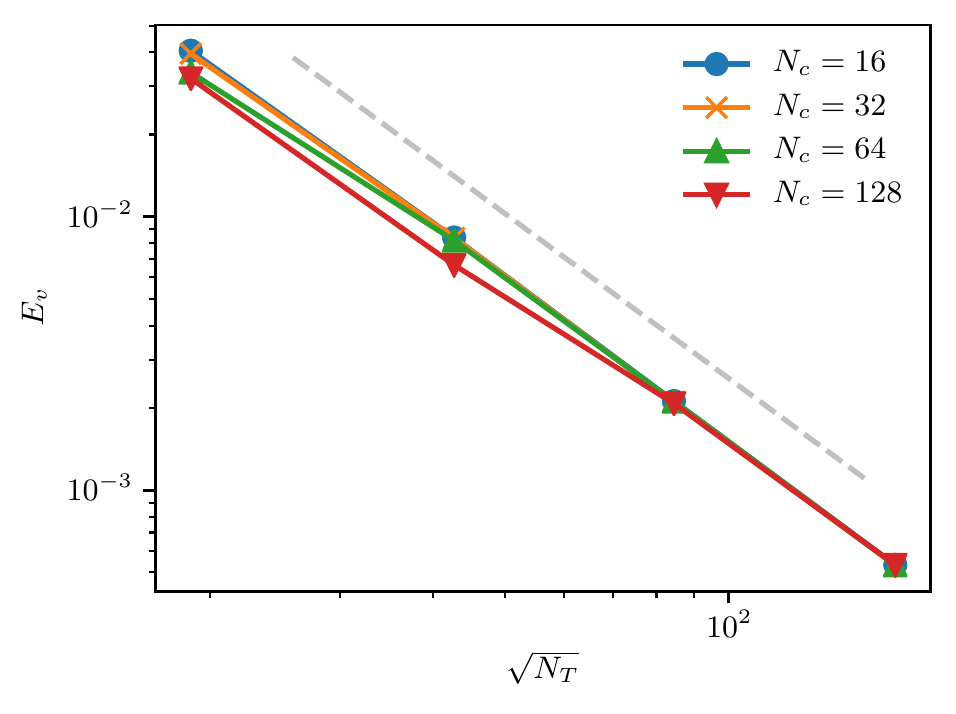}
        \caption{Irregular hexahedral mesh.}
        \label{fig:smca:err:sphere:perturbed}
    \end{subfigure}
    \caption{$E_v$ errors of the \Pvof{} algorithm for the sphere. The grey dashed line indicates second order convergence.} 
    \label{fig:smca:err:sphere}
\end{figure}

\begin{figure}[!htb]
    \centering
    \begin{subfigure}{0.49\textwidth}
        \includegraphics[width=\textwidth]{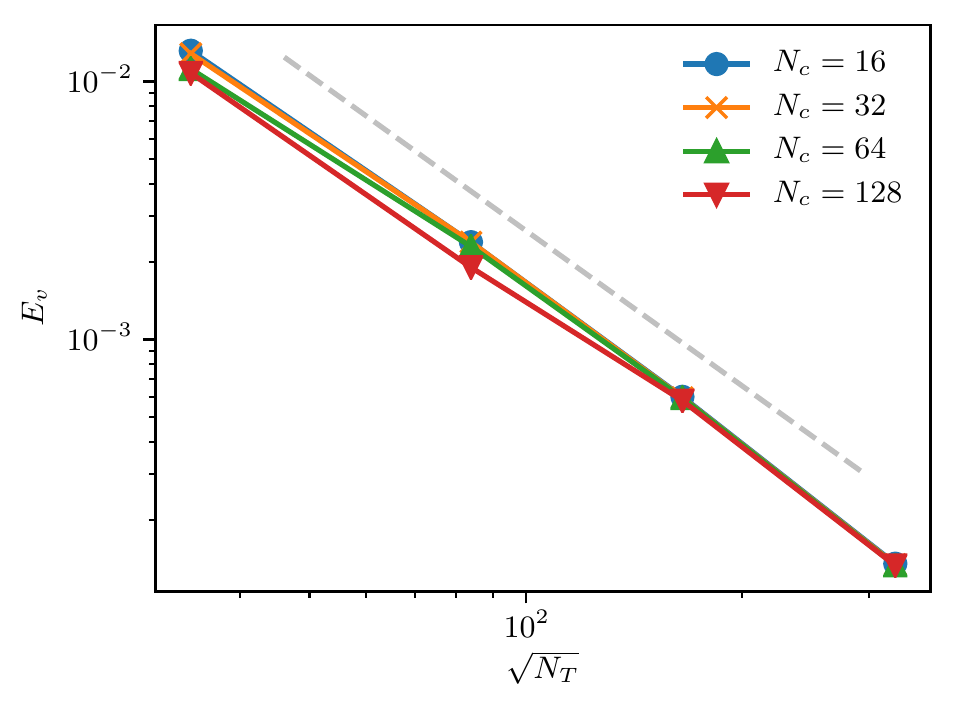}
        \caption{Equidistant mesh.}
        \label{fig:smca:err:ellipsoid:cartesian}
    \end{subfigure}
    \begin{subfigure}{0.49\textwidth}
        \includegraphics[width=\textwidth]{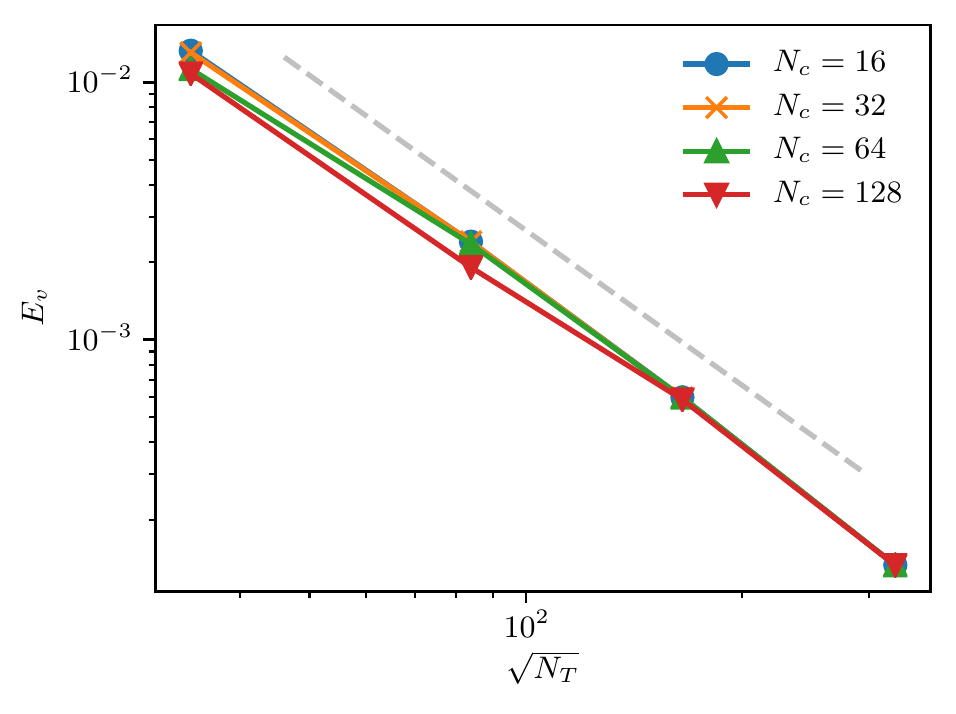}
        \caption{Irregular hexahedral mesh.}
        \label{fig:smca:err:ellipsoid:perturbed}
    \end{subfigure}
    \caption{$E_v$ errors of the \Pvof{} algorithm for the ellipsoid. The grey dashed line indicates second order convergence.} 
    \label{fig:smca:err:ellipsoid}
\end{figure}

Having verified the refinement procedure, accuracy of the \Pvof{} algorithm 
and its convergence with respect to surface mesh resolution is assessed in the following.
As for the SMCI algorithm, a sphere and an ellipsoid are used for this purpose.
Results for the sphere in terms of the global volume error $E_v$
(\cref{eq:Ev}) are shown in \cref{fig:smca:err:sphere} for cubic cells
(\cref{fig:smca:err:sphere:cartesian}) and perturbed hexahedral cells
(\cref{fig:smca:err:sphere:perturbed}). Domain size, sphere centre and radius are
identical to the SMCI setup as well as the perturbation factor $\alpha_e=0.25$.
The maximum refinement level is computed according
to \cref{eq:refinement-inequality}. Both mesh types yield nearly identical
results and show second-order convergence. Resolution of the volume mesh
$N_c$ has a minor influence for coarser surface meshes which vanishes
for $\sqrt{N_T} > 100$.
For the ellipsoidal interface, the errors $E_v$ are shown in 
\cref{fig:smca:err:ellipsoid}. The results are qualitatively and quantitatively
similar to those of the spherical interface.
\begin{figure}[!htb]
    \centering
    \includegraphics{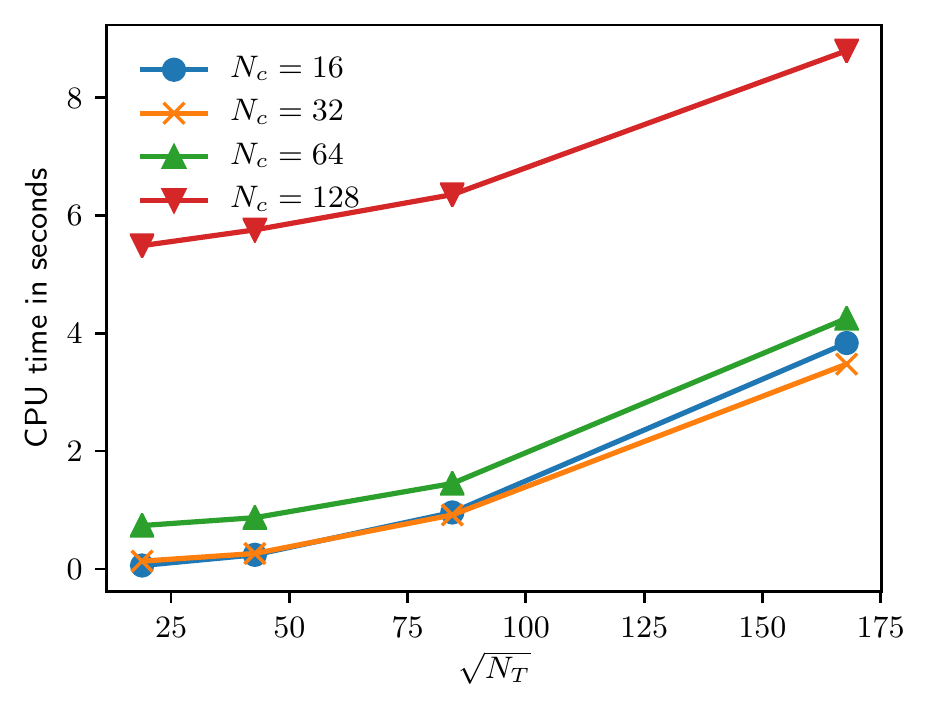}
    \caption{CPU times of the \Pvof{} algorithm for the sphere initialized on a cubic unstructured mesh.} 
    \label{fig:smca:cpu:sphere}
\end{figure}
Absolute computational times required for the initialization of a sphere with the
\Pvof{} algorithm are displayed in \cref{fig:smca:cpu:sphere}. Run times have
been measured on the architecture listed in \cref{tab:testingarch}. As the 
implementation SMCI algorithm, our implementation of the \Pvof{} algorithm has
not yet been optimized for performance.

\textcolor{Reviewer1}{Because of the algebraic calculation of volume fractions from signed distances, the SMCA algorithm allows a direct comparison with volume fraction initialization methods on unstructured meshes that represent the fluid interface using function composition. Considering \cref{sec:intro}, logical choices for the comparison are the methods of \citet{Ahn2007,Fries2016,Jones2019}. However, \citet{Ahn2007} do not provide convergence results for the 3D initialization and \citet{Fries2016} integrate a function that is $\ne 1$ within their 3D surface, so the result of the quadrature does not correspond to the volume enclosed by the surface. We therefore provide a direct comparison with \citet{Jones2019}, specifically \citet[table 3]{Jones2019}.}

\begin{figure}[!htb]
    \centering
    \captionsetup{justification=justified}
    \includegraphics{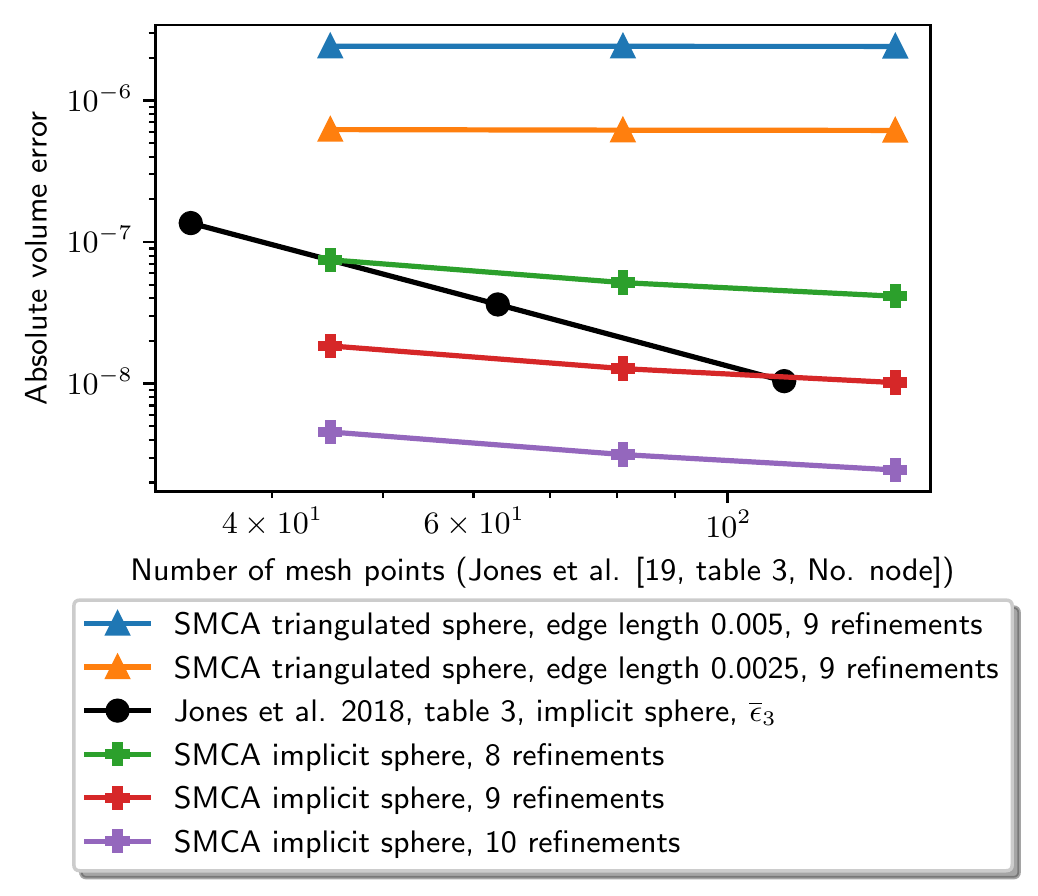}
    \caption{\textcolor{Reviewer1}{Comparing the SMCA algorithm and \citet[table 3]{Jones2019} on tetrahedral meshes. }} 
    \label{fig:smca:jones:sphere}
\end{figure}

\textcolor{Reviewer1}{Absolute volume errors are computed for an octant of a sphere with radius $R=0.5$, placed at $(0, 0, 0)$ within a unit-length cubical domain, and are shown in \cref{fig:smca:jones:sphere}. Tetrahedral unstructured meshes are generated using the Delaunay algorithm in \texttt{gmsh} \cite{gmsh}, by providing a discretization length that results in a number of mesh points comparable to \citet[table 3, No Nodes]{Jones2019}.
As shown in \cref{fig:smca:jones:sphere}, the accuracy of the SMCA algorithm depends on the
volume mesh resolution and the number of refinement levels when an implicit (exact)
sphere is used as interface description. This is expected since both parameters influence the size of the refined tetrahedra which are used to approximate the volume fraction. Consequently,
the achievable accuracy is not limited by the volume mesh resolution and can be controlled through the number of refinement levels.
%
The lowest absolute errors are in the order of magnitude of $10^{-9}$, achieved by SMCA using $10$ refinement levels, and correspond to relative errors in the order of magnitude of $10^{-8}$, which is around $4$ orders of magnitude lower than minimal VOF advection errors reported so far in the literature \citep{Maric2020vofrev}, and are therefore admissible as initial volume fraction values. Even higher levels of absolute accuracy, comparable to \citet[table 3, $\overline{\epsilon}_6,\overline{\epsilon}_9$]{Jones2019}, can be achieved with further refinement, with substantially increased computational expense. However, such further increase in accuracy is without significance to the volume fraction advection \citep{Maric2020vofrev}. Contrary to the implicit (exact) sphere, resolving a sphere using a triangular mesh is more challenging, as the absolute accuracy depends on the resolution of the surface mesh. Results for spheres triangulated using the Frontal Algorithm in \texttt{gmsh} \citep{gmsh} are shown in \cref{fig:smca:jones:sphere}. Doubling the resolution of the surface mesh, as expected, doubles the accuracy of SMCA with triangulated surfaces as input. This approach of course does not make sense for a sphere, whose implicit (exact) function is easily defined. For geometrically complex surfaces shown below, it is important to have in mind that the resolution of the surface mesh together with the refinement level determine the absolute accuracy and computational costs.} 

\subsection{Surface of a fluid from an experiment} 
\label{subsec:experiment}

\begin{figure}[!htb]
    \centering
    \begin{subfigure}[c]{0.49\textwidth}
        \includegraphics[width=\columnwidth]{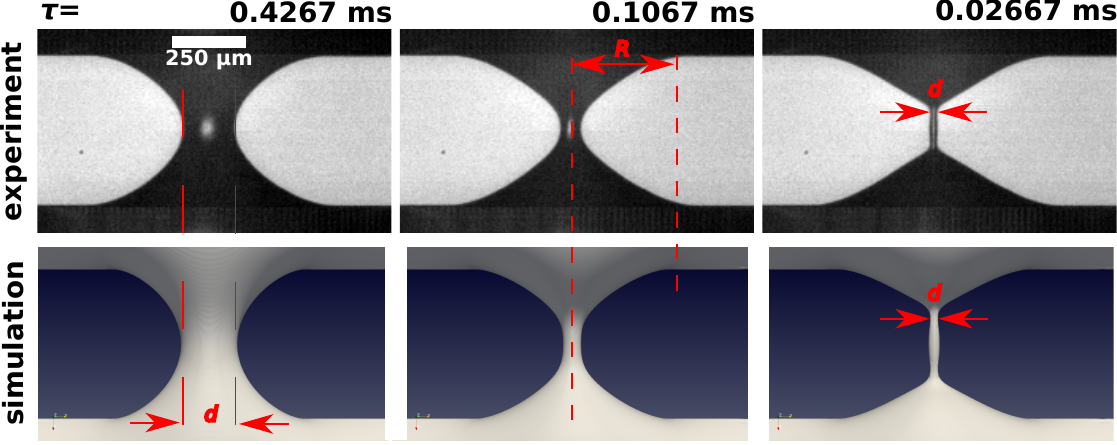}
        \caption{Qualitative comparison with experiment, image from \citep{Hartmann2021}. } 
        \label{fig:hartmann:comp}
    \end{subfigure}
    \begin{subfigure}[c]{0.49\textwidth}
    \def\svgwidth{\columnwidth}
    {\footnotesize
\begingroup%
  \makeatletter%
  \providecommand\color[2][]{%
    \errmessage{(Inkscape) Color is used for the text in Inkscape, but the package 'color.sty' is not loaded}%
    \renewcommand\color[2][]{}%
  }%
  \providecommand\transparent[1]{%
    \errmessage{(Inkscape) Transparency is used (non-zero) for the text in Inkscape, but the package 'transparent.sty' is not loaded}%
    \renewcommand\transparent[1]{}%
  }%
  \providecommand\rotatebox[2]{#2}%
  \newcommand*\fsize{\dimexpr\f@size pt\relax}%
  \newcommand*\lineheight[1]{\fontsize{\fsize}{#1\fsize}\selectfont}%
  \ifx\svgwidth\undefined%
    \setlength{\unitlength}{922.55406658bp}%
    \ifx\svgscale\undefined%
      \relax%
    \else%
      \setlength{\unitlength}{\unitlength * \real{\svgscale}}%
    \fi%
  \else%
    \setlength{\unitlength}{\svgwidth}%
  \fi%
  \global\let\svgwidth\undefined%
  \global\let\svgscale\undefined%
  \makeatother%
  \begin{picture}(1,0.65418633)%
    \lineheight{1}%
    \setlength\tabcolsep{0pt}%
    \put(0,0){\includegraphics[width=\unitlength,page=1]{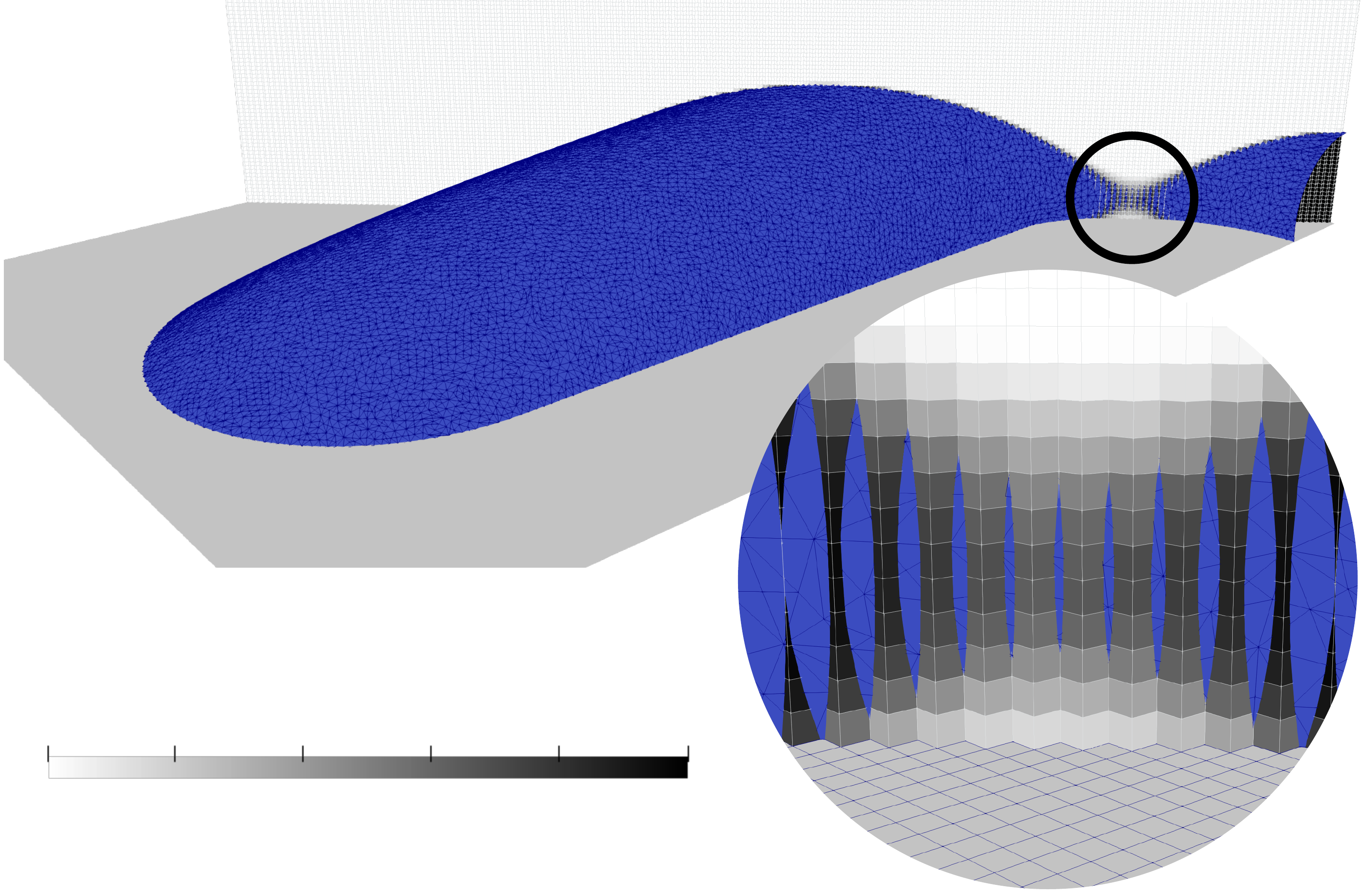}}%
    \put(0.24809907,0.15061447){\makebox(0,0)[lt]{\smash{\begin{tabular}[t]{l}$f$\end{tabular}}}}%
    \put(0.01718314,0.11919682){\makebox(0,0)[lt]{\smash{\begin{tabular}[t]{l}$0.0$\end{tabular}}}}%
    \put(0.47668818,0.11919682){\makebox(0,0)[lt]{\smash{\begin{tabular}[t]{l}$1.0$\end{tabular}}}}%
    \put(0.10865048,0.11919682){\makebox(0,0)[lt]{\smash{\begin{tabular}[t]{l}$0.2$\end{tabular}}}}%
    \put(0.19750036,0.11919682){\makebox(0,0)[lt]{\smash{\begin{tabular}[t]{l}$0.4$\end{tabular}}}}%
    \put(0.29371909,0.11919682){\makebox(0,0)[lt]{\smash{\begin{tabular}[t]{l}$0.6$\end{tabular}}}}%
    \put(0.38299895,0.11919682){\makebox(0,0)[lt]{\smash{\begin{tabular}[t]{l}$0.8$\end{tabular}}}}%
    \put(0,0){\includegraphics[width=\unitlength,page=2]{drop_breakup_1.pdf}}%
  \end{picture}%
\endgroup%

    }
    \caption{Initialization of volume fractions $f$ for the wetting experiment, image adapted from \citep{Hartmann2021}. } 
    \label{fig:smci:breakup}
    \end{subfigure}
    \caption{Simulation of the wetting experiment with the fluid interface given as a triangular surface mesh \citep{Hartmann2021}.}
\end{figure}

Some methods that are surveyed in \cref{sec:intro} can initialize volume fractions from exact implicit surfaces, such as a sphere or an ellipsoid, analyzed in \cref{subsec:sphere-ellips}. One novelty of SMCI/A algorithms is their ability to compute volume fractions from arbitrary surfaces on arbitrary unstructured meshes. For example, volume fractions given by an experimental surface were calculated by the SMCI algorithm in \citet{Hartmann2021} for studying breakup dynamics of a capillary bridge on a hydrophobic stripe between two hydrophilic stripes. In \citep{Hartmann2021}, the experimental setup involves a liquid bridge that is formed between two larger droplets across a hydrophobic stripe. 
The hydrophobic stripe drives the collapse of this liquid bridge, that is observed experimentally and in a simulation in \citep{Hartmann2021}. The quantitative comparison of the simulation and the experiment from \citep{Hartmann2021} is shown in \cref{fig:hartmann:comp}. The experimental surface from \citet{Hartmann2021}, used to initialize volume fractions, is shown in \cref{fig:smci:breakup}. The SMCI algorithm computes the volume vractions of the experimental fluid interface from \citep{Hartmann2021} with the volume error $E_v = 7.789e-06$. As shown in \cref{subsec:sphere-ellips}, the accuracy of the initialization depends on the quality of the surface mesh, not on the resolution of the volume mesh, that is chosen in this case to appropriately resolve the hydrodynamics in \citep{Hartmann2021}.

\subsection{CAD model} 
\label{subsec:cad}
\begin{figure}[!htb]
    \centering
    \begin{subfigure}{0.36\textwidth}
        \includegraphics[width=\textwidth]{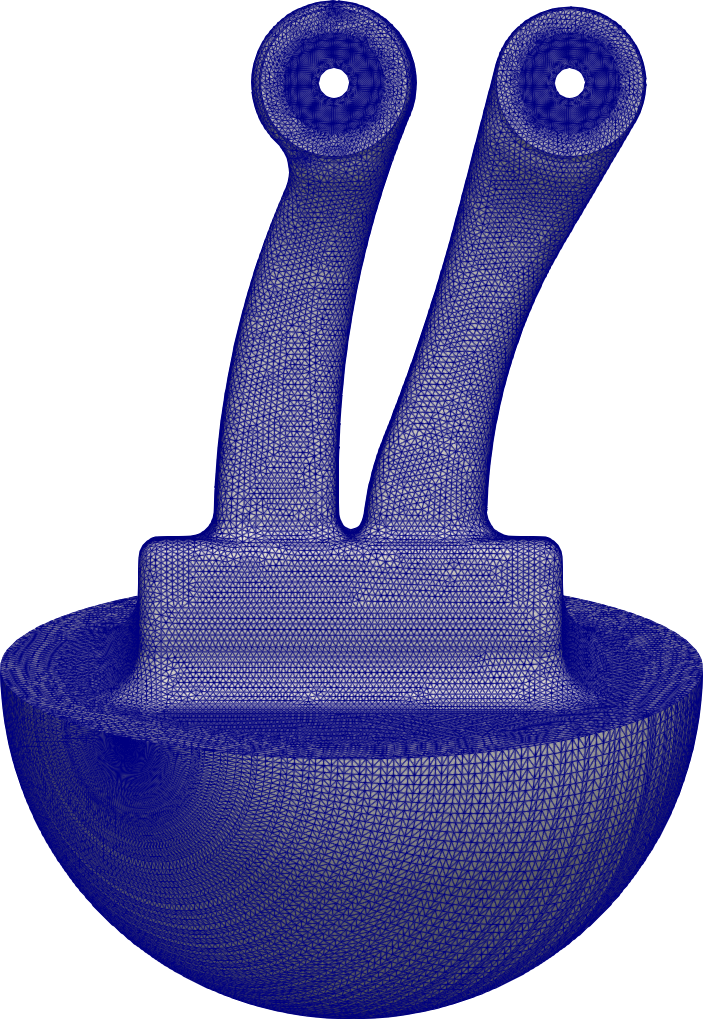}
        \caption{Surface mesh from a CAD model.}
        \label{fig:cad:surfacemesh}
    \end{subfigure}
    \hspace{3em}
    \begin{subfigure}{0.5\textwidth}
        \includegraphics[width=\textwidth]{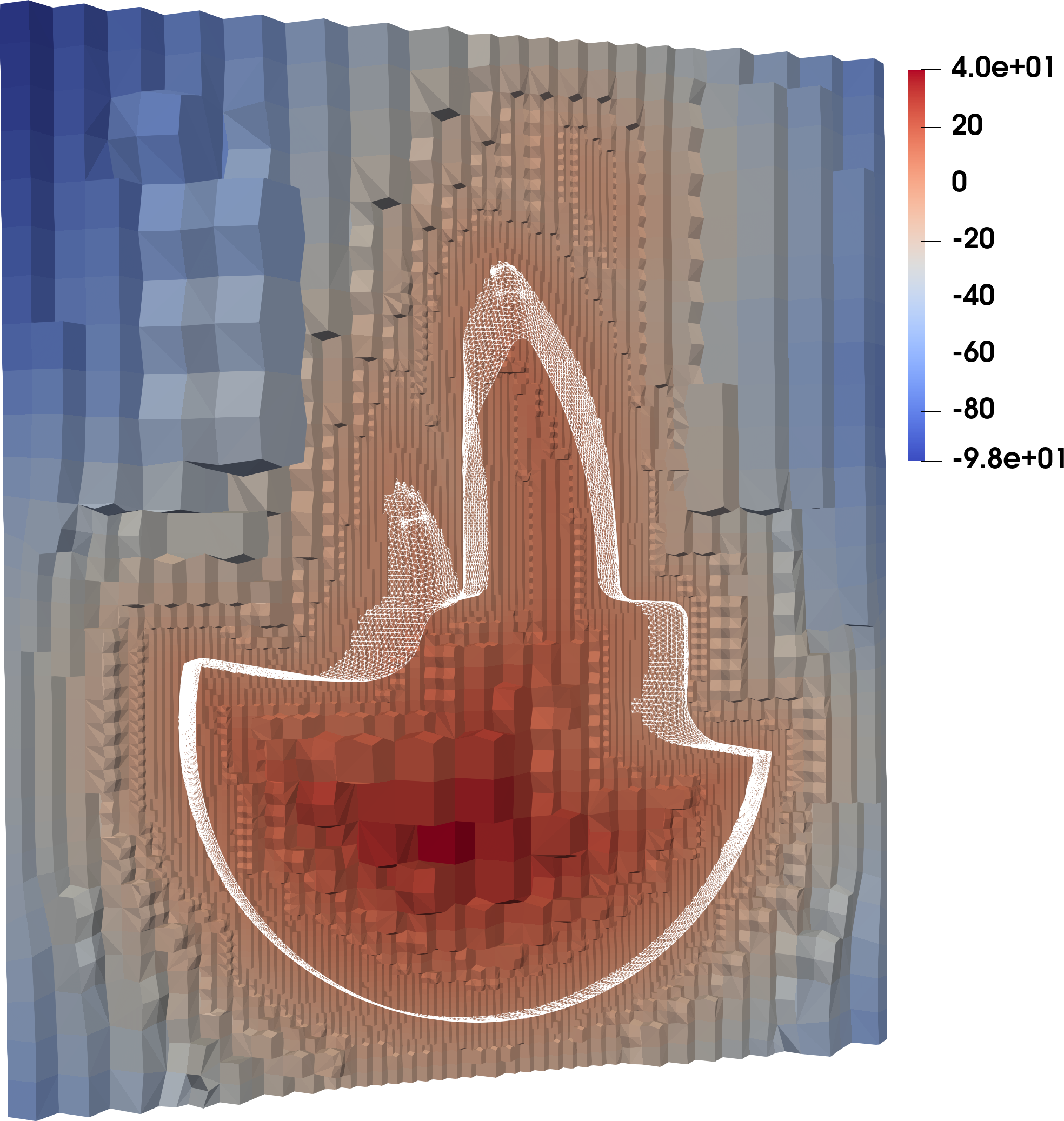}
        \caption{Cross section of the volume mesh with part of the surface mesh,
            colored by signed distance.
        }
        \label{fig:cad:meshcrosssection}
    \end{subfigure}
    \caption{Surface and volume mesh of the CAD model test case.}
    \label{fig:cad:overview}
\end{figure}

To demonstrate that the SMCI/A algorithms are able to handle interfaces more
complex than shown in \cref{subsec:sphere-ellips} and 
\cref{subsec:experiment}, the surface mesh from a CAD model displayed in
\cref{fig:cad:surfacemesh} is used. 
In contrast to the previous interfaces, this one
features sharp edges and geometric features of distinctly different sizes.
The mesh for this test case has been generated with the
\emph{cartesianMesh} tool of cfMesh \cite{cfMesh}. Refinement is used in the
vicinity of the interface. This meshing procedure is chosen to obtain a mesh
that closer resembles that of an industrial application than a uniform cubic
mesh. A cross section of the mesh is depicted in \cref{fig:cad:meshcrosssection}. 
Before examining the computed volume fractions for this case, the signed
distance calculation (\cref{subsec:sigdistcalc}) and sign propagation
(\cref{subsec:sigdistpropagation}) are verified. The presence of sharp edges
(see \cref{fig:cad:surfacemesh}) makes this test case more prone to false 
inside/outside classifications than the others shown so far. 
\begin{figure}[!htb]
    \centering
    \begin{subfigure}{0.35\textwidth}
        \includegraphics[width=\textwidth]{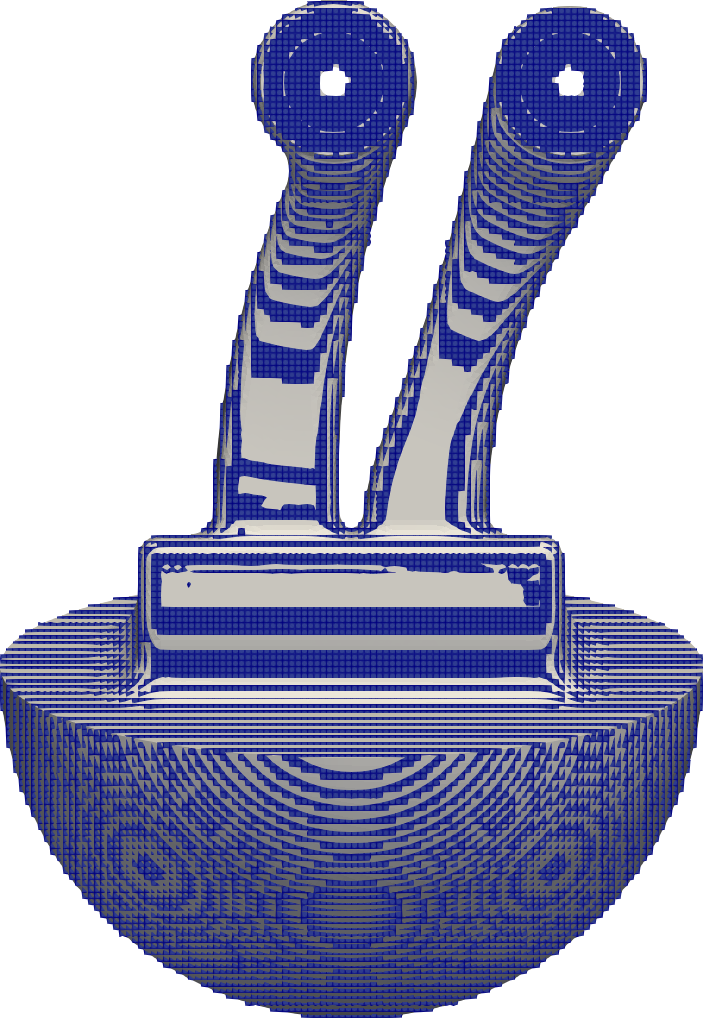}
        \caption{Cells for which $\SignedDistance_c \geq 0$ (blue) overlayed
            with the surface mesh (grey).
        }
        \label{fig:cad:insideoutside}
    \end{subfigure}
    \hspace{3em}
    \begin{subfigure}{0.35\textwidth}
        \includegraphics[width=\textwidth]{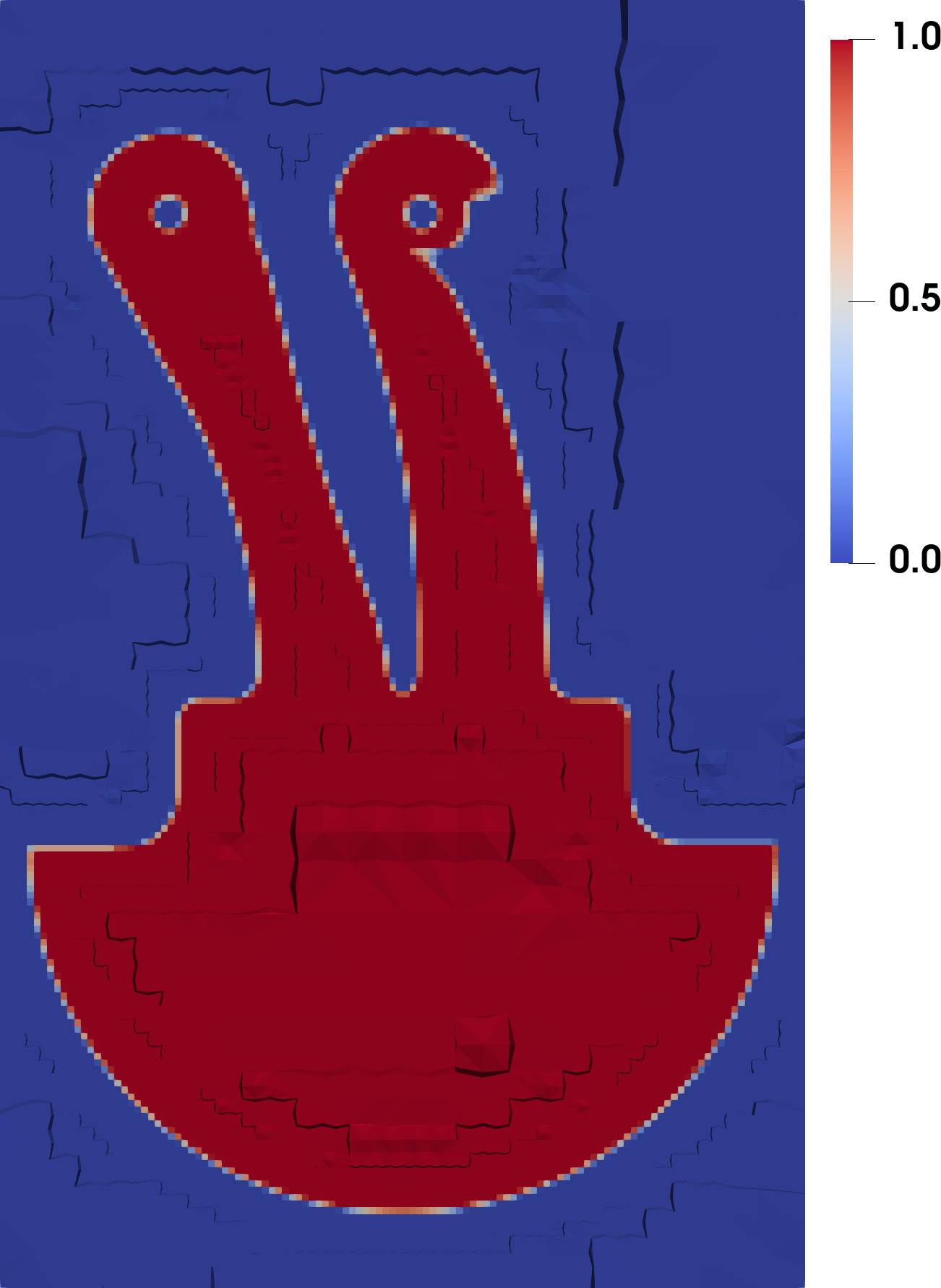}
        \caption{Cross section through the mesh with cells colored by
            volume fraction.
        }
        \label{fig:cad:volumefraction}
    \end{subfigure}
    \caption{Inside/outside computation and resulting volume fractions for the 
        CAD geometry.
    }
    \label{fig:cad:results}
\end{figure}
Yet our procedure yields the correct sign for the distance in all cells as
shown in \cref{fig:cad:insideoutside}. The enclosed volume of the surface mesh is 
considered as $\Omega^+$, thus $\SignedDistance>0$ for all points
$\x \in \Omega^+$. As displayed in \cref{fig:cad:insideoutside} and confirmed by
further manual inspection of the results, the proposed signed distance calculation
correctly classifies all cells within the narrow band and robustly propagates 
this information to the entire domain. This is reflected in the volume fractions as computed, shown in \cref{fig:cad:volumefraction}. Bulk cells are assigned values of either $1$ or $0$, depending on whether 
they are located in $\Omega^+$ or $\Omega^-$ and mixed cells with $0 < \VolFrac_c < 0$ are only found where
the surface mesh is located. Accuracy-wise, the global errors $E_v$ depicted in \cref{fig:cad:convergence}
have been obtained with the \Pvof{} algorithm using different refinement levels. As for the spherical interface
(see \cref{fig:smca:err:sphere:refinement}), second-order convergence is achieved, even though the surface mesh
approximates a non-smooth interface here.
\begin{figure}[!htb]
    \centering
    \includegraphics{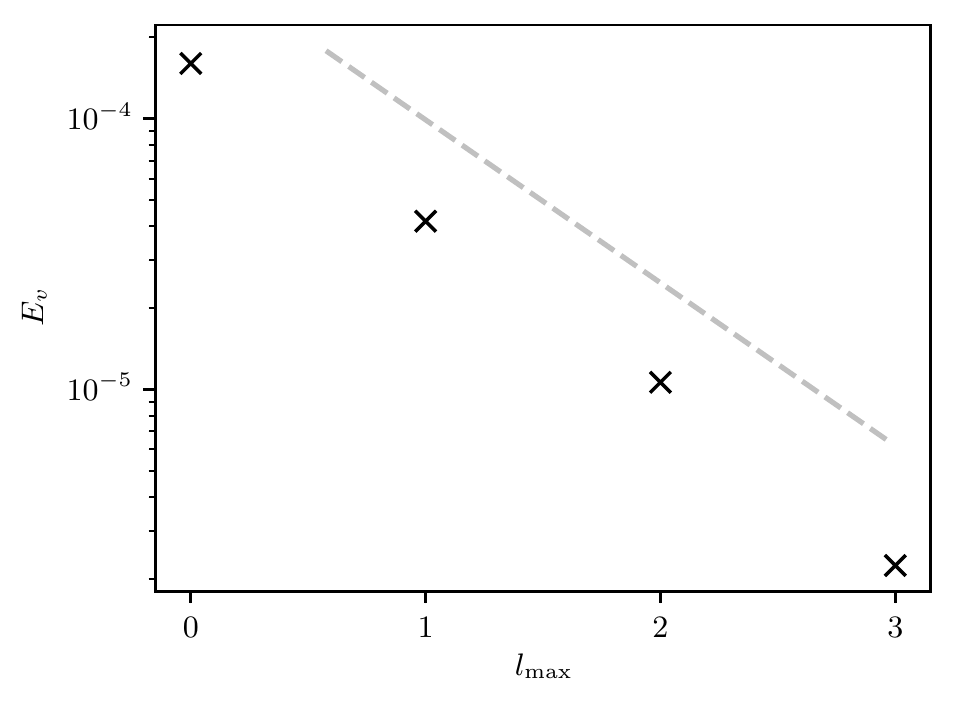}
    \caption{$E_v$ errors of the \Pvof{} algorithm using different refinement levels
        $\Lmax$ for the CAD model with the reference volume $V_e$ computed by \cref{eq:Ve-expanded}.
        The grey dashed line indicates second order convergence.
    }
    \label{fig:cad:convergence}
\end{figure}

\section{Conclusions}
\label{sec:conclusions}

The proposed Surface-Mesh Cell Intersection / Approximation algorithms accurately compute signed distances from arbitrary surfaces intersecting arbitrary unstructured meshes. Geometrical calculations ensure the accuracy of signed distances near the discrete surface. The signed distances (actually their inside / outside information) are propagated into the bulk using the approximate solution of a Laplace equation. Once the signed distances are available in the full simulation domain, the SMCI algorithm computes volume fractions by intersecting arbitrarily-shaped mesh cells with the given surface mesh, while the SMCA algorithm approximates volume fractions using signed distances stored at cell corner points. Both algorithms are robust and show second-order convergence for exact surfaces and arbitrarily shaped surface meshes. The SMCI algorithm scales linearly with a small number of surface triangles per cut-cell. Since a small number of triangles per cell is a requirement for Front Tracking, this linear-complexity makes SMCI an interesting candidate for computing volume fractions in the 3D unstructured Level Set / Front Tracking method \citep{Maric2015, Tolle2020}, which will be the subject of future investigations.

\section{Acknowledgments}

Calculations for this research were conducted on the Lichtenberg high performance computer of the TU Darmstadt.

Funded by the German Research Foundation (DFG)
– Project-ID 265191195 – SFB 1194, Projects B02, B01 and Z-INF.

We are grateful for the discussions on the phase-indicator calculation in the LCRM method on structured meshes \citep{Shin2011} with Prof. Dr. Seungwon Shin, Dr. Damir Juric, and Dr. Jalel Chergui within the project "Initiation of International Cooperations" MA 8465/1-1.

\bibliography{bibliography}

\end{document}